\documentclass[aps,pra,reprint,superscriptaddress,floatfix]{revtex4-2}
\usepackage{amsmath,mathtools}

\usepackage{array}[=2016-10-06]
\usepackage{dcolumn}%
\usepackage{bm}%

\usepackage{braket}
\usepackage{booktabs}  %
\usepackage{wrapfig}
\usepackage[normalem]{ulem}
\usepackage[caption=false]{subfig}
\usepackage{graphicx}
\usepackage{makecell}
\usepackage{adjustbox}
\usepackage{tabularx}
\usepackage{float}

\usepackage{placeins}
\usepackage{booktabs}
\usepackage{multirow}

\usepackage{comment}
\usepackage[pagebackref]{hyperref}
\hypersetup{
    colorlinks,
    linkcolor={red!100!black},
    citecolor={blue!100!black},
}
\usepackage[capitalize,nameinlink,noabbrev]{cleveref}

\usepackage{amssymb}
\usepackage{amsthm}
\usepackage{thmtools}

\Crefname{equation}{Eq.}{Eqs.}
\Crefname{figure}{Fig.}{Figs.}
\Crefname{section}{Sec.}{Secs.}
\Crefname{assumption}{Asm.}{Asms.}
\Crefname{appendix}{App.}{Apps.}

\usepackage{tikz}
\usetikzlibrary{quantikz2,external}
\usetikzlibrary{circuits.logic.US}
\usetikzlibrary{shapes.geometric}

\tikzset{
    iSWAP/.style={
            thickness, %
            filling, %
            draw,
            path picture={%
                    \draw[black]
                    (path picture bounding box.north west) -- (path picture bounding box.south east)
                    (path picture bounding box.south west) -- (path picture bounding box.north east);
                }, %
            inner sep=3.5pt
        }}
\DeclareExpandableDocumentCommand{\iSWAP}{O{}{m}}{|[iSWAP,#1]| \qw}

\definecolor{Joao}{rgb}{0.4,0.3,0.9}
\definecolor{Ale}{rgb}{0.1,0.5,0.5}
\definecolor{Konsta}{rgb}{0.9,0.0,0.0}
\definecolor{Harshit}{rgb}{0.1,0.8,0.2}

\newcommand{\Fid}{\mathcal{F}}                       %
\newcommand{\epsns}{\varepsilon_{\mathrm{ns}}}        %
\newcommand{\epsone}{\varepsilon_{\mathrm{1qb}}}      %
\newcommand{\epstwo}{\varepsilon_{\mathrm{2qb}}}      %
\newcommand{\tquery}{t_{\mathrm{qry}}}                %
\newcommand{\Pavg}{\overline{P}_{\mathrm{dyn}}}       %
\newcommand{\Ppeak}{P^{\max}_{\mathrm{dyn}}}          %
\newcommand{\Edyn}{E_{\mathrm{dyn}}}                  %
\newcommand{\cfgCold}{\textsc{Cold-Heavy}}            %
\newcommand{\cfgGeo}{\textsc{Geometric}}
\newcommand{\cfgLogT}{\textsc{LogT}}

\newtheorem{assumption}{Assumption}

\providecommand{\keywords}[1]{%
    \par\smallskip\noindent\textbf{Keywords:} #1\par\smallskip
}

\makeatletter
\newcommand{\equalnote}{\frontmatter@footnote{All authors contributed equally. Names alphabetically ordered.}}
\makeatother

\begin{document}

\title{The energetic cost of loading classical data in quantum computers}

\author{Joao F. Doriguello\equalnote}
\email{doriguello@renyi.hu}
\affiliation{HUN-REN Alfr\'ed R\'enyi Institute of Mathematics, Budapest, Hungary}

\author{Alessandro Luongo\equalnote}
\email{ale@nus.edu.sg}
\affiliation{Centre for Quantum Technologies, National University of Singapore, Singapore}
\affiliation{Inveriant Pte.\ Ltd., Singapore}

\author{Harshit Verma\equalnote}
\email{h.verma@nus.edu.sg}
\affiliation{Centre for Quantum Technologies, National University of Singapore, Singapore}
\affiliation{MajuLab, CNRS-UCA-SU-NUS-NTU International Joint Research Laboratory, Singapore}

\date{\today}%

\begin{abstract}
	Quantum algorithms for machine learning, chemistry, and cryptanalysis often assume coherent access to classical data at negligible physical cost. We instead treat data loading as a hardware process and quantify the dynamic energetic cost of a quantum random access memory (QRAM) query on a superconducting processor. Using the metric-noise-resource framework, we relate query fidelity and power to gate duration, relative two-qubit drive strength, and the distribution of attenuation across a multi-stage refrigerator, while accounting for thermal photons and gate errors. We analyze the linear-depth multiplexer and the noise-resilient bucket-brigade architectures, adapting the latter to a native superconducting gate set and refining its fidelity analysis using the fully decomposed gate schedule. For a representative bucket-brigade QRAM query with a $12$-qubit address register we obtain a query fidelity of $0.969$. The query consumes $3.51\times10^{-4}\,\mathrm{J}$ of dynamic energy, corresponding to an average dynamic power of $4.38\,\mathrm{W}$ and a peak of $64.4\,\mathrm{W}$. Our optimization shows that, at fixed target fidelity, the operating point and attenuation profile can change the average power by a factor of two or more.
\end{abstract}

\maketitle

\section{Introduction}

Accessing classical data in a quantum computer is key to many algorithms, subroutines, and applications, therefore being worth studying from both theoretical and practical perspectives. For example, quantum algorithms for dihedral hidden subgroup problems~\cite{kuperberg2005subexponential}, collision finding~\cite{brassard1997quantum}, element distinctness~\cite{ambainis2007quantum}, unstructured search~\cite{grover1996fast,grover1997quantum}, optimization~\cite{ambainis2019quantum,van2019quantum,kerenidis2020quantuminterior,Kerenidis2021quantumalgorithms,Augustino2023quantuminterior,huang2022faster,chen2023quantum,Doriguello2025quantumalgorithms}, phase estimation for quantum chemistry~\cite{babbush2018encoding}, pattern recognition and machine learning~\cite{trugenberger2002phase,schutzhold2003pattern,schaller2006quantum,kerenidis2017quantum,kerenidis2020quantum,cornelissen2023you}, state preparation~\cite{grover2002creating}, cryptanalysis~\cite{chailloux2021lattice,doriguello2024practicality}, among others, all require access to classical data.

Translating the algorithmic promise of quantum advantage into a practical speedup requires treating data access as a physical process, executed on real hardware, rather than as an ideal, cost-free oracle: whether an asymptotic advantage survives outside of the oracle model depends strongly on how the underlying device is physically implemented. In this work, we take this physical implementation cost to be the energetic cost, i.e., the energy consumed by the hardware and its supporting infrastructure when performing a query.

Quantum processors operate within a multi-layered stack spanning cryogenic hardware, control electronics, compilation, and user-facing software. In many of these settings, data access is modeled through a quantum random access memory (QRAM)~\cite{giovannetti2008quantum,giovannetti2008architectures,Allcock2024constantdepth,Jaques2025qramsurveycritique}, a memory primitive that enables coherent access to classical data and which, at a high level, performs the unitary map $\ket{i}\ket{0}\mapsto \ket{i}\ket{m_i}$, where $i$ denotes the address of the memory cell being queried, held in a dedicated address register, and $m_i$ the corresponding word stored in memory, i.e., the content located at address $i$.

A physical realization of a QRAM can be regarded as an external memory device sharing registers with the quantum processor. The usefulness of a QRAM architecture is to be determined not only by the asymptotic complexity of the algorithms using it, but also by how circuit depth, active qubits, operating conditions, and classical control apparatus overhead shape the total energetic cost of a query.

\paragraph{Energetics of quantum circuits and the MNR framework.}

The Metric-Noise-Resource (MNR) framework~\cite{Fellous-Asiani2023Oct} is a methodology that jointly models the performance, noise, and resource consumption of quantum devices, and can be used to identify configurations that minimize a resource of interest, e.g., energy or power, for a target level of performance. Building on this perspective, previous works have linked specific hardware configurations to their performance and energetic  profiles~\cite{Chiribella2021fundamental,Fellous-Asiani2023Oct,kotevaetal, boson_energetic, yasser_trapped_ion, yasser_cat_qubits, yasser_rydberg, yasser_spin_qubit}. In this manuscript, we adopt the MNR framework to study the performance and resource cost---in terms of power or energy---of QRAM implementations. Our emphasis is on relating a circuit-level performance metric to the physical resources required to realize a query, so that fidelity and energy can be compared on the same footing across different hardware operating points.

We focus on hardware-specific realizations of QRAM with transmon-based superconducting qubits as a representative low-temperature platform. We model both the physical layer implementing the QRAM, along with the enabling technologies (multi-level cryo-cooling systems, attenuation in the control lines, etc...). We do not assume fault-tolerant construction or architecture, a choice motivated by the noise resilience of several QRAM architectures. Rather than assuming QRAM to be free of correction overheads, we exploit the fact that these architectures exhibit nontrivial noise resilience and may admit milder correction requirements than a generic fault-tolerant treatment would suggest~\cite{arunachalam2015robustness,hann2021resilience,mehta2024analysis,hann2021practicality,dalzell2025distillation}. Our analysis therefore focuses on the energetic cost of a single QRAM query. This cost provides a building block for an end-to-end example later studied related to block-encoding oracles that require repeated QRAM accesses in linear-algebra routines~\cite{BELLANTE2025104341,bellante2022quantum,gilyen2019quantum}.

\paragraph{Related works.}
The QRAM literature includes both architectural proposals and increasingly detailed resource analyses. Foundational works introduced the QRAM abstraction and early circuit constructions \cite{giovannetti2008quantum,giovannetti2008architectures}, while later studies quantified fault-tolerant resource requirements for QRAM queries \cite{di2020fault}, analyzed the circuit resources needed to load or block-encode classical data~\cite{Clader2022blockencode}, and revisited architectural practicality through constant-depth constructions and broader critical surveys~\cite{Allcock2024constantdepth,Jaques2025qramsurveycritique}. In parallel, recent studies have emphasized full-stack resource optimization and explicit space-time-energy accounting for quantum algorithms~\cite{Fellous-Asiani2023Oct,kotevaetal,Tu2025towards}. Our work connects these directions by examining how QRAM circuit structure and hardware operating conditions jointly determine energetic efficiency.
Meier and Yamasaki~\cite{meier2025energy} study computational energy
consumption from a theoretical-computer-science perspective, adopting the standard query-complexity model and deriving energy bounds for quantum
and classical computation. We adopt a related query-complexity viewpoint in
\Cref{sec:applications}, where application-level oracle requirements are
translated into QRAM resource costs. Establishing a formal connection between
their computational energy framework and the hardware-aware MNR model used
here is left for future work.

\subsection{Main results}

\paragraph{Superconducting-optimized bucket-brigade QRAM.}
Our first result is a hardware-aware refinement of the bucket-brigade (BB) QRAM fidelity analysis of Hann \emph{et al.}~\cite{hann2021resilience}. Rather than assigning a uniform error to every circuit layer and conflating one- and two-qubit gates, we evaluate the fidelity bound by distinguishing idle, one-qubit, and two-qubit errors, and track when the routers and their incident modes are acted upon. %
This results in a much more optimized fidelity: under the optimistic assumption that two-qubit gates can reach an infidelity of the order of $10^{-7}$, a QRAM of $12$ address qubits has a fidelity of $\approx 0.969$. By compiling such BB-QRAM circuit on a superconducting transmon-qubit architecture, we reach a depth of $1{,}214$ layers, comprising $207{,}192$ (non-virtual) one-qubit gates and $171{,}750$ two-qubit gates.%

We find that the query infidelities required by the target applications can be achieved within physically plausible parameter ranges for superconducting hardware, as detailed in \Cref{sec:MNR_superc}. The importance of choosing the control parameters is illustrated by comparing two operating points. Here, $g_{\mathrm{fac}}=P_{\mathrm{2qb}}/P_{\mathrm{1qb}}$ is the two-qubit gate power relative to the single-qubit gate power, while $f_{\mathrm{fac}}=\tau_{\mathrm{2qb}}/\tau_{\mathrm{1qb}}$ is their relative duration and, in our model, relative infidelity. Changing $(g_{\mathrm{fac}},f_{\mathrm{fac}})$ from $(1,4)$ to $(16,1)$ reduces the query infidelity from $0.031$ to $0.012$, but increases the average dynamic power from $4.38\,\mathrm{W}$ to $38.2\,\mathrm{W}$, the peak dynamic power from $64.4\,\mathrm{W}$ to $644\,\mathrm{W}$, and the dynamic energy per query from $3.51\times10^{-4}\,\mathrm{J}$ to $1.16\times10^{-3}\,\mathrm{J}$. A less than three-fold reduction of the infidelity thus costs about an order of magnitude in power, which demonstrates the importance of jointly optimizing these parameters within the MNR framework. Comparing cryogenic attenuation profiles further shows that concentrating attenuation at the coldest stages is systematically more expensive than the \cfgGeo{} or \cfgLogT{} allocations.
Moreover, we link the achievable query fidelity ranges to the idling noise parameter, showing that for an order of magnitude change in the idling noise parameter, the power requirement for gate implementation increases dramatically by more than an order of magnitude in many instances, for a (high) target of the query fidelity.

These results are obtained under rather optimistic %
assumptions. The coherence time $t_1 = 1/\gamma = 1\,\text{s}$ of \Cref{tab:fixed_values} anticipates roughly three
orders of magnitude of improvement over the longest reported transmon
lifetimes; the Carnot cooling efficiency of
\Cref{eq:P_dressed} exceeds what dilution refrigerators achieve
at mK temperatures by a comparable factor; $\alpha = 4$ assumes a two-qubit gate of better energy-infidelity quality than the calibration of \Cref{sec:MNR_CR_gates} supports for cross-resonance gates ($\alpha\sim 16$--$200$); and the static per-qubit power is set to zero, so
that the reported cost is the dynamic fraction of the budget rather than
the total.

\section{The MNR Framework}
\label{sec:MNR}
The MNR framework, introduced in~\cite{Fellous-Asiani2023Oct}, provides a systematic methodology for characterizing and optimizing the efficiency of quantum computations by explicitly modeling the interplay between three fundamental components: a user-specified \textbf{M}etric (M) of choice to evaluate the quality of the computation, the \textbf{N}oise (N) characteristics of the quantum computer, and the \textbf{R}esources (R) employed in the computation. The presence of noise---a distinct feature of quantum computing---complicates the relationship between the resource invested and the quality of results, as noise directly affects the fidelity of operations and the overall accuracy of the computation.

The MNR framework enables the analysis of quantum computation across all layers of the quantum stack (which can be applied to the physical, compilation, and control layers, and enabling technologies). It allows for a parametric exploration of (tunable) control parameters $\{C\}$ that influence the Metric, Noise, and Resource (spanning the aforesaid stack), thereby identifying optimal configurations that maximize performance while minimizing resource consumption. Therefore, the MNR framework formalizes the problem as: minimize the resource cost $R(C)$ while ensuring that the performance metric $M(C, N)$ meets or exceeds a predefined target value $M_0$, considering the influence of noise $N$. Mathematically, the problem is expressed as:
\begin{equation}
    \min_C R(C) \quad \text{subject to} \quad M(C,N) \geq M_0.
\end{equation}

\paragraph{Metric.}
The Metric $M$ quantifies the performance or quality of the quantum computation. In the MNR framework, the choice of metric is flexible and depends on the context and layer of the quantum stack being analyzed. At the physical layer, the metric could be gate fidelity, representing the accuracy of individual quantum operations. At the circuit level, it could be circuit fidelity, reflecting the overall accuracy of a sequence of operations. For user-oriented applications, the metric may represent time-to-solution (TTS)~\cite{Ronnow2014defining,king2015benchmarking}, quantum volume~\cite{Moll2018quantum,cross2019validating,Baldwin2022reexaminingquantum,miller2022improved}, or Q-score~\cite{martiel2021benchmarking}, which benchmark the computational power and efficiency of the quantum device. In general, the metric is a function of the control parameters and noise, $M = M(C, N)$. For our purposes, we adopt the query fidelity as the metric of performance.

\paragraph{Noise.}
Noise $N$ is an inherent feature of quantum devices, arising from imperfections in control operations, environmental interactions, and decoherence processes. It directly affects the metric by introducing errors that degrade the fidelity and accuracy of quantum operations. The MNR framework explicitly models noise as a function of the control parameters and other system variables, enabling a quantitative analysis of its impact on performance. Noise is typically described by a noise model, such as depolarizing noise, dephasing noise, or amplitude damping, depending on the physical realization of the quantum hardware. In this context, the control parameters influence both the system's dynamics and the noise characteristics, establishing a complex relationship between $M$, $N$, and $R$. In our context, noise affects the QRAM circuits and decreases the query fidelity.

\paragraph{Resource.}
The Resource $R$ quantifies the cost associated with achieving a target performance metric. They can include physical resources (e.g., the number of physical qubits, the amount of quantum gates, or the circuit depth required to perform a computation), the energy consumption, or similarly the power consumed by the quantum device during operation, including control electronics and cooling systems, but also resources like the \emph{classical computation} required for error correction, or the \emph{wall-clock time}. The MNR framework defines resources as cost functions $R(C)$ that depend on the control parameters. We consider the dynamic resource cost, i.e., the resource cost underlying the implementation of quantum primitives such as gates (supported by the enabling technologies), and neglect the static costs which are typically independent of the application---here QRAM implementation.

\paragraph{Control.}
Control parameters $C$ are the tunable inputs that influence the Metric, Noise, and Resource. They can be parameters that can be adjusted dynamically by the user, such as gate pulse amplitude, gate frequency, qubit temperature, or cooling power, or pre-determined by the hardware fabrication (such as qubit architecture, coupling strengths, and device layout).

\paragraph{Efficiency.}
Efficiency in the MNR framework is a user-defined quantity that represents how costly is to achieve the metric of choice with respect to the resources employed. A common choice of efficiency is therefore the ratio of the performance metric to the total resource cost: $\eta = \frac{M}{R}$. The efficiency allows for the comparison of different quantum computing architectures, algorithms, and noise models.

\section{Superconducting platforms}

Processors based on superconductivity have emerged as one of the main physical platforms to realize a quantum computer and have been deployed by several of the main quantum computing companies~\cite{arute2019quantum,kjaergaard2020superconducting,bravyi2022future,mckay2023benchmarking,acharya2025quantum,abughanem2025ibm}. Qubits in a superconducting platform are implemented using electronic circuits at very low temperature, typically containing Josephson junctions, with designs ranging from transmons~\cite{X_mon_1, X_mon_2} to fluxonium~\cite{Fluxonium_intro}.

It is instructive to look at the following aspects of superconducting platforms for the MNR analysis: (a) native gate set for determining the (compiled) circuit structure, and (b) relation between energy and performance of primitives. We cover these aspects in the subsequent subsections.

\subsection{Native quantum gates}\label{sec:native_gates}

The traditional single-qubit gates in quantum computation~\cite{Nielsen_Chuang_2010} are the Pauli matrices $\mathrm{X},\mathrm{Y},\mathrm{Z}$, the Hadamard gate $\mathrm{H}$, the phase gate $\mathrm{S}$, and the $\mathrm{T}$ gate:
\begin{equation*}
	\begin{gathered}
		\mathrm{X} = \begin{pmatrix}
			0 & 1 \\ 1 & 0
		\end{pmatrix}, \quad
		\mathrm{Y} = \begin{pmatrix}
			0 & -i \\ i & 0
		\end{pmatrix}, \quad
		\mathrm{Z} = \begin{pmatrix}
			1 & 0 \\ 0 & -1
		\end{pmatrix}, \\
		\mathrm{H} = \frac{1}{\sqrt{2}}\begin{pmatrix}
			1 & 1 \\ 1 & -1
		\end{pmatrix}, \quad
		\mathrm{S} = \begin{pmatrix}
			1 & 0 \\ 0 & i
		\end{pmatrix}, \quad
		\mathrm{T} = \begin{pmatrix}
			1 & 0 \\ 0 & e^{i \pi / 4}
		\end{pmatrix}.
	\end{gathered}
\end{equation*}
More generally, any single-qubit can be written as $\mathrm{R}_{\bm{n}}(\theta) = \exp(-i \frac{\theta}{2} (n_x \mathrm{X} + n_y \mathrm{Y} + n_z \mathrm{Z}))$ for a real unit vector $\bm{n} = (n_x,n_y,n_z)$. The typical two-qubit gates, on the other hand, are the $\mathrm{CNOT}$, $\mathrm{CZ}$, $\mathrm{SWAP}$, and $\mathrm{iSWAP}$ gates:
\begin{eqnarray}
	\mathrm{CNOT} &= \begin{pmatrix}
		1 & 0 & 0 & 0 \\
		0 & 1 & 0 & 0 \\
		0 & 0 & 0 & 1 \\
		0 & 0 & 1 & 0
	\end{pmatrix}, \quad
	\mathrm{CZ} = \begin{pmatrix}
		1 & 0 & 0 & 0  \\
		0 & 1 & 0 & 0  \\
		0 & 0 & 1 & 0  \\
		0 & 0 & 0 & -1
	\end{pmatrix}, \nonumber \\
	\mathrm{SWAP} &= \begin{pmatrix}
		1 & 0 & 0 & 0 \\
		0 & 0 & 1 & 0 \\
		0 & 1 & 0 & 0 \\
		0 & 0 & 0 & 1
	\end{pmatrix}, \quad
	\mathrm{iSWAP} = \begin{pmatrix}
		1 & 0 & 0 & 0 \\
		0 & 0 & i & 0 \\
		0 & i & 0 & 0 \\
		0 & 0 & 0 & 1
	\end{pmatrix}. \nonumber
\end{eqnarray}
Regarding three-qubit gates, we mention the Toffoli (controlled-CNOT) and controlled-iSWAP gates:
\begin{align*}
	\text{Toffoli}:|x\rangle|y\rangle|b\rangle & \mapsto |x\rangle\mathrm{CNOT}^x(|y\rangle|b\rangle), \quad x\in\{0,1\},  \\
	\text{C-iSWAP}:|x\rangle|y\rangle|b\rangle & \mapsto |x\rangle\mathrm{iSWAP}^x(|y\rangle|b\rangle), \quad x\in\{0,1\}.
\end{align*}

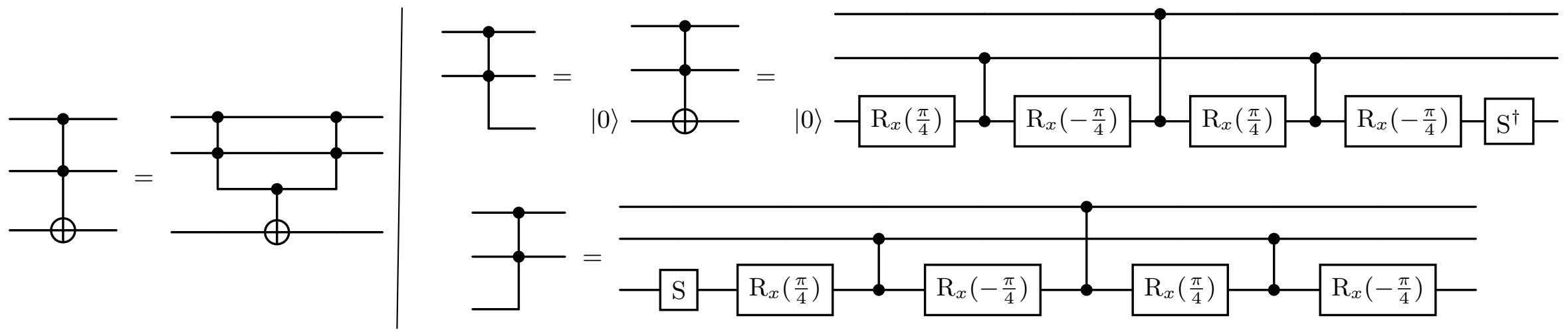
\begin{figure*}[t]
	\centering
	\begin{tikzpicture}
		\draw (-5.95,0.7) -- (-6,-2.5);
		\node[scale=0.8] at (-8,-1) {
			\begin{quantikz}
				& \ctrl{2} & \\
				& \ctrl{0} & \\
				& \targ{} &
			\end{quantikz} = \begin{quantikz}[row sep=0.3cm]
				& \ctrl{2} & & \ctrl{2} & \\ %
				& \ctrl{0} & & \ctrl{0} & \\
				&  \wireoverride{n} & \ctrl{1} &  & \wireoverride{n} \\
				& & \targ{} & &
			\end{quantikz}};
		\node[scale=0.8] at (0,0) {
			\begin{quantikz}[row sep=0.4cm]
				& \ctrl{2} & \\
				& \ctrl{0} & \\[0.18cm]
				& \wireoverride{n} &
			\end{quantikz} = \begin{quantikz}[row sep=0.4cm]
				& \ctrl{2} & \\
				& \ctrl{0} & \\
				\lstick{$\ket{0}$} & \targ{} &
			\end{quantikz} = \begin{quantikz}[row sep=0.4cm, column sep=0.3cm]
				& & & & \ctrl{2} & & &&&\\ %
				& & \ctrl{1} & & &  &  \ctrl{1} &&& \\
				\lstick{$\ket{0}$} & \gate{\mathrm{R}_x(\frac{\pi}{4})} & \ctrl{0} & \gate{\mathrm{R}_x(-\frac{\pi}{4})} & \ctrl{0} & \gate{\mathrm{R}_x(\frac{\pi}{4})} & \ctrl{0} & \gate{\mathrm{R}_x(-\frac{\pi}{4})} & \gate{\mathrm{S}^\dagger} & %
			\end{quantikz}};
		\node[scale=0.8] at (-0.3,-1.8) {
			\vspace{0.5cm}
			\begin{quantikz}[row sep=0.4cm]
				& \ctrl{2} & \\
				& \ctrl{0} & \\[0.18cm]
				&  & \wireoverride{n}
			\end{quantikz} = \begin{quantikz}[row sep=0.4cm, row sep=0.25cm]
				& & & & & \ctrl{2} & & & &  \\
				& & & \ctrl{1} & & & & \ctrl{1} & & \\
				& \gate{\mathrm{S}} & \gate{\mathrm{R}_x(\frac{\pi}{4})} & \ctrl{0} & \gate{\mathrm{R}_x(-\frac{\pi}{4})} & \ctrl{0} & \gate{\mathrm{R}_x(\frac{\pi}{4})} & \ctrl{0} & \gate{\mathrm{R}_x(-\frac{\pi}{4})} &
			\end{quantikz}};
	\end{tikzpicture}
	\caption{Decomposition for the Toffoli gate into native gates using AND-gadgets. The target qubit in the AND-gadgets is usually drawn as a ``corner'' in order to highlight that it will be (or was) OFF after (before) the operation.}
	\label{fig:toffoli}
\end{figure*}

Within superconducting platforms, single-qubit gates are usually performed via microwave irradiation, which couples to the qubit's transition frequency to drive Rabi oscillations in the computational subspace. By controlling the phase and amplitude of the drive, a rotation about an arbitrary axis in the $x$-$y$ plane can be implemented. On the other hand, $z$-axis rotations $\mathrm{R}_z(\theta)$ are performed in a virtual manner, i.e., as a classical update of the phase reference, $\phi \mapsto \phi + \theta$,  applied to all subsequent $x$-$y$-plane drives rather than as a physical pulse on the qubit~\cite{McKay2017efficientZ}. Consequently, $R_z(\theta)$ has zero duration and, within our resource model, zero energetic cost.
Single-qubit gates are now routinely reported with fidelities $\gtrapprox 99.9\%$~\cite{Barends2014superconducting,reagor2018demonstration}. Superconducting implementations for two-qubit gates, on the other hand, vary widely---from cross-resonance (CR) gates~\cite{Chow2011,Rigetti2010} to direct $\mathrm{CZ}$~\cite{Barends2014superconducting,CZ_gate_2,kjaergaard2020superconducting,Chen2014qubit}, $\mathrm{CNOT}$~\cite{Rosenblum2018cnot} or $\sqrt{\mathrm{iSWAP}}$ gates~\cite{Dewes2012characterization,McKay2016universal}, and tunable-coupler schemes~\cite{tunable_coupler_1,Sung2021}---each with distinct performance-cost characteristics.

In some commercially available superconducting processors~\cite{Ganzhorn2020benchmarking,abughanem2025ibm}, the native gate set comprises $\mathrm{R}_z(\theta)$, $\mathrm{X}$, $\mathrm{R}_x(\frac{\pi}{2})$, and $\mathrm{CZ}$ gates. Other companies~\cite{arute2019quantum} use a gate set that includes $\mathrm{R}_{x,y}(\theta_x,\theta_y)$, $\mathrm{R}_z(\theta)$, and the two-qubit tunable gate
\begin{equation*}
	\mathrm{fSim}(\theta,\phi) = \begin{pmatrix}
		1 & 0            & 0            & 0         \\
		0 & \cos\theta   & i \sin\theta & 0         \\
		0 & i \sin\theta & \cos\theta   & 0         \\
		0 & 0            & 0            & e^{i\phi}
	\end{pmatrix},
\end{equation*}
which includes $\mathrm{CZ}$ and $\mathrm{iSWAP}$ gates as special cases. Another superconducting platform~\cite{kusyk2021survey} provides a native gate set with $\mathrm{R}_z(\theta)$, $\mathrm{X}$, $\mathrm{R}_x(\pm\frac{\pi}{2})$, $\mathrm{CZ}$, and the two-qubit tunable gate
\begin{equation*}
	\mathrm{XY}(\theta) = \begin{pmatrix}
		1 & 0            & 0            & 0 \\
		0 & \cos\theta   & i \sin\theta & 0 \\
		0 & i \sin\theta & \cos\theta   & 0 \\
		0 & 0            & 0            & 1
	\end{pmatrix},
\end{equation*}
which includes the $\mathrm{iSWAP}$ gate. Based on this information, we make the following assumption.
\begin{assumption}\label{ass:native_gates}
	Our native gate set for the superconducting platform is $\{\mathrm{R}_z(\theta), \mathrm{R}_x(\theta), \mathrm{CZ}, \mathrm{iSWAP}, \mathrm{iSWAP}^\dagger\}$. Furthermore, the gates $\mathrm{R}_z(\theta)$ are implemented virtually.%
\end{assumption}

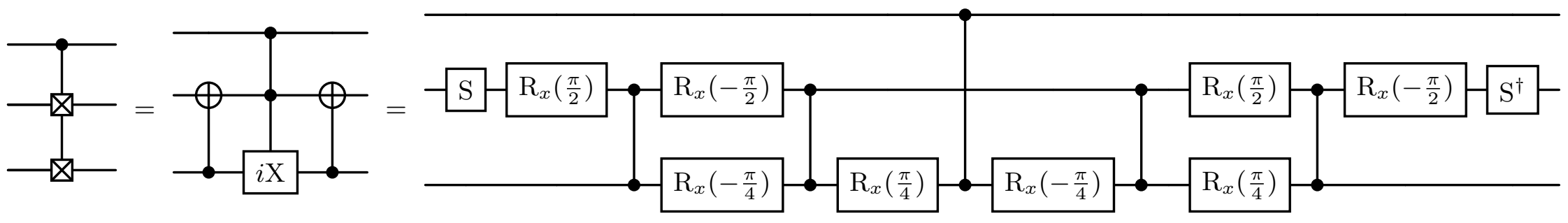
\begin{figure*}[t]
	\centering
	\begin{tikzpicture}
		\node[scale=0.8] {
			\begin{quantikz}
				& \ctrl{2} & \\
				& \iSWAP{} & \\
				& \iSWAP{} &
			\end{quantikz} =
			\begin{quantikz}[column sep=0.25cm]
				& & \ctrl{2} & &\\
				& \targ{} & \ctrl{0} & \targ{} & \\
				& \ctrl{-1} & \gate{i\mathrm{X}} & \ctrl{-1} &
			\end{quantikz} =
			\begin{quantikz}[column sep=0.25cm]
				& & & & & & & \ctrl{2} & & & & & & & & \\
				& \gate{\mathrm{S}} & \gate{\mathrm{R}_x(\frac{\pi}{2})} & \ctrl{1} & \gate{\mathrm{R}_x(-\frac{\pi}{2})} & \ctrl{1} & & & & \ctrl{1} & & \gate{\mathrm{R}_x(\frac{\pi}{2})} & \ctrl{1} & \gate{\mathrm{R}_x(-\frac{\pi}{2})} & \gate{\mathrm{S}^\dagger} & \\
				& & & \ctrl{0} & \gate{\mathrm{R}_x(-\frac{\pi}{4})} & \ctrl{0} &  \gate{\mathrm{R}_x(\frac{\pi}{4})} & \ctrl{0} &  \gate{\mathrm{R}_x(-\frac{\pi}{4})} & \ctrl{0} & & \gate{\mathrm{R}_x(\frac{\pi}{4})} & \ctrl{0} & & &
			\end{quantikz}};
	\end{tikzpicture}
	\caption{Decomposing the controlled-iSWAP gate ($\boxtimes$------$\boxtimes$) into native gates. The gadget is symmetric under the exchange of its two targets, so the outer pair $\mathrm{S}\,\mathrm{R}_x(\frac{\pi}{2})\cdots\mathrm{R}_x(-\frac{\pi}{2})\,\mathrm{S}^\dagger$ may be placed on either of them; the bucket-brigade compilation uses this freedom to cancel outer rotations between consecutive gadgets (\Cref{app:bb_counts}).}
	\label{fig:controlled_iswap}
\end{figure*}

All $n$-qubit gates need to be decomposed into the native gates from \Cref{ass:native_gates}. In particular, the Hadamard gate can be implemented as $\mathrm{H} = \mathrm{S}\cdot \mathrm{R}_x(\frac{\pi}{2})\cdot \mathrm{S}$, while an open-control $\mathrm{CZ}$ gate can be implemented as
\begin{align*}
	(\mathrm{X}\otimes \mathrm{I})\mathrm{CZ}(\mathrm{X}\otimes \mathrm{I}) = |0\rangle\langle 0|\otimes \mathrm{Z} + |1\rangle\langle 1|\otimes \mathrm{I}
	= \mathrm{CZ}(\mathrm{I}\otimes \mathrm{Z}),
\end{align*}
i.e., without the need for Pauli-$\mathrm{X}$ gates. Likewise, the $\mathrm{Z}^{m_i}$ gates that read the memory in the bucket-brigade circuit (\Cref{subsec:bb-cswaps}) and the two $\mathrm{S}$ gates of each Hadamard are virtual, so that reading the memory costs no physical gate and a Hadamard costs a single $\mathrm{R}_x(\frac{\pi}{2})$.
The Toffoli gate, on the other hand, can be implemented using the AND-gadgets from~\cite{Jones2013low,Gidney2018halvingcostof,babbush2018encoding} as shown in \Cref{fig:toffoli}. The AND-gadget simply computes the product $xy$ of two input qubits $|x\rangle$ and $|y\rangle$ using $4$ $\mathrm{R}_x(\pm\frac{\pi}{4})$ gates (equivalent to $\mathrm{T}$ gates) in depth $8$. While there is a measurement-based uncomputation scheme for the AND-gadget~\cite{Jones2013low}, in this paper we do not consider measurements in the energetic analysis for simplicity. Finally, the controlled-iSWAP gate can be implemented according to \Cref{fig:controlled_iswap} by adapting a gadget from~\cite{Selinger2013quantum}.

When studying circuit complexity one is usually concerned with the circuit's size (total number of gates) and its depth. However, in order to analyze the energetic profile of QRAM circuits we require extra measures summarized in \Cref{tab:measurecomplexity} like fidelity and the number of gates at each layer or time step. These metrics allow for a better understanding of the power profile and energetic consumption of a quantum circuit.
\begin{table}[t]
	\centering
	\small
	\setlength{\tabcolsep}{3pt}
	\caption{Circuit metrics used to characterize the energetic profile of a QRAM circuit.}
	\label{tab:measurecomplexity}
	\begin{tabularx}{\columnwidth}{@{}l|X@{}}
		\toprule
		\textbf{Metric} & \textbf{Description}                                                                                                    \\
		\midrule \midrule

		Size            & Number of one- and two-qubit gates                                                                                      \\
		\hline
		Depth           & Minimum number of sequential layers into which the circuit can be partitioned                                           \\
		\hline
		Qubit count     & Total qubits used in the circuit                                                                                        \\
		\hline
		Gates at depth  & Number of gates that are executed at a given depth                                                                      \\
		\hline
		Qubits at depth & Number of qubits that are acted upon by a gate in a given depth                                                         \\
		\hline
		Fidelity        & $F(|\psi\rangle,\rho) = \langle \psi|\rho|\psi\rangle$ between the ideal state $|\psi\rangle$ and the real state $\rho$ \\
		\bottomrule
	\end{tabularx}
\end{table}

\subsection{Energetics of superconducting platform}
\label{sec:MNR_superc}

A quantum computer consists of a full technological stack covering compilation, cooling, post-processing, and the physical implementation of computational primitives. Rather than adopting a full-stack approach, we focus on the energetics of implementing computational primitives on a superconducting platform and the tightly integrated enabling technologies that support it, while contributions such as compilation are not considered. Thus, the energetic analysis assumes that compilation to native gates shown in \Cref{sec:native_gates} is outside its scope.

Among these enabling technologies, the cooling system plays a central role, as
it must continuously evacuate heat generated by implementing quantum primitives (idling, gates) to maintain the temperatures at which coherence persists.  The cryogenic energy budget has two distinct components: a static
contribution that is independent of the computation being performed, and a
dynamic contribution---the cost of cooling to implement
primitives---governed by physical parameters such as pulse
duration, coupling strength, etc. The latter contribution can, in principle, be optimized through the choice of physical control parameters, such as the ones given above. We
therefore focus on the latter contribution as it offers actionable handles
on the overall energy expenditure.

Before developing the model for this dynamic contribution, we first state explicitly how the static contribution is treated.

\subsubsection{Static power consumption}
\label{sec:static}
Enabling subsystems such as those for signal generation, amplification, and cable thermal conduction contribute a per-qubit static resource cost that is independent of the computation being performed. Heat conduction along control and readout cables is a representative example; such costs are fixed for a given architecture and can only be reduced through structural redesign, e.g., by reducing the number of cables. These are set to zero throughout.

Estimates place the static contribution at roughly 50--75\% of the total
wall-plug power in full-stack models of superconducting fault-tolerant
machines~\cite{Fellous-Asiani2023Oct}, with the always-on control
electronics alone accounting for approximately $1\,\text{mW}$ of a total
$1.3$--$2\,\text{mW}$ per physical qubit. Similarly, other platforms such as photonic Boson Sampling architecture~\cite{boson_energetic} exhibit a domination of static over dynamic costs in near-term devices.
However, since the static contribution does not
constitute an optimizable parameter (without extensive redesign or
architectural modification), we adhere to the gate-based model introduced in
\Cref{sec:MNR_superc}, which focuses on the dynamic resource cost.

\subsubsection{Dynamic power consumption}
Following~\cite{Fellous-Asiani2023Oct, boson_energetic}, we adopt a simple
model that relates the device's energetic footprint directly to the
energetics of individual primitives such as a gate. Such power is termed as dynamic power consumption, on account of its root in the execution of computational primitives---it is incurred only while gates are actively driven, and therefore scales with the number and duration of operations in the circuit.

For a given primitive such as a quantum gate, %
the dressed power consumption $P_{\mathrm{drs}}$, which accounts for cooling overhead, is
\begin{eqnarray}\label{eq:cooling_relation}
	P_{\mathrm{drs}} = \frac{A \times P_{\mathrm{prim}}}{\text{Cooling efficiency}},
\end{eqnarray}
where $A$ represents the attenuation of the control line used for implementing the gate, the cooling efficiency is associated with the pump that evacuates the heat generated, and $P_{\mathrm{prim}}$ denotes the driving power required for implementing the computation primitive on the level of the quantum processor. Note that while $P_{\mathrm{prim}}$ might be very small, the power required to ensure $P_{\mathrm{prim}}$ is available on the level of the processor is amplified by the attenuation factor and the cooling efficiency of the heat pump. For a circuit consisting of many gates, say $N_{\mathrm{g}}$, the above relation can be scaled to obtain the  total energetic footprint by replacing $P_{\mathrm{prim}}$ with $N_{\mathrm{g}} \times P_{\mathrm{gate}}$. Note that the same expression can also be used for calculating the energy consumption by multiplying it with the relevant timescale of implementing the primitive.

Assuming the efficiency of the cooler evacuating the heat to be the theoretical upper bound, i.e., equal to Carnot efficiency, the dressed power consumption can be written as
\begin{align}\label{eq:P_dressed}
	P_{\mathrm{drs}} (P_{\mathrm{prim}}, T_{\mathrm{qb}}, A) = A \times P_{\mathrm{prim}}\times\left(\frac{T_{\mathrm{ext}}}{T_{\mathrm{qb}}}-1 \right),
\end{align}
where $T_{\mathrm{ext}}$ is the ambient temperature and $T_{\mathrm{qb}}$ is the temperature of the qubits---essentially the operating temperature of the quantum device. Note that these are the temperatures between which a pump evacuating the heat operates.

Taking the cooler to operate at Carnot efficiency is a very optimistic setting as current dilution refrigerators fall short of this bound by three to four orders of magnitude
at mK temperatures and increasingly so as the temperature
decreases.

\Cref{eq:P_dressed} corresponds to an idealized cryostat with a single attenuator; it serves as the limiting case ($K=2$) of the multi-stage expression derived next, which is the form used in all subsequent calculations.
Cooling to sub-Kelvin temperatures is typically achieved through a
multi-stage cryostat, where each stage $i$ operates at a characteristic
temperature $T_i$. While the coldest temperature is maintained at the qubit stage, the highest stage operates at the room temperature. The control line passes through these stages sequentially,
with $A_i$ denoting the attenuation applied between stages $i$ and $i+1$.
The total attenuation of the control line is then the cumulative product
across all stages. Following~\cite{Fellous-Asiani2023Oct}, this staged
picture can be incorporated into \Cref{eq:P_dressed} via the
following modification:
\begin{equation}\label{eq:stages_stack}
	P_{\mathrm{drs}}(P_{\mathrm{prim}}, \{T_i\}, \{A_i\}) = P_{\mathrm{prim}}
	\sum_{i=1}^{K-1} \frac{T_{\mathrm{ext}} - T_i}{T_i}
	(\mathcal{A}_i - \mathcal{A}_{i-1}),
\end{equation}
where $K$ is the total number of cooling stages and $\mathcal{A}_i =
	\prod_{j=1}^{i} A_j$ is the cumulative attenuation from the qubit stage up
to stage $i$, and $\mathcal{A}_0 := 0$, so that for $K=2$ the sum reduces to \Cref{eq:P_dressed}. While this model
omits certain fixed per-qubit costs, it captures the two key classes of
parameters relevant to our analysis: the operational parameter
$P_{\mathrm{prim}}$, which reflects the cost of the primitive itself, and the
architectural parameters $\{A_i\}$ and $\{T_i\}$, which characterize the
cryogenic stack and are, in principle, optimizable.
We now introduce the functional form of $P_{\mathrm{prim}}$ that can be used in \Cref{eq:stages_stack}.

\subsubsection{Single-qubit gates}
\label{subsec:single_qubit_gates}
A fundamental trade-off exists between gate performance and resource cost
in quantum computing, yet models explicitly linking these trade-offs to
tunable physical parameters remain scarce. The modeling of such trade-off is further compounded by the diversity of superconducting architectures. Here, we
adopt a simplified model for transmon-based superconducting qubits
presented in~\cite{Fellous-Asiani2023Oct}. The model treats spontaneous
emission as the dominant decoherence mechanism, characterized by a decay
rate $\gamma$, and assumes the drive amplitude couples to the qubit
through the same dissipative channel, giving $\Omega = \sqrt{4\gamma P/\hbar\omega}$ where $\Omega$ is the Rabi frequency and $\omega$ is the
qubit frequency. Under these assumptions, the power consumed by a
single-qubit $\pi$-rotation is given in~\cite{Fellous-Asiani2023Oct}:
\begin{equation}\label{eq:single_qubit_power}
	P_{\pi}(\tau_{\mathrm{1qb}}) = \frac{\hbar\omega\pi^2}{4\gamma
		\tau_{\mathrm{1qb}}^2},
\end{equation}
where $\tau_{\mathrm{1qb}}$ is the gate time. We take the average power of all single-qubit gates (native gates in \Cref{sec:native_gates}) to be the same, and therefore set $P_{\mathrm{1qb}} = P_\pi$ uniformly. Furthermore, the corresponding gate
infidelity is
\begin{equation}\label{eq:single_qubit_infidelity}
	\epsone(\tau_{\mathrm{1qb}}, T_{\mathrm{qb}})
	= \gamma\tau_{\mathrm{1qb}}(1+\bar{n}),
\end{equation}
where $\bar{n}$ denotes the number of thermal photons directly affecting the achievable fidelity and $\tau_{\mathrm{1qb}}$ the gate duration. It is useful to note here that a trade-off exists between the gate power and
infidelity, and can be clearly seen from
\Cref{eq:single_qubit_power,eq:single_qubit_infidelity}. If the power is
increased, for instance by decreasing the gate duration $\tau_{\mathrm{1qb}}$,
the infidelity decreases. Writing $E_{\mathrm{1qb}} = P_{\mathrm{1qb}}\tau_{\mathrm{1qb}}$ for the energy consumed per
gate, the trade-off takes a particularly simple form,
\begin{equation}\label{eq:1qb_invariant}
	E_{\mathrm{1qb}} \epsone = \frac{\hbar \omega \pi^{2}}{4}(1 + \bar{n}),
\end{equation}
which is independent of both the gate duration $\tau_{\mathrm{1qb}}$ and the decay rate $\gamma$.
The gate duration therefore fixes where along this curve the gate is
operated, not the energy required for a given accuracy, which is set by
$\omega$ and $\bar{n}$ alone.

The role of the control line is central to the above expressions as the thermal
photons propagating along it degrade gate fidelity, and thus,
attenuation is used to suppress this exposure. For a single attenuator with factor $A$ placed between the ambient environment at temperature $T_{\mathrm{ext}}$ and the qubit at temperature $T_{\mathrm{qb}}$, the effective thermal photon number is
\begin{equation}
	\bar{n} = \frac{(A-1)\,n(T_{\mathrm{qb}})}{A} + \frac{n(T_{\mathrm{ext}})}{A},
	\label{eq:eff_thermal}
\end{equation}
where $n(T) = 1/\bigl(e^{\hbar\omega/k_BT}-1\bigr)$ is the Bose-Einstein
distribution.
This single-attenuator expression is the $K=2$ special case of the multi-stage expression below (one attenuator between two temperature stages).

For a multi-stage setup, while the dressed power consumption is governed
by \Cref{eq:stages_stack} with $P_{\mathrm{prim}}$ from \Cref{eq:single_qubit_power}, the fidelity expression in
\Cref{eq:single_qubit_infidelity} must be revised. The effective
number of noise photons seen by the qubit in a multi-stage setup becomes
\begin{equation}
	\bar{n} = n_{\mathrm{BE}}(T_1)
	+ \sum_{i=1}^{K-1}
	\frac{n_{\mathrm{BE}}(T_{i+1}) - n_{\mathrm{BE}}(T_i)}{\mathcal{A}_i},
	\label{eq:effective_noise}
\end{equation}
where $T_1 = T_{\mathrm{qb}}$ places the qubit at the coldest stage and again $\mathcal{A}_i = \prod_{j=1}^{i} A_j$. Thus, the gate infidelity is obtained from \Cref{eq:single_qubit_infidelity} by using $\bar{n}$ from \Cref{eq:effective_noise} .
While it is desirable to concentrate attenuation at the colder stages to suppress $\bar{n}$, the Carnot factor
$(T_{\mathrm{ext}} - T_i)/T_i$ in \Cref{eq:stages_stack} grows orders of magnitude larger at lower temperatures (at lower $i$), making such attenuation energetically costly.

We adopt this formalism---treating thermal photons from the control line
as the main noise source---as a zeroth-order approximation, acknowledging
that other error mechanisms are neglected. However, these simplifications
preserve the essential trade-off structure: faster gates reduce exposure
to thermal noise and improve fidelity, but at the cost of increased power
consumption.

\subsubsection{Two-qubit gates}
\label{subsec:two_qubit_gates}
Multiple hardware approaches exist for realizing two-qubit gates in superconducting processors, each with distinct performance-resource cost characteristics. While it is desirable to derive the metric-noise-resource model for multi-qubit gates such as in \Cref{eq:single_qubit_infidelity,eq:single_qubit_power}, variety of implementations motivates a more abstract approach. Here, we introduce the following relations for the power and infidelity of two-qubit gates in terms of the same quantities associated with a single-qubit gate:
\begin{equation}\label{eq:two_qb_facs}
	\begin{aligned}
		P_{\mathrm{2qb}} = g_{\mathrm{fac}} P_{\mathrm{1qb}}, \\
		\epstwo = f_{\mathrm{fac}} \epsone.
	\end{aligned}
\end{equation}
Here, $P_{\mathrm{2qb}}$ and $ \epstwo$ refer to the two-qubit power consumption and infidelity, respectively. In order to preserve the trade-off between fidelity and power explicitly laid out for single-qubit gates in \Cref{eq:single_qubit_infidelity,eq:single_qubit_power}, we now relate $g_{\mathrm{fac}}$ and $f_{\mathrm{fac}}$ for two-qubit gates.

The two factors introduced in \Cref{eq:two_qb_facs} above are not independent. Treating them as independent would admit two-qubit gates that are simultaneously arbitrarily cheap (energetically) and arbitrarily accurate, which no physical implementation permits. Typically, accuracy is increased with drive strength, and drive strength is what is dissipated, contributing to the resource cost. We therefore impose the constraint
\begin{equation}\label{eq:gfac_ffac_link}
	g_{\mathrm{fac}}\, f_{\mathrm{fac}}^2 = \alpha^2 ,
\end{equation}
so that a single hardware parameter $\alpha$ determines $f_{\mathrm{fac}}$ from
$g_{\mathrm{fac}}$, which is the parameter we take to be free. Note that the model is physically meaningful only for
$g_{\mathrm{fac}} \leq \alpha^{2}$, which is equivalent to $f_{\mathrm{fac}} \geq 1$.
This simply means that a two-qubit gate is slower,
therefore less accurate, than a single-qubit gate.

\Cref{eq:gfac_ffac_link} is the constraint behind any gate that requires a control amplitude, here called $\lambda$, and whose effective two-qubit interaction rate is therefore, to leading order, linear in $\lambda$, while the dissipated power is quadratic in $\lambda$. The gate duration therefore obeys $\tau_{\mathrm{2qb}} \propto 1/\lambda$ and the power $P_{\mathrm{2qb}} \propto \lambda^2$, so that $P_{\mathrm{2qb}} \propto \tau_{\mathrm{2qb}}^{-2}$. The single-qubit gate obeys the same scaling, since its Rabi frequency is likewise linear in the drive amplitude, and \Cref{eq:single_qubit_power} gives $P_{\mathrm{1qb}} \propto \tau_{\mathrm{1qb}}^{-2}$. The ratio of the two powers is therefore fixed up to a constant prefactor,
\begin{equation}\label{eq:gfac_duration}
	g_{\mathrm{fac}} = \alpha^2 \left(\frac{\tau_{\mathrm{1qb}}}{\tau_{\mathrm{2qb}}}\right)^2 ,
\end{equation}
where the dimensionless constant $\alpha^2$ connects the proportionality constants
of the two relations. Equivalently, $\alpha$ is the duration ratio $\tau_{\mathrm{2qb}}/\tau_{\mathrm{1qb}}$ at which the two gates would consume equal power. We also assume that the same decoherence mechanism governs the error incurred during one- and two-qubit gate operations such that the infidelity accumulates at the same rate. Formulating this phenomenology, we obtain
\begin{equation}\label{eq:ffac_duration}
	f_{\mathrm{fac}} = \frac{\tau_{\mathrm{2qb}}}{\tau_{\mathrm{1qb}}} ,
\end{equation}
and eliminating the duration ratio between \Cref{eq:gfac_duration,eq:ffac_duration} returns \Cref{eq:gfac_ffac_link}.

\Cref{eq:gfac_ffac_link} is also the condition under which the two-qubit gate
inherits the energy-infidelity trade-off of the single-qubit gate. Using
\Cref{eq:two_qb_facs,eq:gfac_ffac_link}, the product of
\Cref{eq:1qb_invariant} for the two-qubit gate is
\begin{align*}
	E_{\mathrm{2qb}} \epstwo = \alpha^{2}\, E_{\mathrm{1qb}} \epsone,
\end{align*}
which is likewise independent of gate durations and $g_{\mathrm{fac}}$. The parameter $\alpha^{2}$ can therefore be interpreted as the energy-infidelity quality of the two-qubit gate relative to the single-qubit gate, and it is the single number by which the hardware approaches are distinguished in this model.

The assumptions underlying \Cref{eq:gfac_duration,eq:ffac_duration}
restrict the class of two-qubit gates to those that are
drive-activated, perturbative, and decoherence-limited. All-microwave drives on
fixed-frequency qubits (cross-resonance scheme~\cite{Rigetti2010,Chow2011}) and
parametrically modulated flux-tunable elements~\cite{McKay2016universal,reagor2018demonstration,Ganzhorn2020benchmarking}
satisfy these conditions, since to leading order their effective interaction
rate is linear in the control amplitude~\cite{Sheldon2016,Malekakhlagh2020,Didier2018},
with $\lambda$ being the drive amplitude on the control qubit in the former
and the flux-modulation amplitude in the latter. The same cannot be said for gates activated by bringing
qubits into resonance at a fixed coupling~\cite{Barends2014superconducting,CZ_gate_2,Rol2019}, since their duration is set at fabrication and cannot be traded against
power; for these the ($\alpha$-based) model describes a given operating point rather than a
trade-off curve. We further assume that both gates are delivered through control
lines with the same attenuation profile, so that they share the dressing factor
of \Cref{eq:stages_stack}. The value of $\alpha$ appropriate to one gate family is
calibrated in \Cref{sec:MNR_CR_gates}.

\begin{table}
	\centering
	\small
	\setlength{\tabcolsep}{3pt}
	\caption{Parameters of the superconducting platform: symbols, meanings, and adopted values, grouped by tunability (cf.\ the control-parameter taxonomy in the text).}
	\label{tab:fixed_values}
	\begin{tabularx}{\columnwidth}{%
		@{}c|
		>{\hsize=0.75\hsize\centering\arraybackslash}X|
		>{\hsize=0.45\hsize\centering\arraybackslash}X@{}}
		\toprule
		\textbf{Symbol}     & \textbf{Description}                   & \textbf{Value}                          \\
		\midrule
		\multicolumn{3}{c}{\textit{Fixed hardware parameters}}                                                 \\
		\midrule
		$\omega / 2\pi$     & Qubit frequency                        & $6\,\mathrm{GHz}$                       \\
		$\gamma$            & Spontaneous emission rate              & $1\,\mathrm{Hz}$                        \\
		$K$                 & \makecell{Number of temperature                                                  \\ stages}                     & $6$ \\
		$\{T_i\}$           & Stage temperatures                     & $\mathbf{T}$ in \Cref{sec:super_specs}  \\
		$T_\mathrm{qb}$     & Qubit temperature ($T_1$)              & $6\,\mathrm{mK}$                        \\
		$T_\mathrm{ext}$    & \makecell{External (room)                                                        \\ temperature ($T_K$)}              & $293.15\,\mathrm{K}$ \\
		$P_\mathrm{static}$ & Static per-qubit power                 & $0$ (assumption)                        \\
		\midrule
		\multicolumn{3}{c}{\textit{Design-time parameters}}                                                    \\
		\midrule
		$\{A_i\}$           & Attenuation per stage                  & \makecell{\cfgCold{},                   \\ \cfgGeo{},\\ \cfgLogT{}} \\
		$\prod_j A_j$       & \makecell{Total attenuation                                                      \\ (fixed across configs)}          & \makecell{${\approx}\,9500$ ($\approx 40\,\text{dB}$),\! \\ $\mathbf{A}$ in \Cref{sec:super_specs}} \\
		\midrule
		\multicolumn{3}{c}{\textit{Run-time (optimized) parameters}}                                           \\
		\midrule
		$\tau_\mathrm{1qb}$ & Single-qubit gate duration             & varied                                  \\
		$\tau_\mathrm{2qb}$ & Two-qubit gate duration                & varied based on \Cref{eq:gfac_duration} \\
		$g_\mathrm{fac}$    & \makecell{Relative power cost                                                    \\ (2-qubit vs.\ 1-qubit gate)}   & varied \\
		$f_\mathrm{fac}$    & \makecell{Relative infidelity                                                    \\ (2-qubit vs.\ 1-qubit gate)}   & varied based on \Cref{eq:gfac_ffac_link} \\
		\midrule
		\multicolumn{3}{c}{\textit{Circuit-dependent quantities}}                                              \\
		\midrule
		$N_{\mathrm{1qb}}$  & \makecell{Number of single-qubit gates                                           \\ (total; $N_{\mathrm{1qb}}(l)$ in layer $l$)} & circuit dependent \\
		$N_{\mathrm{2qb}}$  & \makecell{Number of two-qubit gates                                              \\ (total; $N_{\mathrm{2qb}}(l)$ in layer $l$)} & circuit dependent \\
		\bottomrule
	\end{tabularx}
\end{table}

\subsubsection{Idling}
\label{subsec:idling}

An idle qubit is not driven, so no power is delivered down its control line
and no heat is dissipated. Within the dynamic-cost model discussed here, its
cost is the static per-qubit contribution of \Cref{sec:static}, which we
set to zero throughout. The power and energy of a QRAM query are therefore
charged entirely to its gates.

Idling does, however, cost fidelity. We assign to an undriven qubit an error (or infidelity)
$\epsns$ per circuit layer, describing decoherence and residual
environmental coupling in the absence of a drive. Since the error accumulates
with the time the qubit is active---either idling or driven---and a layer
lasts as long as its slowest gate, $\epsns$ is quoted per a
reference duration (here set to be $\tau_{\mathrm{ref}}=100\,$ns) and rescaled to the layer executed,
\begin{equation}
	\epsns(l) = \epsns\,\frac{\tau_l}{\tau_{\mathrm{ref}}},
	\quad
	\tau_l = \max\{\tau \text{ of the gates in layer } l\},
	\label{eq:eps-ns-layer}
\end{equation}
so that idling is characterized by a rate $\Gamma_{\mathrm{ns}} =
	\epsns/\tau_{\mathrm{ref}}$, playing for undriven qubits the role
that $\gamma(1+\bar n)$ plays for driven ones in
\Cref{eq:single_qubit_infidelity}.

The three values $\epsns \in \{10^{-6},10^{-7},10^{-8}\}$ used in the results correspond to $\Gamma_{\mathrm{ns}}\in\{10,1,0.1\}\,\text{Hz}$ and are set relative to the baseline $\gamma = 1\,\text{Hz}$ of
\Cref{tab:fixed_values}. The sweep is thus a direct comparison between
environment-induced decoherence and the drive-induced infidelity of
\Cref{eq:single_qubit_infidelity}, the central value places the two on
an equal footing, while the other values make the environment
respectively dominant over and negligible against the intrinsic decay.

Although not completely unrealistic, our choices for $\epsns$ are very optimistic. Ramsey coherence times reaching $t_2=2.0\,\text{ms}$ in superconducting qubits have been reported~\cite{somoroff2023millisecond,Tuokkola2025methods,Bland2025} and correspond, for the reference layer duration of $\tau_{\mathrm{ref}}$, to a phase-flip error of roughly
\begin{align*}
	\epsns \approx \frac{\tau_{\mathrm{ref}}}{2 t_2} = 2.5\times 10^{-5},
\end{align*}
which is more than one order of magnitude larger than the higher sweep end. In actual quantum computers, one should expect much lower coherence times, e.g., Google's Willow experiment~\cite{acharya2025quantum} reports coherence times of $t_1 = 68\,\mu\text{s}$ and $t_2 = 89\,\mu\text{s}$, which puts the above idling error close to $5.6\times 10^{-4}$. We resume the discussion on the choice and feasibility of $\epsns \in \{10^{-6},10^{-7},10^{-8}\}$ in \Cref{sec:discussion}.

Note that the idle term is present at every layer, including layers in which
the qubit is acted upon. %
Note also that $\Gamma_{\mathrm{ns}}$ is the one
noise parameter not related to the control line and thus, unlike the gate
errors, carries no dependence on $\omega$, attenuation profile or stage
temperatures.

\subsubsection{Platform specifications and control parameters}
\label{sec:super_specs}

\paragraph{Control parameters.}
While all the parameters describing energy and fidelity---such as $\{T_i\}$, $\{A_i\}$, $\omega$, $\tau_{\mathrm{1qb}}$, $g_{\mathrm{fac}}$, etc.---are in principle controllable, many are tied to the physical hardware and are not freely tunable. For instance, the spontaneous emission rate $\gamma$ is fixed upon qubit fabrication, and improving it constitutes a multi-year technological roadmap. The temperature stages $\{T_i\}$ are similarly constrained: they are dictated by the dilution-refrigerator design and, once the cryostat is selected, are effectively fixed. By contrast, the gate duration $\tau_{\mathrm{1qb}}$ and the factor $g_{\mathrm{fac}}$
offer genuine operational flexibility. The attenuation profile $\{A_i\}$ occupies an intermediate regime---its allocation across cryogenic stages can be reconfigured to a certain extent. Thus, while the attenuation allocation is not a real-time degree of freedom, it is nonetheless a design variable that can be revisited, placing it on a timescale of days to weeks, between the in-situ tunability of $\tau_{\mathrm{1qb}}$ and the years-long horizon associated with improving $\gamma$.

\paragraph{Qubit characteristics.}
The qubit frequency is fixed at $\omega /2\pi = 6\,\mathrm{GHz}$. The spontaneous
emission rate is set to $\gamma = 1\,\mathrm{Hz}$, corresponding to a
coherence time of $1\,\mathrm{s}$. This is very optimistic with respect
to current state-of-the-art hardware with the longest reported two-dimensional
transmon lifetimes reaching $\simeq 1.7\,\mathrm{ms}$~\cite{Bland2025}, following earlier devices at $0.3$--$0.5\,\mathrm{ms}$~\cite{Place2021,Wang2022}, while
qubits in deployed multi-qubit processors typically operate an order of
magnitude below these figures. The chosen $\gamma$ therefore anticipates
roughly three orders of magnitude of improvement over the best reported devices, and four over deployed processors.

\paragraph{Multi-stage cooling and attenuation.}
The cooling chain comprises $K = 6$ temperature stages (5 attenuators),
with operating temperatures fixed as
\begin{align*}
	\textbf{T} = [0.006,\,0.1,\,0.8,\,4.0,\,70.0,\,293.15] \quad\text{(in Kelvin)},
\end{align*}
so that the qubit is situated at $T_\mathrm{qb} = 6\,\mathrm{mK}$ and the
room temperature is $T_\mathrm{ext} = 293.15\,\mathrm{K}$. In all the
calculations, a total attenuation of
$\prod_j A_j \approx 9500$ (${\approx}~40\,\mathrm{dB}$) is assumed,
below the ${\approx}\,60\,\mathrm{dB}$ of state-of-the-art setups~\cite{Krinner_2019}, which results in a correspondingly larger $\bar{n}$,
differing only in how the attenuation is distributed across stages. The
distribution, rather than the total attenuation, becomes an important
factor for resource considerations because attenuation at the lower
stages is more impactful noise-wise, but leads to enormous resource
cost. Thus, the choice of such distribution influences the operating
efficiency of the QRAM circuit to a large extent. To explore this
tradeoff, we consider the following configurations, with the attenuators
placed between consecutive stages:
\begin{itemize}
	\item{\cfgCold.}
	      $\mathbf{A} = [20,\,10,\,6,\,4,\,2]$.
	      The coldest attenuator operates between 6\,mK and 100\,mK and carries
	      the largest factor $A_1 = 20$, while the chain tapers monotonically to
	      $A_5 = 2$ near room temperature.

	\item{\cfgGeo.}
	      $\mathbf{A} = [6.25,\,6.25,\,6.25,\,6.25,\,6.25]$.
	      Each of the five stages receives equal log-attenuation,
	      $A_k = A_{\mathrm{total}}^{1/5} \approx 6.25$.
	      This is the assumption-free baseline, i.e., no stage is preferentially
	      loaded, and the distribution is uniform.

	\item{Log-temperature-weighted (\cfgLogT{}).}
	      $\mathbf{A} = [11,\,6,\,4,\,12,\,3]$.
	      Attenuation is allocated in proportion to the log-temperature span of
	      each inter-stage interval,
	      \begin{equation}\label{eq:logT_split}
		      \log A_k \;=\; \log A_{\mathrm{total}}
		      \cdot \frac{\log(T_{k+1}/T_k)}{\sum_{j}\log(T_{j+1}/T_j)},
	      \end{equation}
	      motivated by the per-stage temperature-ratio reference of Krinner \textit{et al.}~\cite{Krinner_2019}, with values rounded to integers while keeping the total near $9500$.
\end{itemize}

In summary, once the hardware-fixed values (\Cref{tab:fixed_values}) and attenuation configuration are chosen, the remaining freely tunable parameters are $\tau_{\mathrm{1qb}}$ and $g_{\mathrm{fac}}$, which together span the optimization landscape considered next.

\subsubsection{MNR optimization}
Having established the MNR model for the superconducting platform and its
control parameters in \Cref{sec:MNR_superc,sec:super_specs}, we now
state how resource-fidelity optimization proceeds in the presence
of noise. The optimization proceeds in two steps:
\begin{itemize}
	\item identifying optimal operating points in the
	      $(g_{\mathrm{fac}},\tau_{\mathrm{1qb}})$-landscape, which minimize the
	      average power for a given circuit fidelity;
	\item comparing the minimum power required for a targeted fidelity
	      across the attenuation configurations: \cfgCold{}, \cfgGeo{}, and \cfgLogT{}.
\end{itemize}
The results for this two-step optimization are presented in \Cref{sec:results}.

\subsubsection{Scope and limitations}
\label{sec:scope}

The relations in \Cref{subsec:single_qubit_gates,subsec:two_qubit_gates,subsec:idling} express the cost of primitive operations that will serve as building blocks for a
QRAM query. We note here three
limitations of this accounting before proceeding with the MNR optimization to generate the results and data.

First, the model is not meant to predict the energy consumption of an
existing device. Some of the values adopted in \Cref{tab:fixed_values} are
optimistic with respect to current hardware, and are indicated as such
where they are introduced. The purpose of the model is instead to identify
which parameters control the metric-noise-resource trade-offs for a
platform where the gate errors and drive powers are governed by the
same dissipative channel.

Second, attributing the resource entirely to gates accounts for the dynamic
fraction of the budget and not the total. This distinction was introduced
in \Cref{sec:static} and is particularly relevant for a QRAM, typically consisting of multiple qubits idling. The figures further below in \Cref{sec:results} therefore
quantify the part of the budget on which the control parameters act---which is the part relevant for optimization.

Third, the metric and the resource do not count the same qubits. Every
qubit contributes to the infidelity at every layer through
\Cref{eq:eps-ns-layer}, while only the driven qubits contribute to the
energy. Such an asymmetry is unavoidable in a dynamic-cost model, but it
implies that the parameter binding the attainable fidelity, namely
$\epsns$ as shown in \Cref{sec:results}, is not accounted for
by the resource model.

\subsection{Error model}

In order to compute the fidelity between the final state $\rho$ of a QRAM circuit (or any circuit for that matter) and its error-free version $|\psi_{\mathrm{ideal}}\rangle$, we first specify our error model. We assume that every qubit in a circuit is subject to some \emph{mixed-unitary error channel}, i.e., an error channel of the form
\begin{equation}
	\mathcal{E}_\varepsilon(\rho)
	= (1-\varepsilon)\rho + \sum_{j\geq 1} K_j \rho K_j^\dagger \quad\text{with}~K_j^\dagger K_j\propto I,
\end{equation}
for some error parameter $\varepsilon\in[0,1]$ at \emph{every discrete time step}, i.e., after every layer of one- and two-qubit gates. The error parameter $\varepsilon$ is dependent on the specific qubit and circuit layer:
\begin{enumerate}
	\item if no gate is applied onto a qubit and it is idling, then it suffers only environment noise described by $\varepsilon = \epsns$ from \Cref{subsec:idling};
	\item if a single-qubit gate is applied, then the qubit is perturbed by a single-qubit-gate error $\epsone$ from \Cref{subsec:single_qubit_gates} to a total error $\varepsilon = \epsns + \epsone$;
	\item if a two-qubit gate is applied, then the qubit is perturbed by a two-qubit-gate error $\epstwo$ from \Cref{subsec:two_qubit_gates} to a total error $\varepsilon = \epsns + \epstwo$.

\end{enumerate}

Our choice for mixed-unitary error channels is without loss of generality and is purely to simplify the fidelity analysis behind some of the QRAM architectures. Mixed-unitary channels include bit-flip, dephasing, and depolarizing channels~\cite{watrous2018theory}.

\section{QRAM circuits}
\label{sec:qrams}

A QRAM is a device or operation that allows access to classical or quantum stored information in superposition. From an architecture perspective, a QRAM is comprised of a memory register $\mathtt{M}$ that stores either classical or quantum information, an address register $\mathtt{A}$ that points to the memory cell to be addressed, and a target register $\mathtt{T}$ into which the content of the addressed memory cell is copied. If necessary, it also includes an auxiliary register supporting the overall operation, which is reset to its initial state at the end of the computation.
Formally, for all $x_1,\dots,x_{2^n},b \in \{0,1\}^m$ and $i\in\{1,\dots,2^n\}$, a call to a QRAM of size $n$ implements the following unitary
\begin{equation}\label{eq:qram}
	\ket{i}_{\mathtt{A}}\ket{b}_{\mathtt{T}}\ket{x_1, \dots, x_{2^n}}_{\mathtt{M}} \mapsto \ket{i}_{\mathtt{A}}\ket{b\oplus x_i}_{\mathtt{T}} \ket{x_1, \dots, x_{2^n}}_{\mathtt{M}}.
\end{equation}
The bit-strings $x_1,\dots,x_{2^n}$ represent the data to be accessed in superposition, which are separate from the qubits in the \emph{work register} of a fully programmable quantum computer. Several architectures for implementing the QRAM operation from \Cref{eq:qram} have been proposed in the past~\cite{giovannetti2008architectures,giovannetti2008quantum,park2019circuit,niu2022entangling,phalak2022approximate,chen2021scalable,zoufal2019quantum,agliardi2022optimized,Allcock2024constantdepth}. Below we focus on two main implementations: the \emph{multiplexer} and the \emph{bucket-brigade}.

In what follows, the overhead for error correcting QRAM circuits is not taken into account. While this primarily stems from wanting a clearer analysis, our choice is justified by the \emph{noise-resilience} of certain QRAM architectures~\cite{arunachalam2015robustness,hann2021resilience,mehta2024analysis,hann2021practicality} and from recent works showing that QRAM might only require mild quantum error correction codes~\cite{dalzell2025distillation}.

\begin{figure}[t]
	\centering
	\resizebox{0.5\textwidth}{!}{
		\input{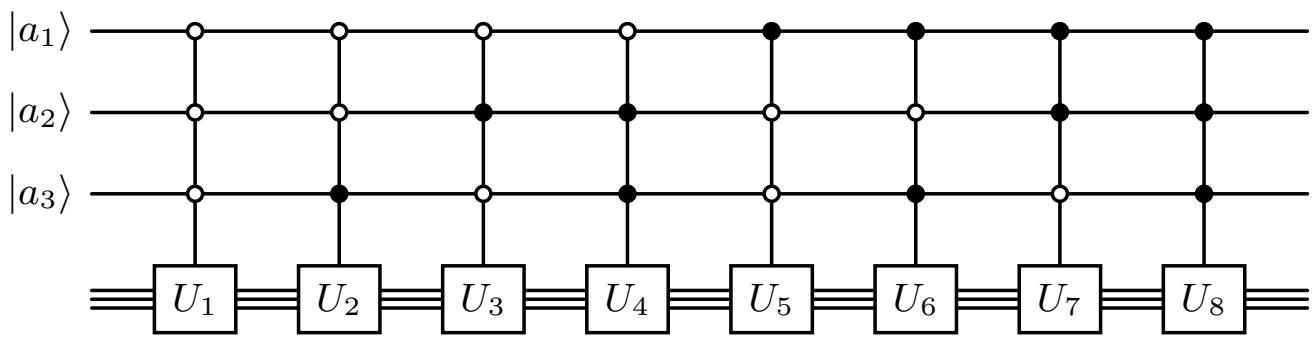}}
	\caption{A multiplexer with memory size $8$ and $n=3$ address qubits $|a_3, a_2, a_1\rangle$.}
	\label{fig:multiplexer}
\end{figure}

\subsection{Multiplexer Architecture}\label{ss:multiplexer}
The simplest way of implementing a QRAM query is with a \emph{multiplexer} (also known as \emph{data-lookup oracle} or \emph{select operator}), a circuit of linear depth in the number of memory elements that sequentially checks whether the address register is $i=1,2,\dots,2^n$. An example with $n=3$ address qubits and memory of size $2^n=8$ is shown in \Cref{fig:multiplexer}. The unitary $U_i$---which writes $x_i$ into the target register via a tensor product of some Pauli-$\mathrm{X}$ operators---is applied onto the target register controlled on the index register being in state $|i\rangle$.

\begin{figure*}[t]
	\centering
	\begin{tikzpicture}
\node[scale=1]{%
\begin{quantikz}[column sep = 0.2cm]
    \lstick{$\ket{a_1}$} & \ctrl[open]{2} & & & & & \ctrl[open]{2} & & & & & \ctrl{2} & & & & & & \ctrl{2} & & & & & \ctrl{2} & \\
    \lstick{$\ket{a_2}$} & \ctrl[open]{0} & & & & & & & & & & & \ctrl{1} & & & & & & & & & & \ctrl{0} & \\
    & \wireoverride{n} & \ctrl{2} & \ctrl{3}  &   & \ctrl{2} & \targ{} & \ctrl{2} & \ctrl{3} &  & \ctrl{2} & \targ{} & \targ{} & \ctrl{2} & \ctrl{3}  &   & \ctrl{2} & \targ{} & \ctrl{2} & \ctrl{3} &  & \ctrl{2} & & \wireoverride{n} \\
    \lstick{$\ket{a_3}$} & & \ctrl[open]{0} & & & \ctrl[open]{0} &  & \ctrl[open]{0} & & & \ctrl[open]{0} & & & \ctrl[open]{0} & & & \ctrl[open]{0} &  & \ctrl[open]{0} & & & \ctrl[open]{0} & & \\
    & \wireoverride{n} &  \wireoverride{n} &  & \ctrl{1} & & \wireoverride{n} & \wireoverride{n} &  & \ctrl{1} & & \wireoverride{n} & \wireoverride{n} & \wireoverride{n} &  & \ctrl{1} & & \wireoverride{n} & \wireoverride{n} &  & \ctrl{1} & & \wireoverride{n} & \wireoverride{n} \\
    \lstick{$\ket{\psi}$} & \qwbundle{m} & & \gate{\mathrm{X}^{\otimes S_1}} & \gate{\mathrm{\mathrm{X}}^{\otimes S_2}} & & & & \gate{\mathrm{X}^{\otimes S_3}} & \gate{\mathrm{\mathrm{X}}^{\otimes S_4}} & & && & \gate{\mathrm{X}^{\otimes S_5}} & \gate{\mathrm{\mathrm{X}}^{\otimes S_6}} & & & & \gate{\mathrm{X}^{\otimes S_7}} & \gate{\mathrm{X}^{\otimes S_8}} & & &
\end{quantikz}};
\end{tikzpicture}
	\caption{The improved multiplexer of \cite{babbush2018encoding} with memory size $8$ and $n=3$ address qubits $|a_3,a_2,a_1\rangle$. By $\mathrm{X}^{\otimes S_i}$ we mean a tensor product of Pauli-$\mathrm{X}$ gates indexed by the set $S_i\subseteq[m]$, i.e., $\bigotimes_{j\in S_i}\mathrm{X}_j$.}
	\label{fig:multiplexer_improved}
\end{figure*}
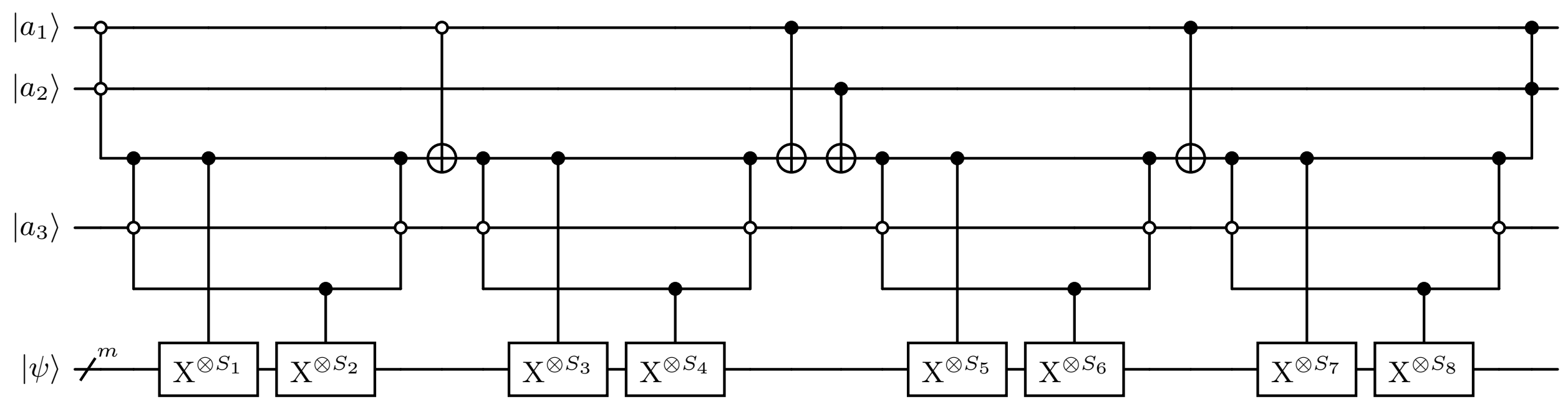

The multiplexer is a very simplistic circuit that has seen several variants in the past~\cite{babbush2018encoding,childs2018toward,Low2024tradingtgatesdirty}. A successful variant has been proposed by Babbush \emph{et al.}~\cite{babbush2018encoding} in the context of quantum chemistry and relies on implicitly and progressively computing (and uncomputing) the one-hot encoding of the address register (the one-hot encoding of $m$ is a binary string whose $(m+1)$-th entry is $1$ and the remaining are $0$). Once such an encoding is constructed, a series of CNOTs are applied onto the target qubits. An improved version of Babbush \emph{et al.}'s multiplexer is shown in \Cref{fig:multiplexer_improved} for $n=3$, where by $\mathrm{X}^{\otimes S_i}$ we represent a tensor product of Pauli-$\mathrm{X}$ gates indexed by a set $S_i$ describing which qubits should be flipped. The main advantage of their construction is a smaller $\mathrm{T}$-count, in this case $4\cdot 2^{n}-4$ in the memory size $2^n$, compared to $O(2^n n)$ for the simplistic multiplexer from \Cref{fig:multiplexer}, and is built on the AND-gadgets from \Cref{fig:toffoli}. Our slight improvement over their original construction is noticing that one need not compute the right child's one-hot encoding at the bottom of the tree since its indicator function $r$ is simply the XOR of the parent node's and left child's indicator functions $p$ and $l$, respectively. In other words, if the left and right child's writing operators are $\mathrm{X}^{\otimes l\cdot S_i}$ and $\mathrm{X}^{\otimes  r\cdot S_{i+1}}$, respectively, then $\mathrm{X}^{\otimes  l\cdot S_i}\mathrm{X}^{\otimes  r\cdot S_{i+1}} = \mathrm{X}^{\otimes  l\cdot(S_i\oplus S_{i+1})}\mathrm{X}^{\otimes  p\cdot S_{i+1}}$.

In \Cref{tab:resources_table} we summarize the various resources required to implement the multiplexer from~\cite{babbush2018encoding} under our native gate set from \Cref{ass:native_gates}. %
Regarding the fidelity, a simple lower bound can be obtained from the probability that no errors occur throughout a query to the multiplexer on any of the $2n+m-1$ qubits (out of which $n-1$ are ancillae). Let $L_i^{\mathrm{1qb}}\subseteq[\operatorname{depth}]$ be the set of layers at which a single-qubit gate is applied onto qubit $i$, and similarly for $L_i^{\mathrm{2qb}}\subseteq[\operatorname{depth}]$ where single-qubit gate is replaced with two-qubit gate. Then the error $\varepsilon_{i,l}$ that qubit $i$ suffers at layer $l$ is
\begin{align*}
	\varepsilon_{i,l} & = \epsns(l) + \epsone\mathbf{1}[l \in L_i^{\mathrm{1qb}}] + \epstwo\mathbf{1}[l \in L_i^{\mathrm{2qb}}],
\end{align*}
where $\epsns(l)$ is the layer-rescaled idling error of \Cref{eq:eps-ns-layer}. The fidelity of the multiplexer is thus
\begin{equation*}
	\begin{aligned}
		\Fid(|\psi_{\mathrm{ideal}}\rangle, \rho_{\mathrm{multiplexer}})
		 & \geq \prod_{i=1}^{2n+m-1} \prod_{l=1}^{\operatorname{depth}} (1-\varepsilon_{i,l}),
	\end{aligned}
\end{equation*}
where $\mathrm{depth}$ is the multiplexer's depth given in \Cref{tab:resources_table}. If $\varepsilon_{i,l} = O(\varepsilon)$ for all $i,l$, then the multiplexer's infidelity scales as $1-\Fid = O(\varepsilon 2^n n)$, linearly in the memory size $2^n$, which is usually referred to as an unfavorable scaling~\cite{hann2021resilience}.

\subsection{Bucket-Brigade Architecture}\label{subsec:bb-cswaps}

One of the main drawbacks of the multiplexer and other QRAM implementations like the Fan-Out is their poor resilience against noise as evident by the unfavorable infidelity scaling with the memory size. In order to tackle this issue, the \emph{bucket-brigade} architecture has been proposed~\cite{giovannetti2008quantum,giovannetti2008architectures}. In the original proposal, a binary tree of ancillary three-level quantum systems---\emph{qutrits} initialized in an idle state $|W\rangle$---are used as \emph{routers} to sequentially guide the address and target qubits down the tree to the memory contents. Recent works~\cite{hann2021resilience,mehta2024analysis}, however, have shown that not only two-level routers can be employed instead, but the two-level routers can be initialized in any arbitrary state while still preserving the noise resilience properties of the original bucket-brigade architecture.

In a two-level bucket-brigade QRAM (referred simply as bucket-brigade QRAM from now on), the $i$-th address qubit is guided down to the $i$-th level of the ancillary tree, where its state is swapped with the encountered router, i.e., $|a_i,0\rangle \mapsto |0,a_i\rangle$ for $a_i\in\{0,1\}$ assuming all routers are initialized in the state $|0\rangle$. Once the $i$-th address qubit has been placed into the $i$-th level, it will serve as an active router for subsequent address and target qubits. More precisely, the next address qubit $|a\rangle$ that reaches an active router is then routed to the left (swapped with a left output mode $|l\rangle$) if the router is in state $|0\rangle$ or to the right (swapped with a right output mode $|r\rangle$) if the router is in state $|1\rangle$, i.e.,
\begin{equation*}
	\begin{aligned}
		 & \forall a,l,r\in\{0,1\}:                                                                           \\
		 & |a,\mathrm{router},l,r\rangle \mapsto \begin{cases}
			                                         |l,\mathrm{router},a,r\rangle & \text{if}~\mathrm{router}=0, \\
			                                         |r,\mathrm{router},l,a\rangle & \text{if}~\mathrm{router}=1.
		                                         \end{cases}
	\end{aligned}
\end{equation*}
The address qubits are then continuously routed down the tree to their proper levels. After all $n$ address qubits have been routed and properly placed, a path is effectively carved out in the tree for the target qubits down to some leaf, where the memory content of, e.g., a tape, can be copied.

As clear from the above discussion, the only two quantum operations required to route and set up a routing path for target qubits are SWAP and controlled-SWAP gates. In order to implement the bucket-brigade circuit within the superconducting platform, each SWAP and controlled-SWAP gate must be decomposed using the native gate set from \Cref{ass:native_gates}. Even though such a decomposition exists, it is possible to adapt the bucket-brigade circuit to fit within superconducting platforms more suitably by \emph{crucially observing that all SWAP and controlled-SWAP gates can be replaced with iSWAP and controlled-iSWAP gates, respectively!} This is because the action of SWAP and iSWAP gates on the computation basis only differs by relative phases, which do not interfere with the routing operation. The target qubits are left unentangled with the routers after uncomputing the routing process by using either SWAP or iSWAP gates. Another way of understanding such fact is by leveraging the arbitrariness of the router's state initialization. The difference in phases from SWAP and iSWAP gates can be interpreted simply as different initial router's states.

\begin{figure*}[t]
	\centering
	\begin{tikzpicture}
		\node[scale=0.85] {
			\begin{quantikz}[column sep = 0.5cm, row sep = 0.35cm]
				\lstick{$\ket{bus}$} & \gate{\mathrm{H}}\gategroup[25,steps=1,style={dashed,rounded corners,fill=blue!20, inner xsep=3pt},background,label style={label position=below,anchor=north,yshift=-0.2cm}]{{\sc Route $a_1$}} && \gategroup[25,steps=3,style={dashed,rounded corners,fill=red!20, inner xsep=3pt},background,label style={label position=below,anchor=north,yshift=-0.2cm}]{{\sc Route $a_2$}} &&&& \gategroup[25,steps=5,style={dashed,rounded corners,fill=green!20, inner xsep=3pt},background,label style={label position=below,anchor=north,yshift=-0.2cm}]{{\sc Route $a_3$}}&&&&&& \iSWAP{}\wire[d][15]{q} \gategroup[25,steps=6,style={dashed,rounded corners,fill=yellow!20, inner xsep=3pt},background,label style={label position=below,anchor=north,yshift=-0.2cm}]{{\sc Route Bus}} & \iSWAP{}\wire[d][14]{q} &&&&&& \gategroup[25,steps=1,style={dashed,rounded corners,fill=orange!20, inner xsep=3pt},background,label style={label position=below,anchor=north,yshift=-0.2cm,xshift=-0.2cm}]{{\sc Read}} && \push{\;\cdots\;}\gategroup[25,steps=2,style={dashed,rounded corners,fill=gray!15, inner xsep=3pt},background,label style={label position=below,anchor=north,yshift=-0.2cm,xshift=0.3cm}]{{\sc Unroute}} & \gate{\mathrm{H}} & \\[0.1cm]
				\lstick{$\ket{a_3}$} &&&&&&& \iSWAP{}\wire[d][14]{q} & \iSWAP{}\wire[d][13]{q} &&&&&&&&&&&&&& \push{\;\cdots\;} && \\[-0.2cm]
				\lstick{$\ket{a_2}$} &&& \iSWAP{}\wire[d][13]{q} & \iSWAP{}\wire[d][12]{q} & &&&&&&&&&&&&&&&&& \push{\;\cdots\;} && \\[-0.2cm]
				\lstick{$\ket{a_1}$} & \iSWAP{}\wire[d][11]{q} && & & &&&&&&&&&&&&&&&&& \push{\;\cdots\;} && \\[0.2cm]
				\lstick[3]{$\ket{0}^{\otimes 3}$} && & & & &&&&&&&&&&&&& \iSWAP{}\wire[d][3]{q} && \gate{\scriptstyle \mathrm{Z}^{m_1}} && \push{\;\cdots\;} && \\[-0.3cm]
				& & & & &&&&&&& \iSWAP{}\wire[d][2]{q} &&&&&& \ctrl[open]{2} & \ctrl{0} & &&& \push{\;\cdots\;} && \\[-0.3cm]
				& & & & &&&&&&&&&&&&& \iSWAP{} &&& \gate{\scriptstyle \mathrm{Z}^{m_2}} && \push{\;\cdots\;} && \\[0.2cm]
				\lstick[3]{$\ket{0}^{\otimes 3}$} && & && & &&&& \iSWAP{}\wire[d][6]{q} & \iSWAP{} &&&&& \iSWAP{}\wire[d][6]{q} & \iSWAP{} & \iSWAP{} &&&& \push{\;\cdots\;} && \\[-0.2cm]
				& && & & \iSWAP{}\wire[d][5]{q} &&&& \ctrl[open]{5} & \ctrl{0} &&&&& \ctrl[open]{5} & \ctrl{0} &&&&&& \push{\;\cdots\;} && \\[-0.2cm]
				& & & && &&&& \iSWAP{} && \iSWAP{}\wire[d][2]{q} &&&& \iSWAP{} && \iSWAP{}\wire[d][3]{q} & \iSWAP{}\wire[d][2]{q} &&&& \push{\;\cdots\;} && \\[0.25cm]
				\lstick[3]{$\ket{0}^{\otimes 3}$} & & & && &&&&&&&&&&&&& \iSWAP{} && \gate{\scriptstyle \mathrm{Z}^{m_3}} && \push{\;\cdots\;} && \\[-0.3cm]
				& & & & &&&&&&& \iSWAP{} &&&&&& \ctrl[open]{0} & \ctrl{0} &&&& \push{\;\cdots\;} && \\[-0.3cm]
				& & & & &&&&&&&&&&&&& \iSWAP{} &&& \gate{\scriptstyle \mathrm{Z}^{m_4}} && \push{\;\cdots\;} && \\[0.2cm]
				\lstick[3]{$\ket{0}^{\otimes 3}$} & && & \iSWAP{} & \iSWAP{} && & \iSWAP{} & \iSWAP{} & \iSWAP{} &&&& \iSWAP{} & \iSWAP{} & \iSWAP{} &&&&&& \push{\;\cdots\;} && \\[-0.2cm]
				& \iSWAP{} && \ctrl[open]{0} & \ctrl{0} && & \ctrl[open]{0} & \ctrl{0} &&&&& \ctrl[open]{0} & \ctrl{0} &&&&&&&& \push{\;\cdots\;} && \\[-0.2cm]
				& && \iSWAP{} & & \iSWAP{}\wire[d][5]{q} && \iSWAP{} && \iSWAP{}\wire[d][6]{q} & \iSWAP{}\wire[d][5]{q} &&& \iSWAP{} && \iSWAP{}\wire[d][6]{q} & \iSWAP{}\wire[d][5]{q} &&&&&& \push{\;\cdots\;} && \\[0.2cm]
				\lstick[3]{$\ket{0}^{\otimes 3}$} & & & && &&&&&&&&&&&&& \iSWAP{}\wire[d][3]{q} &&  \gate{\scriptstyle \mathrm{Z}^{m_5}} && \push{\;\cdots\;} && \\[-0.3cm]
				& & & & &&&&&&& \iSWAP{}\wire[d][2]{q} &&&&&& \ctrl[open]{2} & \ctrl{0} &&&& \push{\;\cdots\;} && \\[-0.3cm]
				& & & & &&&&&&&&&&&&& \iSWAP{} &&& \gate{\scriptstyle \mathrm{Z}^{m_6}} && \push{\;\cdots\;} && \\[0.2cm]
				\lstick[3]{$\ket{0}^{\otimes 3}$} & && & && &&&& \iSWAP{} & \iSWAP{} &&&&& \iSWAP{} & \iSWAP{} & \iSWAP{} &&&& \push{\;\cdots\;} && \\[-0.2cm]
				& & && & \iSWAP{} &&&& \ctrl[open]{0} & \ctrl{0} &&&&& \ctrl[open]{0} & \ctrl{0} &&&&&& \push{\;\cdots\;} && \\[-0.2cm]
				& & & && &&&& \iSWAP{} && \iSWAP{}\wire[d][2]{q} &&&& \iSWAP{} && \iSWAP{}\wire[d][3]{q} & \iSWAP{}\wire[d][2]{q} &&&& \push{\;\cdots\;} && \\[0.2cm]
				\lstick[3]{$\ket{0}^{\otimes 3}$} & & && & &&&&&&&&&&&&& \iSWAP{} &&  \gate{\scriptstyle \mathrm{Z}^{m_7}} && \push{\;\cdots\;} && \\[-0.3cm]
				& & & & &&&&&&& \iSWAP{} &&&&&& \ctrl[open]{0} & \ctrl{0} &&&& \push{\;\cdots\;} && \\[-0.3cm]
				& & & & &&&&&&&&&&&&& \iSWAP{} &&& \gate{\scriptstyle \mathrm{Z}^{m_8}} && \push{\;\cdots\;} &&
			\end{quantikz}};
	\end{tikzpicture}
	\caption{An adapted version of the bucket-brigade QRAM from \cite{giovannetti2008quantum} using iSWAP and controlled-iSWAP gates. This example uses $n=3$ address qubits $|a_3,a_2,a_1\rangle$. The bus and address qubits are indicated at the top of the circuit, while the routers, initialized in $|0\rangle$, are indicated by the remaining rails. Each router is described by a set of three qubits: the middle qubit represents the router itself, while the top qubit is its left mode and the bottom qubit its right mode. The symbol $\boxtimes$---$\boxtimes$ represents an iSWAP gate. The memory is read in the phase basis by applying a Hadamard on the bus before the routing and after the unrouting, and employing virtual Pauli-$\mathrm{Z}$ gates on every leaf mode (see \Cref{app:clean_query}); the unrouting (not drawn) is the mirror image of the routing.}%
	\label{fig:bucket-brigade-cswap}
\end{figure*}
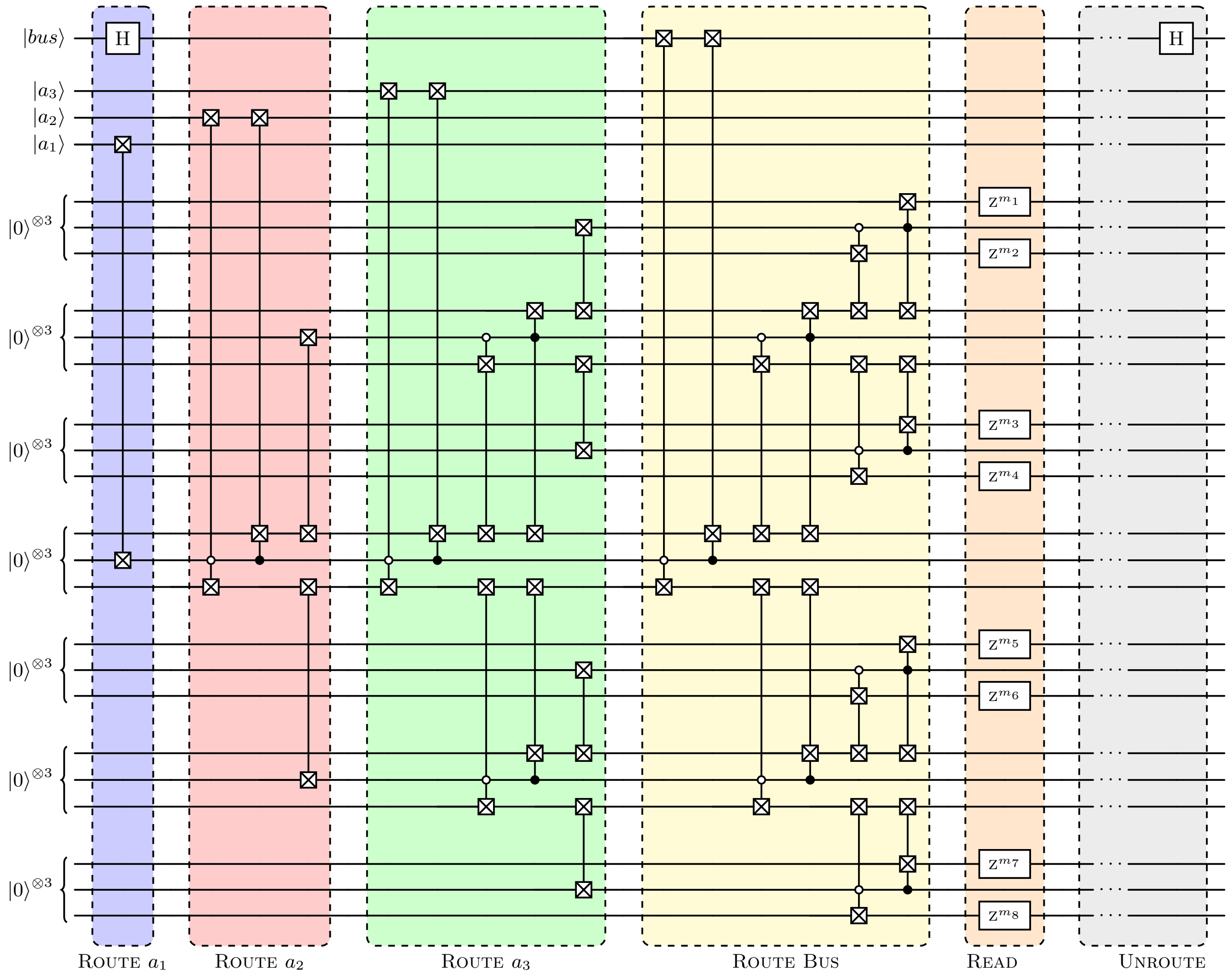

Therefore, we propose an alternative circuit for the bucket-brigade QRAM employing only iSWAP and controlled-iSWAP gates, an example of which is shown in \Cref{fig:bucket-brigade-cswap} for memory size $8$ ($n=3$ address qubits). The main reason for our new proposal is due to these gates being more easily decomposed in the native gate set from superconducting platforms according to \Cref{sec:native_gates}: iSWAP gates are already native, while controlled-iSWAP gates can be decomposed according to \Cref{fig:controlled_iswap}.

The memory is read in the phase basis. In the original construction of~\cite{giovannetti2008quantum,hann2021resilience}, the memory bit $m_i$ is copied onto the bus once it has reached leaf $i$ by a bit flip $\mathrm{X}^{m_i}$ on the output mode of the leaf. When the modes are qubits, however, a gate $\mathrm{X}^{m_i}$ cannot know whether the bus is present: it also excites the empty leaf modes with $m_i=1$, and since the unrouting only reverses the routing, these excitations are left behind in a pattern that depends on the address, so that the tree remains entangled with the address register after the query (\Cref{app:clean_query}). We therefore apply a Hadamard gate to the bus before the routing and after the unrouting, and read the memory with $\mathrm{Z}^{m_i}$ on every leaf mode. Since $\mathrm{Z}|0\rangle = |0\rangle$, the empty modes are untouched whatever the memory contains, while on the leaf holding the bus, $\mathrm{Z}^{m_i}$ maps $\mathrm{H}|b\rangle$ to $\mathrm{H}|b\oplus m_i\rangle$. The unrouting is then the exact mirror image of the routing and returns every router and mode to $|0\rangle$, and the final Hadamard yields $|b\oplus m_i\rangle$. The $\mathrm{Z}^{m_i}$ are virtual gates (\Cref{ass:native_gates}), so the read costs neither physical gates nor a layer, while the two Hadamards cost one physical $\mathrm{R}_x(\frac{\pi}{2})$ each. For $m>1$ the same holds bit by bit, with $\mathrm{Z}^{x_{i,j}}$ on the $j$-th data mode of leaf $i$.

A naive reading of \Cref{fig:bucket-brigade-cswap} would lead to a circuit depth of $O(n^2)$ since the $\ell$-th address qubit requires $2\ell + 1$ layers of (controlled) iSWAP gates and the bus $2n$, to a total depth of $\sum_{\ell = 0}^{n-1} (2\ell+1) + 2n = (n+1)^2 - 1$ for the routing (the memory read is virtual and takes no layer). However, as pointed out by Hann \emph{et al.}~\cite{hann2021resilience}, it is not necessary to wait for an address qubit to reach its final destination before subsequent qubits are routed down the tree. Notice that routing operations that involve routers located at even levels of the tree act on disjoint qubits, and therefore commute (the same is true for routing operations involving routers on odd levels). This means that once an address qubit has reached level $\ell=2$, the next address qubit can be sent down the tree at level $\ell=0$. Hence the target must only wait a circuit depth of $4n$ before being sent down the tree, which requires an extra circuit depth of $2n$ to a total of $6n$ for the routing, and of $12n$ for the whole query once the routing is uncomputed (in terms of iSWAPs and controlled-iSWAPs; the circuit depth with respect to the native gates from \Cref{ass:native_gates} can be found in \Cref{tab:resources_table}).

\begin{table*}[t]
	\centering
	\caption{Resources required to implement the QRAM operation with memory size $2^n$ ($n$-qubit address register) and $m$-qubit target state using the multiplexer and bucket-brigade architectures. We assume that half of the memory entries are $1$ (memory density $1/2$), which only affects the virtual $\mathrm{Z}$ gates of the bucket-brigade read and the write gates of the multiplexer. The single-qubit gate counts for the bucket-brigade evaluate to an integer despite the fractions. The $\mathrm{R}_z(\theta)$ gates are virtual (\Cref{ass:native_gates}): for the bucket-brigade they are the $\mathrm{Z}$ of the open controls, the $\mathrm{S}$, $\mathrm{S}^\dagger$ of the controlled-iSWAP gadgets of \Cref{fig:controlled_iswap}, the $\mathrm{Z}^{m_i}$ of the memory read, and the four $\mathrm{S}$-gates of the two Hadamards ($\mathrm{H}=\mathrm{S}\,\mathrm{R}_x(\frac{\pi}{2})\,\mathrm{S}$). The bucket-brigade $\mathrm{R}_x(\theta)$-count includes the outer $\mathrm{R}_x(\pm\frac{\pi}{2})$ of the gadgets that survive the cancellation of adjacent inverse rotations, placed on the incident or on the output mode of each gadget depending on the tree level; its depth is the level-synchronous depth of the compiled schedule, in which the $2^\ell$ routers of a level act in the same layers (see \Cref{app:bb_counts} for both). %
    For simplicity we quote the fidelity expressions with every qubit subject to the same error $\varepsilon$ at each time step. The general expressions can be found in \Cref{sec:qrams}. In the expressions, $\mathbf{1}[n\,\text{even}] = 1$ if $n$ is even and $0$ otherwise.}%
	\label{tab:resources_table}
	\small
	\setlength{\tabcolsep}{3pt}
	\renewcommand{\arraystretch}{1.4}
	\begin{tabularx}{\textwidth}{@{}>{\bfseries}p{0.3\textwidth}|>{\centering\arraybackslash}p{0.25\textwidth}|>{\centering\arraybackslash}X@{}}
		\toprule
		Resource                                         & Multiplexer                                          & Two-level Bucket-Brigade                                                                          \\ \midrule \midrule
		One-qubit $\mathrm{R}_z(\theta)$ gates (virtual) & $7\cdot 2^n+4m-14$                                   & $(2.5m+4+2^{\mathbf{1}[n\,\text{even}]}/3)2^n-2n+2m-\frac{20}{3}$                                   \\ \hline
		\textbf{One-qubit $\mathrm{R}_x(\theta)$ gates}  & $9\cdot 2^n + 2m - 26$                               & $(24m + 26 + 2^{\mathbf{1}[n\,\text{even}]}/3) 2^n - 24n - 22m - \frac{86}{3}$                   \\ \hline
		Two-qubit gates                                  & $(0.5m+6.5)2^{n}-18$                                 & $(20m+22)2^n-20(n+m)-22$                                                                          \\ \hline
		Depth (no $\mathrm{R}_z(\theta)$ gates)          & $(0.5m+11)2^{n}-24$                                  & $110n+72m-178$                                                                                    \\ \hline
		Max. active qubits per layer                     & $4$                                                  & $\frac{5}{4}2^n$ ($m=1$, $n\geq 3$)                                                               \\ \hline
		Ancillae                                         & $n-1$                                                & $(m+2)2^n-3$                                                                                      \\ \hline
		Fidelity bound                                   & $\geq (1-\varepsilon)^{(2n+m-1)\cdot\mathrm{depth}}$ & $\geq ((2-\varepsilon n(n-1)\cdot\mathrm{depth})(1-\varepsilon)^{(2n+1)\cdot\mathrm{depth}}-1)^2$ \\
		\bottomrule
	\end{tabularx}
\end{table*}

The main motivation behind the introduction of the bucket-brigade architecture was its high resilience against noise, which was first proven in~\cite{arunachalam2015robustness} for some restricted noise models and later greatly expanded in~\cite{hann2021resilience,mehta2024analysis}. Significant to our paper are the results of Hann \emph{et al.}~\cite{hann2021resilience}, who obtained a simple fidelity bound under the mixed-unitary error model~\footnote{Although~\cite{hann2021resilience} obtained fidelity bounds under more general error channels that depend on $\langle W| K^\dagger_0 K_0|W\rangle$, we focus on mixed-unitary error models for simplicity.} with a single error parameter~$\varepsilon$:
\begin{align}\label{eq:hann_fidelity}
	\Fid(|\psi_{\mathrm{ideal}}\rangle, \rho_{\mathrm{bb}})
	 & \geq (2(1-\delta)(1-\varepsilon)^{n\cdot \mathrm{depth}} - 1)^2,
\end{align}
where $\delta\in[0,1]$. Before addressing the parameter $\delta$, we generalize the above expression by decomposing the controlled-iSWAP gates as in \Cref{fig:controlled_iswap} and taking into account different sources of errors: decoherence from environment noise ($\epsns$), single-qubit gate errors ($\epsone$), and two-qubit gate errors ($\epstwo$) from \Cref{sec:MNR_superc}. Moreover, Hann \emph{et al.}\ disregard decoherence in the router's incident and output modes, which we consider here.

The high-level argument of~\cite{hann2021practicality} is to compute the expected fraction $\mathbb{E}_d[\Lambda]$ of ``good'' (error-free) branches for a level-$d$ tree. For a level-$1$ tree, the expected fraction equals the probability that the lone router never suffers an error, which is $\mathbb{E}_1[\Lambda] = (1-\varepsilon)^{\mathrm{depth}}$ in~\cite{hann2021practicality}. For deeper trees, the ratio $\mathbb{E}_d[\Lambda]/\mathbb{E}_{d-1}[\Lambda]$ equals the expected fraction of error-free routers at level $d$, which is also $(1-\varepsilon)^{\mathrm{depth}}$ in~\cite{hann2021practicality}. Here, such expressions involve both the router and its incident mode never suffering an error. Taking into account that all routers at the same level experience the same gate sequence, let $R^{\mathrm{1qb}}_{d}\subseteq[\mathrm{depth}]$ be the set of layers at which a single-qubit gate is applied onto a level-$d$ router and let $I^{\mathrm{1qb}}_{d}\subseteq[\mathrm{depth}]$ be the set of layers at which a single-qubit gate is applied onto the incident mode of a level-$d$ router. Similarly for $R^{\mathrm{2qb}}_{d},I^{\mathrm{2qb}}_{d}\subseteq[\mathrm{depth}]$ with respect to two- instead of single-qubit gates. These sets are read off the compiled schedule. In the decomposition of \Cref{fig:controlled_iswap} the only gate acting on a router is the central $\mathrm{CZ}$ of each controlled-iSWAP, so $R^{\mathrm{1qb}}_d=\emptyset$. The incident mode of a level-$d$ router takes part in the controlled-iSWAPs of that router as the incoming mode and in those of the parent router as the output mode, and receives from them the remaining $\mathrm{CZ}$ and $\mathrm{R}_x$ gates of the gadget, together with the outer $\mathrm{R}_x(\pm\frac{\pi}{2})$ of those gadgets whose outer pair sits on this mode and survives the cancellation of adjacent inverse rotations (\Cref{app:bb_counts}); the $\mathrm{S}$, $\mathrm{S}^\dagger$ and the $\mathrm{Z}$ of the open controls are virtual and error-free. The bus is the incident mode of the root while it is routed and additionally receives the $\mathrm{R}_x(\frac{\pi}{2})$ of the two Hadamards, in the first and in the last layer, which enter $I^{\mathrm{1qb}}_1$. The error $\varepsilon^{\mathrm{rt}}_{d,l}$ that a level-$d$ router suffers at layer $l$ is
\begin{align*}
	\varepsilon_{d,l}^{\mathrm{rt}} = \epsns(l) + \epsone\mathbf{1}[l\in R^{\mathrm{1qb}}_{d}] + \epstwo\mathbf{1}[l\in R^{\mathrm{2qb}}_{d}]
\end{align*}
and similarly for the error $\varepsilon_{d,l}^{\mathrm{ir}}$ for the incident mode of a $d$-th level router, where $\epsns(l)$ is the layer-rescaled idling error of \Cref{eq:eps-ns-layer}. Then (let $\mathbb{E}_0[\Lambda] := 1$)
\begin{align}
	\frac{\mathbb{E}_d[\Lambda]}{\mathbb{E}_{d-1}[\Lambda]} & = \prod_{l=1}^{\mathrm{depth}} \big((1-\varepsilon^{\mathrm{rt}}_{d,l})(1-\varepsilon^{\mathrm{ir}}_{d,l})\big)  \nonumber        \\
	\implies
	\mathbb{E}_n[\Lambda]                                   & = \prod_{d=1}^{n+1}\prod_{l=1}^{\mathrm{depth}} \big((1-\varepsilon^{\mathrm{rt}}_{d,l})(1-\varepsilon^{\mathrm{ir}}_{d,l})\big),
\end{align}
where $\varepsilon_{n+1,l}^{\mathrm{rt}} = 0$ since there are only incident modes at level $n+1$: the leaf modes, which receive the gates of the deepest routing step and the virtual $\mathrm{Z}^{m_i}$ of the memory read (see \Cref{fig:bucket-brigade-cswap}); the latter add no gate error, so at the read the leaf modes only idle.

As proven by Hann \emph{et al.}~\cite{hann2021resilience}, the fidelity lower bound on a qutrit-based QRAM thus follows from
\begin{align}\label{eq:fidelity_bound}
	\Fid \geq (2\mathbb{E}_n[\Lambda] - 1)^2,
\end{align}
assuming that $\mathbb{E}_n[\Lambda] \geq \frac{1}{2}$. However, the QRAM under consideration employs two-level routers instead of qutrits. As shown in~\cite{hann2021resilience}, this effectively reduces the number of good branches and leads to a reduced expected fraction
\begin{align}
	\mathbb{E}_n[\widetilde{\Lambda}] = (1-\delta)\mathbb{E}_n[\Lambda],
\end{align}
where $\delta$ is the aforementioned parameter related to the average probability that errors propagate from ``bad'' branches (branches where an error has occurred) to good branches. In order to estimate $\delta$, suppose that a level-$d$ router $r$ suffers an error $\varepsilon_{r,l}^{\mathrm{rt}} := \varepsilon_{d,l}^{\mathrm{rt}}$ at layer $l$, and let $P_{r\to i}(l)$ be the probability of this error propagating into branch $i$. Then, to leading order in $\varepsilon_{\max} = \max_{d,l}\varepsilon^{\mathrm{rt}}_{d,l}$,
\begin{equation*}
	\delta = \max_{i} \sum_{r,l} \varepsilon_{r,l}^{\mathrm{rt}} P_{r\to i}(l) + O(\varepsilon_{\max}^2).
\end{equation*}
As argued by Hann \emph{et al.}, errors are free to propagate from a router's left output mode to its input since, by default, all routers are initialized in $|0\rangle$. In contrast, an error propagates from the router's right output to its input only if the router's state is first flipped from $|0\rangle$ to $|1\rangle$. Therefore, only errors that propagate upward from left output modes contribute to $\sum_{r,l}P_{r\to i}(l)$~\footnote{Through a QRAM query, as address qubits are loaded into the routing tree, the state of some routers might change from $|0\rangle$ to $|1\rangle$, after which errors will be free to propagate upward from its right output instead of its left output.}. A conservative approach is to set $P_{r\to i}(l) = 1$ for all routers $r$ and layers $l$, so that
\begin{align}\label{eq:delta_bound}
	\delta &\leq \sum_{r}\sum_{l=1}^{\mathrm{depth}} \varepsilon_{r,l}^{\mathrm{rt}} + O(\varepsilon_{\max}^2) \nonumber \\
    &\leq \sum_{d=1}^n\sum_{l=1}^{\mathrm{depth}} (d-1)\varepsilon_{d,l}^{\mathrm{rt}} + O(\varepsilon_{\max}^2),
\end{align}
using that at most $d-1$ routers at level $d$ are connected to a branch $i$ through left routers. By substituting the above expression together with $(1-\delta)\mathbb{E}_n[\Lambda]$ into the fidelity bound \Cref{eq:fidelity_bound}, we arrive at our desired expression. A better scaling can be obtained from a more careful analysis of $\sum_{r,l} \varepsilon_{r,l}^{\mathrm{rt}} P_{r\to i}(l)$, which will necessarily depend on the specific error channel. If $\varepsilon^{\mathrm{rt}}_{d,l}, \varepsilon^{\mathrm{ir}}_{d,l} = O(\varepsilon)$, then the infidelity reduces to $1 - \Fid = O(\varepsilon n \cdot \mathrm{depth} + \varepsilon n^2 \cdot \mathrm{depth}) = O(\varepsilon n^3)$. Small scale simulations from~\cite{hann2021resilience} suggest scalings better than cubic on $n$ for several error channels including depolarizing, bit-flip, dephasing, and damping.

\begin{figure*}
	\centering
	\includegraphics[width=\linewidth, trim = 0cm 0cm 0cm 1cm, clip]{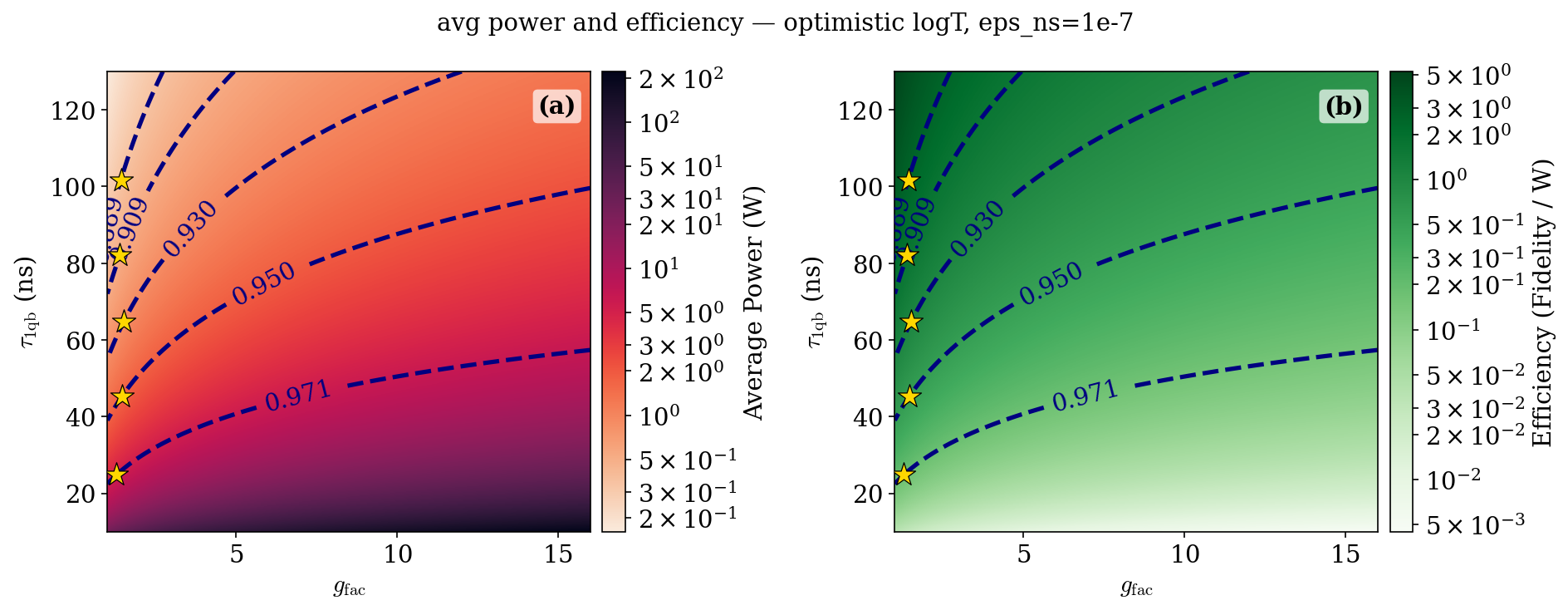}
	\caption{Resource optimization plots for a bucket brigade QRAM with $n =12$, $m=1$. In (a) the heatmap of average power is shown as a function of the control parameters $(g_{\mathrm{fac}}, \tau_{\mathrm{1qb}})$ with \cfgLogT{} configuration for attenuation distribution and superconducting system specifications in \Cref{sec:MNR_superc}, along with $\epsns = 10^{-7}$. The heatmap is overlaid by iso-fidelity blue dashed lines with annotations denoting the fidelities. The optimal operating points---points of minimum average power---are marked as yellow stars. For the provided specifications the optimal points lie at $g_{\mathrm{fac}}\approx 1.4$--$1.5$ along each iso-fidelity contour. In (b), we show the MNR efficiency $\eta = \text{Circuit fidelity}/ \text{Average power}$ as a function of the control parameters. The iso-fidelity contours are overlaid, with the most efficient point for an iso-fidelity again marked as yellow star. The optimal points are the same in both figures as they mark the (a) points of minimal average power subject to fixed fidelities, which is the same as (b) most efficient for a given fidelity. These plots allow for efficiency optimization by either resource minimization for a targeted fidelity or fidelity maximization for a targeted resource cost.}
	\label{fig:MNR_1}
\end{figure*}

\begin{figure*}
	\centering
	\includegraphics[width=\linewidth, trim = 0cm 0cm 0cm 0cm, clip]{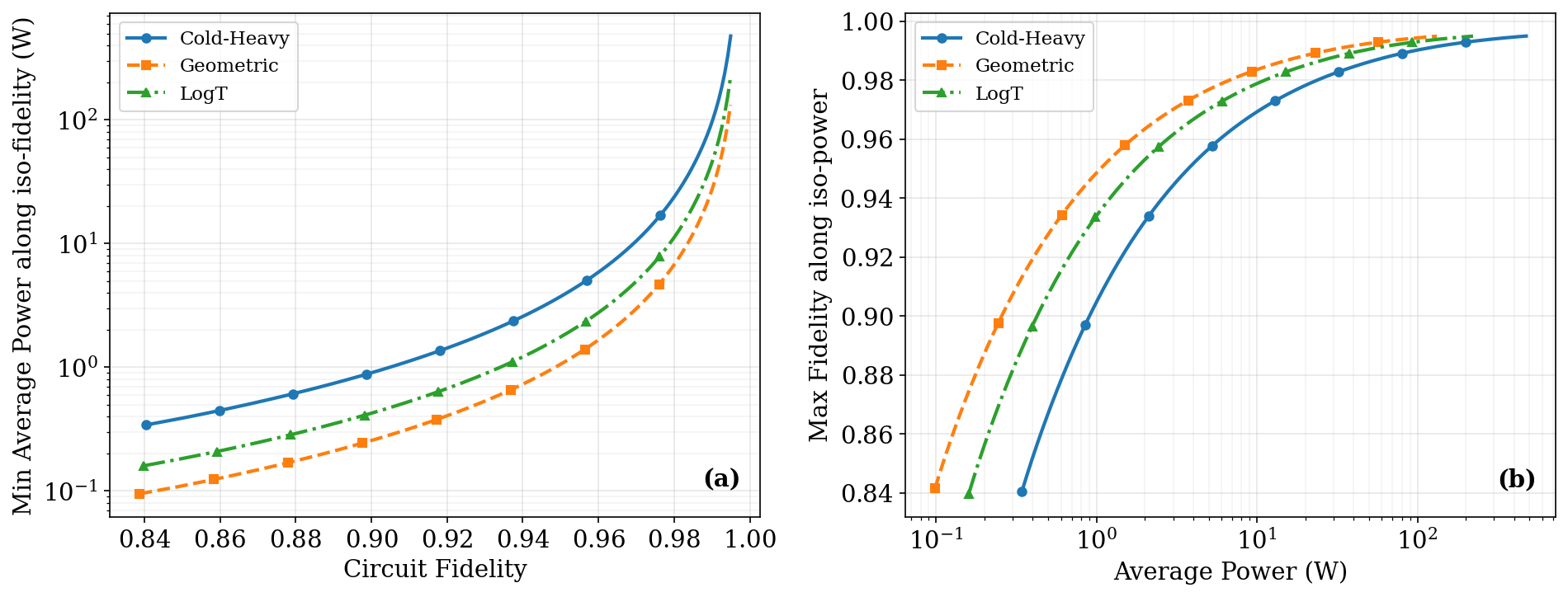}
	\caption{Plots comparing the MNR characteristics of different configurations of a superconducting quantum computer defined in \Cref{sec:super_specs}, namely \cfgLogT{}, \cfgGeo{}, and \cfgCold{}. Here the underlying quantum circuit is one query to a BB-QRAM with $n =12$, $m=1$. (a) shows the minimum power required for achieving a certain circuit fidelity. This corresponds to moving along iso-fidelity contours on the $(g_{\mathrm{fac}},\tau_{\mathrm{1qb}})$-heatmap and identifying the point of minimum average power. For example, see \Cref{fig:MNR_1}(a) with iso-fidelity contours shown as blue dashed lines. In (b) a complementary result is shown: the maximum circuit fidelity obtainable for fixed average power. This is analogous to moving along the heatmap contours (different colored regions) in \Cref{fig:MNR_1}(a) and finding the maximum achievable fidelity.}%
	\label{fig:MNR_2}
\end{figure*}

\begin{figure}
	\centering
	\includegraphics[width = \columnwidth]{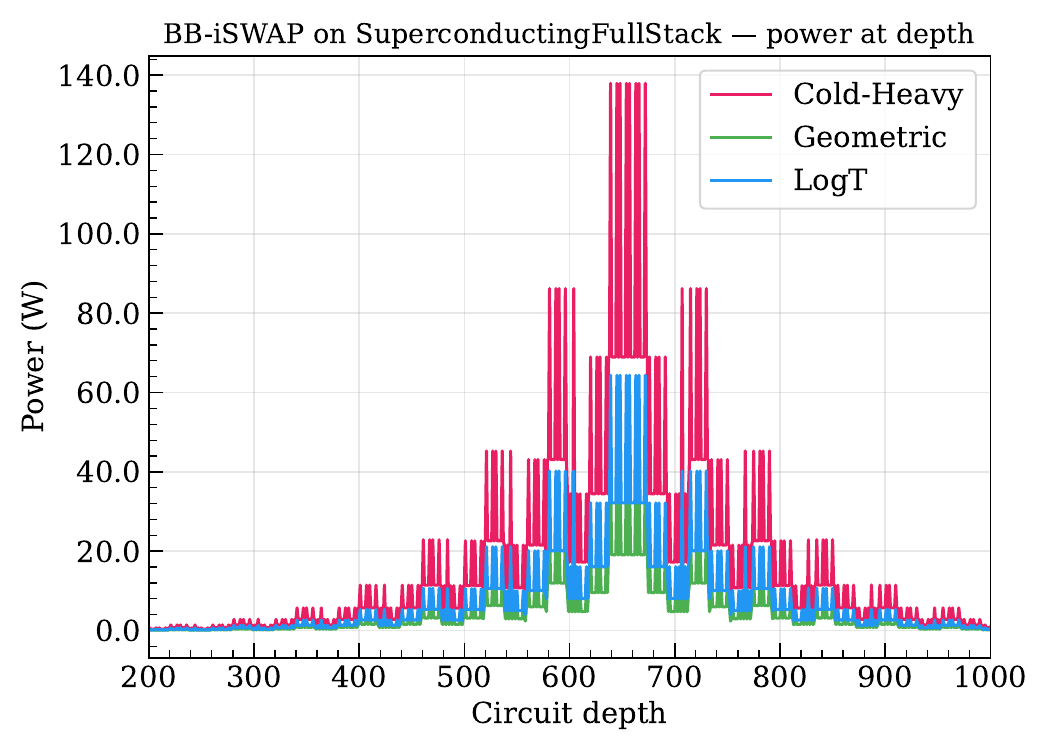}
	\caption{Power as a function of circuit depth (BB-QRAM with $n=12$, $m=1$) for three attenuation configurations of the superconducting quantum computer defined in \Cref{sec:super_specs}. Apart from the fixed superconducting specifications, the controllable parameters have been kept identical for all three configurations: $g_{\mathrm{fac}} = 1$, $\tau_{\mathrm{1qb}} = 25\,\mathrm{ns}$, $\alpha = 4$, hence $\tau_{\mathrm{2qb}} = 100\,\mathrm{ns}$. The horizontal axis shows circuit depths $200$--$1000$ of the $1214$ layers}
	\label{fig:power_at_depth}
\end{figure}

\section{Numerical simulation}
\label{sec:results}
In~\Cref{sec:MNR_superc} we expressed the power and energy consumption and
the one- and two-qubit gate fidelities as functions of the control parameters, and we refer to \Cref{tab:fixed_values} for the values of such parameters. Recall that the qubit has a frequency
$\omega / 2\pi = 6\,\mathrm{GHz}$ and a spontaneous emission rate
$\gamma = 1\,\mathrm{Hz}$, and sits at the coldest stage of the $K = 6$ cooling
chain, with the total attenuation fixed at $\prod_j A_j \approx 9500$
($\approx 40\,\mathrm{dB}$). The attenuation is distributed according to one of
the three configurations enumerated in \Cref{sec:super_specs}, namely
\cfgCold{}, \cfgGeo{}, and \cfgLogT{}. On the circuit side,
we take the BB-QRAM (already decomposed into the native gate set from \Cref{sec:native_gates}) with $n = 12$ address qubits. The fidelity of the BB-QRAM is bounded as in \Cref{subsec:bb-cswaps} and that of the multiplexer as in \Cref{ss:multiplexer}; in both cases every qubit is charged the idling error $\epsns(l)$ of \Cref{eq:eps-ns-layer} at every layer, rescaled by the layer duration, in addition to the error of the gates acting on it.
\begin{figure*}
	\centering
	\includegraphics[width=\linewidth, trim = 0cm 0cm 0cm 0cm, clip]{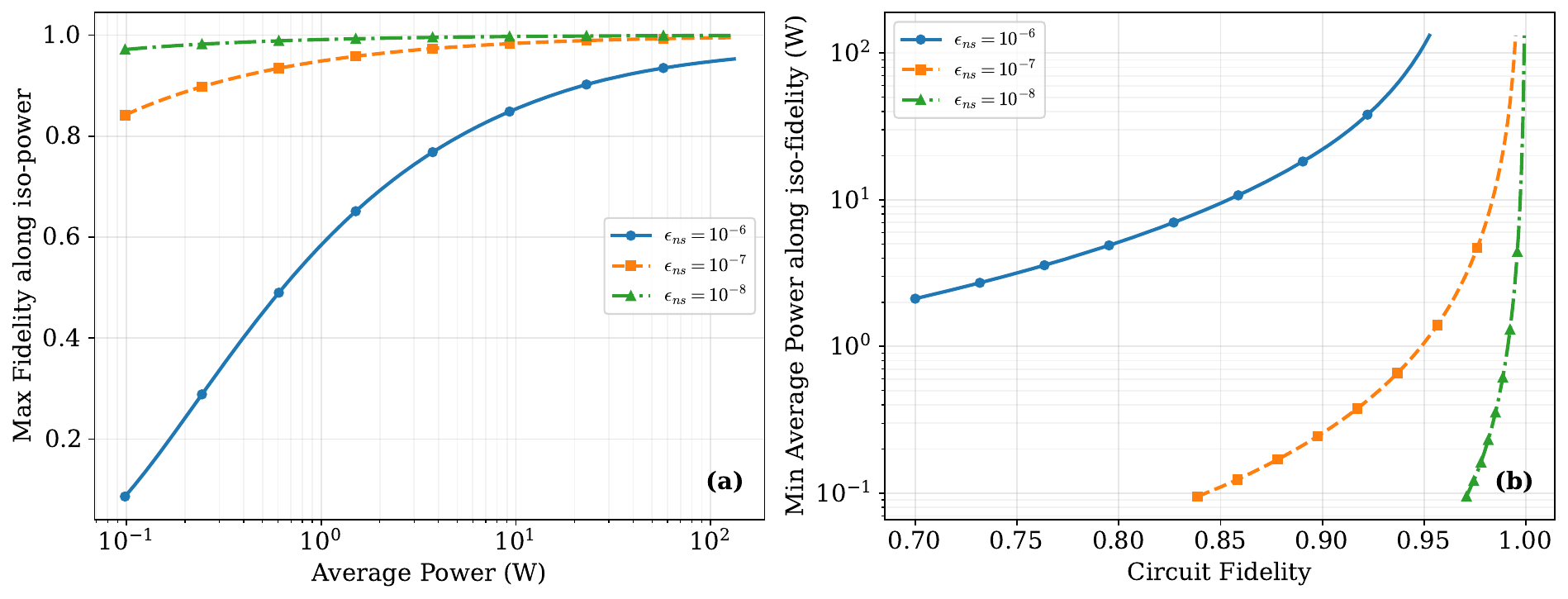}
	\caption{Plots comparing the MNR characteristics for three strengths of the
		idling noise, $\epsns = 10^{-6}$, $10^{-7}$, $10^{-8}$, with
		the \cfgGeo{} configuration for attenuation distribution and
		superconducting system specifications in \Cref{sec:super_specs}. Here the underlying quantum circuit is one query to a BB-QRAM with $n =12$, $m=1$. (a) shows the
		maximum circuit fidelity obtainable for a fixed average power, and in (b) the
		complementary result is shown: the minimum average power required for achieving
		a certain circuit fidelity. As in \Cref{fig:MNR_2}, these correspond to moving
		along the iso-fidelity contours and along the constant-power regions of the
		$(g_{\mathrm{fac}},\tau_{\mathrm{1qb}})$-heatmap of \Cref{fig:MNR_1}(a), now
		repeated for each $\epsns$; the curve for
		$\epsns = 10^{-7}$ is the \cfgGeo{} curve of
		\Cref{fig:MNR_2}. The idling noise sets the band of circuit fidelities accessible
		to the control parameters, so that consecutive curves overlap only marginally.
		Within the overlap, a ten-fold increase in $\epsns$ is worth roughly a factor of
		$33$ in average power at a fixed circuit fidelity. In each case the required
		power rises steeply towards the top of the accessible band, so that the last
		increments of fidelity are the most expensive, and no power budget within the
		sweep range moves one curve into the regime of the next. Fidelities below $0.7$
		are omitted in (b).}%
	\label{fig:MNR_3}
\end{figure*}

A quantity that remains to be fixed is the hardware parameter $\alpha$, linking the two-qubit factors through \Cref{eq:gfac_ffac_link}. Recall that $\alpha$ is the duration ratio $\tau_{\mathrm{2qb}}/\tau_{\mathrm{1qb}}$ at which the two gate types consume equal power, so that it is fixed once from a reference operating point of the hardware. An operating point provides the values of all controllable and fixed parameters mentioned in \Cref{tab:fixed_values}.
We take as reference $\tau_{\mathrm{1qb}} = 25\,\mathrm{ns}$ and
$\tau_{\mathrm{2qb}} = 100\,\mathrm{ns}$ for all configurations, giving $\alpha = 4$ and hence
\begin{equation}\label{eq:ffac_scenario1}
	f_{\mathrm{fac}} = \frac{4}{\sqrt{g_{\mathrm{fac}}}} \, .
\end{equation}

Recall from \Cref{subsec:two_qubit_gates} that $\alpha$ enters the analysis in two ways. It sets how much accuracy is
bought for a given increase in two-qubit power, and also delimits the region
over which the ($g_{\mathrm{fac}},\tau_{\mathrm{1qb}}$)-MNR landscape
is meaningful. Since the model requires
$f_{\mathrm{fac}} \geq 1$, the accessible range is
$g_{\mathrm{fac}} \in (0,\,\alpha^{2}]$, so that the above reference timings place
the boundary at $g_{\mathrm{fac}} = 16$.
The value $\alpha=4$ is optimistic: the calibration of \Cref{sec:MNR_CR_gates} gives its effective value to be
$\alpha_{\mathrm{eff}}$
$\sim 16$--$200$ for cross-resonance gates, so
that $\alpha = 4$ assumes a two-qubit gate of appreciably better
energy-infidelity quality than that family currently
supports.

To provide an expression for the energy consumption in terms of the tunable control parameters $\tau_{\mathrm{1qb}}$ and $g_{\mathrm{fac}}$, we make the following derivation. We assume that a layer
lasts as long as its slowest gate, and the two-qubit gate is the slower of the two. Each of the
$D_{\mathrm{2qb}}$ layers of maximum arity two therefore takes time $\tau_{\mathrm{2qb}}$, and each of
the $D_{\mathrm{1qb}}$ single-qubit layers takes time $\tau_{\mathrm{1qb}}$, giving a total runtime of
\begin{equation}\label{eq:runtime}
	\tquery = D_{\mathrm{1qb}}\,\tau_{1\mathrm{qb}} + D_{\mathrm{2qb}}\,\tau_{2\mathrm{qb}} .
\end{equation}
Each gate is charged its own duration, so that the total energy consumed by a
circuit of $N_{\mathrm{1qb}}$ single- and $N_{\mathrm{2qb}}$ two-qubit gates is
$\Edyn = P_{\mathrm{drs}}\,(\tau_{1\mathrm{qb}} N_{\mathrm{1qb}}
	+ g_{\mathrm{fac}}\,\tau_{2\mathrm{qb}} N_{\mathrm{2qb}})$, where $P_{\mathrm{drs}}$ is the
dressed power of a single-qubit gate from \Cref{eq:stages_stack}. The time-averaged
power dissipation $\Pavg = \Edyn/\tquery$ is then
\begin{equation}\label{eq:avg_power}
	\Pavg = \frac{P_{\mathrm{drs}}(\{T_i\}, \{A_i\}, \tau_{\mathrm{1qb}})
		\bigl(\tau_{1\mathrm{qb}}\,N_{\mathrm{1qb}}
		+ g_{\mathrm{fac}}\,\tau_{2\mathrm{qb}}\,N_{\mathrm{2qb}}\bigr)}
	{D_{\mathrm{1qb}}\,\tau_{1\mathrm{qb}} + D_{\mathrm{2qb}}\,\tau_{2\mathrm{qb}}}.
\end{equation}

\subsection{Fixed-point benchmark}
\label{sec:fixed_point}
Before using the MNR framework to optimize the control parameters, we report a pointwise estimate of the energy consumption of the BB-QRAM circuit with $n=12$ address qubits and $m=1$ target qubit. In particular, we fix $\tau_{\mathrm{1qb}} = 25\,$ns and $g_{\mathrm{fac}} = 1$, so that
$f_{\mathrm{fac}} = 4$ and $\tau_{\mathrm{2qb}} = 100\,$ns follow from
\Cref{eq:gfac_ffac_link,eq:gfac_duration} with $\alpha = 4$. We further pick a \cfgLogT{}
attenuation allocation and $\epsns = 10^{-7}$. The remaining
parameters are chosen according to \Cref{tab:fixed_values}. The resulting
quantities are collected in \Cref{tab:fixed_point}.

One BB-QRAM query uses $12{,}298$ qubits and has depth $1{,}214$, of which
$12{,}285$ qubits are the routers and their modes and the remaining $13$ are
the address and bus registers. The native-gate-set decomposition from
\Cref{sec:native_gates} yields $207{,}192$ $\mathrm{R}_x(\theta)$-gates and $171{,}750$ two-qubit gates, plus $29{,}326$ virtual $\mathrm{R}_z(\theta)$ gates. At most $5{,}120$
qubits are acted upon in any layer, i.e., $42\%$ of the qubits, and the average number of active qubits is $454$. On the other hand, the multiplexer requires only $24$ qubits at the expense of circuit depth $47{,}080$. The native-gate-set decomposition from
\Cref{sec:native_gates} yields $36{,}840$ $\mathrm{R}_x(\theta)$-gates and $28{,}654$ two-qubit gates (plus $28{,}662$ virtual $\mathrm{R}_z(\theta)$ gates), a substantial decrease compared to the bucket-brigade. Only $4$ qubits at most are acted upon in any layer.

At this operating point, the one- and two-qubit gate errors are
$3.09\times10^{-8}$ and $1.24\times10^{-7}$, respectively. These errors are
comparable to the idling error $\epsns(l)=10^{-7}$ incurred by each idle qubit during a two-qubit layer. With a propagation correction
$\delta\approx 5.45\times 10^{-3}$, %
the query infidelity is $1-\Fid\lesssim 0.0313$, corresponding to a
query fidelity $\Fid \gtrsim 0.969$. The query
lasts for $\tquery=80.2\,\mu\mathrm{s}$ and consumes
$\Edyn=3.51\times10^{-4}\,\mathrm{J}$ of dynamic energy. The peak
dynamic power, defined as the maximum over all circuit layers, is
$\Ppeak=64.4\,\mathrm{W}$.

For comparison, at $g_{\mathrm{fac}}=16$ and $f_{\mathrm{fac}}=1$, so that
$\tau_{\mathrm{2qb}}=\tau_{\mathrm{1qb}}=25\,$ns, the one- and two-qubit gate
errors are both $3.09\times10^{-8}$. The propagation correction drops to
$\delta\approx 2.04\times 10^{-3}$ and the query infidelity to $1-\Fid \lesssim 0.0118$ ($\Fid\gtrsim 0.988$). The query is shorter, $\tquery=30.4\,\mu\mathrm{s}$, but the average dynamic power increases to $38.2\,\mathrm{W}$, the peak dynamic power to $644\,\mathrm{W}$, and the dynamic energy per query to
$1.16\times10^{-3}\,\mathrm{J}$.

\begin{table}
	\centering
	\small
	\setlength{\tabcolsep}{3pt}
	\caption{Compiled resources, fidelity, and dynamic energetic cost of the fixed-point bucket-brigade and multiplexer QRAM benchmark with $n=12$ and $m=1$ under a \cfgLogT{} attenuation allocation, $\epsns = 10^{-7}$ (per $\tau_{\mathrm{ref}}=100\,$ns layer, rescaled by the layer duration), $\tau_{\mathrm{1qb}} = 25\,\text{ns}$, $g_{\mathrm{fac}}=1$, and other parameters according to \Cref{tab:fixed_values}. Static system power is excluded.}\label{tab:fixed_point}
	\begin{tabularx}{\columnwidth}{@{}>{\raggedright\arraybackslash}X c c c@{}}
		\toprule
		\textbf{Quantity}            & \textbf{Symbol}              & \makecell{\textbf{Value}                                                     \\ \textbf{BB}}  &  \makecell{\textbf{Value}\\ \textbf{Multiplexer}}               \\
		\midrule
		\multicolumn{4}{c}{\textit{Compiled-query resources}}                                                                                      \\
		\midrule
		Number of qubits             & $N_{\mathrm{qb}}$            & $12{,}298$                          & $24$                                   \\
		Circuit depth                & $D$                          & $1{,}214$                           & $47{,}080$                             \\
		$\mathrm{R}_x$ gates         & $N_{R_x}$                    & $207{,}192$                         & $36{,}840$                             \\
		CZ gates                     & $N_{\mathrm{CZ}}$            & $163{,}560$                         & $28{,}654$                             \\
		iSWAP gates                  & $N_{\mathrm{iSWAP}}$         & $8{,}190$                           & $0$                                    \\
		Virtual $\mathrm{R}_z$ gates & $N_{R_z}^{\mathrm{virt}}$    & $29{,}326$                          & $28{,}662$                             \\
		Max.\ active qubits          & $N_{\mathrm{active}}^{\max}$ & $5{,}120$                           & $4$                                    \\
		\midrule
		\multicolumn{4}{c}{\textit{Fidelity}}                                                                                                      \\
		\midrule
		One-qubit gate error         & $\epsone$                    & $3.09\!\times\!10^{-8}$             & $3.09\times10^{-8}$                    \\
		Two-qubit gate error         & $\epstwo$                    & $1.24\!\times\!10^{-7}$             & $1.24\times10^{-7}$                    \\
		Correction parameter         & $\delta$                     & $5.45\!\times\! 10^{-3}$            & -                                      \\
		Query fidelity               & $\Fid$                       & $0.969$                             & $0.922$                                \\
		Query infidelity             & $1-\Fid$                     & $0.0313$                            & $0.0775$                               \\
		\midrule
		\multicolumn{4}{c}{\textit{Dynamic power and energy}}                                                                                      \\
		\midrule
		Query duration               & $\tquery$                    & $80.2\,\mu\mathrm{s}$               & $3.02\! \times\! 10^{-3}\,\mathrm{s}$  \\
		Avg.\ dynamic power          & $\Pavg$                      & $4.38\,\mathrm{W}$                  & $1.97\! \times\! 10^{-2}\,\mathrm{W}$  \\
		Peak dynamic power           & $\Ppeak$                     & $64.4\,\mathrm{W}$                  & $3.14 \!\times\! 10^{-2}\, \mathrm{W}$ \\
		Dyn.\ energy (query)         & $\Edyn$                      & $3.51\!\times\!10^{-4}\,\mathrm{J}$ & $5.95 \!\times\! 10^{-5}\,\mathrm{J}$  \\
		\bottomrule
	\end{tabularx}
\end{table}

\subsection{Resource optimization using $g_{\mathrm{fac}}$ and $\tau_{1qb}$}

We illustrate the two-step MNR optimization over the control parameter space
constituted by $(g_\mathrm{fac}, \tau_\mathrm{1qb})$, with fixed idling noise parameter, $\epsns = 10^{-7}$, corresponding to $\tau_{\mathrm{ref}} = 100\,\text{ns}$ as mentioned in \Cref{subsec:idling}. The same analysis is repeated in the \Cref{app:further} with a different $\epsns$.

\paragraph{Power and fidelity optimizations.}
For the BB-QRAM circuit, we plot in \Cref{fig:MNR_1}(a) the heatmap of the average power as a function of the control parameters $(g_{\mathrm{fac}},\tau_{\mathrm{1qb}})$ under the \cfgLogT{} attenuation allocation. We also superpose the circuit fidelity contours, obtained from \Cref{subsec:bb-cswaps}, on this heatmap. This presentation of iso-fidelities on resource cost heatmap enables resource optimization with fixed circuit fidelity or fidelity optimization with fixed average power. In the former case, we show the minimum resource consumption points on the iso-fidelity contours as yellow stars. The optimization is performed over $g_{\mathrm{fac}}\in[1,16]$ and $\tau_{\mathrm{1qb}}\in[10,130]\,\mathrm{ns}$. The upper limit $g_{\mathrm{fac}}=\alpha^2=16$ follows from the model ($f_{\mathrm{fac}}\geq 1$). In several cases the optimal points lie on these limits rather than at an interior trade-off. For instance, at the lower limit $\tau_{\mathrm{1qb}}\approx 10\,\mathrm{ns}$ for the highest target at each $\epsns$, and at $g_{\mathrm{fac}}=1$, $\tau_{\mathrm{1qb}}=130\,\mathrm{ns}$ for the entries marked $\dagger$ in \Cref{tab:BB_iswap_MNR_data}.

Next, in \Cref{fig:MNR_1}(b) we plot the MNR efficiency $\eta$ (the ratio between circuit fidelity and average power, see \Cref{sec:MNR}) as a function of the control parameters.

\paragraph{Power and fidelity trade-offs across attenuation configurations.}
To compare the various configurations for attenuation allocation, we plot the minimum average power required for reaching a desired (fixed) circuit fidelity in \Cref{fig:MNR_2}(a), and the maximum fidelity that can be reached as a function of fixed average power in \Cref{fig:MNR_2}(b). The dual plots in \Cref{fig:MNR_2} address complementary constraints
(a) governed by application requirements and (b) governed by resource considerations. The  points shown on these plots are obtained from constrained optimization using average power heatmaps superposed with iso-fidelity contours as shown for the \cfgLogT{} configuration in \Cref{fig:MNR_1}(a).
From the plots in \Cref{fig:MNR_2}, the configurations differ by an approximately constant factor in average power at every circuit fidelity: \cfgCold{} requires about $3.5$ times, and \cfgLogT{} about $1.7$ times, the power of \cfgGeo{}, which is the least expensive in all fidelity regimes (see \Cref{tab:BB_iswap_MNR_data}). Above $\Fid\approx 0.96$ the curves steepen and draw closer, since all configurations approach the same fidelity ceiling set by the idling noise; at a fixed average power their attainable fidelities therefore nearly coincide, while at a fixed fidelity the power ratio is unchanged.
The ordering follows from the Carnot-weighted attenuation sum in \Cref{eq:stages_stack}, which is approximately $5.9\times10^{5}$, $1.0\times10^{6}$, and $2.1\times10^{6}$ for \cfgGeo{}, \cfgLogT{}, and \cfgCold{}, respectively, and alone reproduces the power ratios above. Although \cfgCold{} has the lowest thermal photon number ($\bar{n}\approx 0.15$, against $0.24$ for \cfgLogT{} and $0.35$ for \cfgGeo{}), this advantage barely affects the circuit fidelity here, because the idling noise dominates the error budget; the configuration with the smallest Carnot-weighted sum is therefore the least expensive.
\paragraph{Peak power and total energy consumption.}
To make the comparison between configurations for attenuation allocations rooted in practical constraints,we plot the power as a function of the circuit depth in \Cref{fig:power_at_depth} for the three attenuation configurations at a common operating point ($g_{\mathrm{fac}}=1$, $\tau_{\mathrm{1qb}}=25\,\mathrm{ns}$). This power profile provides an additional constraint of maximum power required for QRAM operations. Alongside the time-averaged power, the instantaneous demand on the control electronics is of independent interest, since it is the peak rather than the mean that the wiring and the cooling budget must accommodate. Gates within a layer act concurrently, so their powers add up;
the power drawn during layer $l$, containing $N_{\mathrm{1qb}}(l)$ single- and $N_{\mathrm{2qb}}(l)$ two-qubit gates, is therefore
\begin{equation}\label{eq:layer_power}
	P(l) = P_{\mathrm{drs}}\bigl(N_{\mathrm{1qb}}(l) + g_{\mathrm{fac}}\,N_{\mathrm{2qb}}(l)\bigr),
\end{equation}
with $\sum_{l} N_{\mathrm{1qb}}(l) = N_{\mathrm{1qb}}$ and $\sum_{l} N_{\mathrm{2qb}}(l) = N_{\mathrm{2qb}}$, and the peak power of the circuit is $\Ppeak = \max_{l \leq D} P(l)$. Note
that $\Ppeak$ is not obtained by re-weighting \Cref{eq:avg_power}: in a
layer of maximum arity two that also contains single-qubit gates, the latter
complete after $\tau_{\mathrm{1qb}}$ and idle for the remainder of the step, so that
\Cref{eq:layer_power} reports the demand at the start of the layer while the energy
in $\Pavg$ charges each gate only its own duration. The two quantities are thus
complementary, $\Pavg$ measuring the load averaged over the query and
$\Ppeak$ the maximum power that the hardware must sustain.

\paragraph{Power and fidelity trade-offs across idling noise.}
The attenuation configuration fixes how much power a given fidelity costs; the
idling noise fixes which fidelities are available at all. To separate the two, in
\Cref{fig:MNR_3} we
repeat the dual presentation of \Cref{fig:MNR_2} for the three idling-noise
values $\epsns = 10^{-6}, 10^{-7}, 10^{-8}$ ($\tau_{\mathrm{ref}} = 100\,\text{ns}$ is fixed for all three), with the attenuation
allocation fixed to be the \cfgGeo{} configuration: (a) the maximum circuit fidelity reachable at a fixed
average power, and (b) the minimum average power required for reaching a desired
circuit fidelity, both optimized over the control parameters
$(g_{\mathrm{fac}},\tau_{\mathrm{1qb}})$. Unlike the attenuation configurations, which rescale every resource by a nearly constant factor and change the gate errors only marginally through $\bar{n}$,
the idling noise shifts the accessible fidelity band itself---$0.080$--$0.953$,
$0.839$--$0.995$ and $0.971$--$0.999$ for the three values---so that
consecutive families overlap only in a narrow window and no fidelity is common to
all three. Where they do overlap, an order of magnitude
in $\epsns$ is worth
roughly a factor $34$ in average power at equal fidelity: a circuit fidelity of
$0.98$ costs $6.7\,\mathrm{W}$ at $\epsns = 10^{-7}$ and
$0.20\,\mathrm{W}$ at $10^{-8}$. Within each family the required power rises
steeply as the top of its band is approached, so that the last increments of
fidelity are by far the most expensive, and no power budget within the sweep
range ($0.095$--$132\,\mathrm{W}$) lifts one family into the regime of the next.

\paragraph{Identification of optimal resource cost.} The three resources considered so far---average power, peak power, and total energy---weigh the gate count and the schedule differently, and are therefore not necessarily minimized at the same operating point. In practice, two of them nearly are. Across the three attenuation configurations and the three idling-noise values studied, the operating point that minimizes the average power increases the energy by at most $0.8\%$ above its own minimum at the same target fidelity, and vice versa, and the point that minimizes the peak power costs at most $0.9\%$ in average power or energy. The converse does not hold: the peak power is set by the widest layer of the schedule, the $2^n$ single-qubit gates of the leaf level, whose power does not depend on $g_{\mathrm{fac}}$, and the widest two-qubit layer takes over only for $g_{\mathrm{fac}} \gtrsim 1.6$. The peak-power optimum therefore sits on the iso-fidelity contour at $g_{\mathrm{fac}}\approx1.6$, with a longer gate, whereas the average-power optimum sits at $g_{\mathrm{fac}}\approx 1.4$ (at $g_{\mathrm{fac}}\approx 2.6$ for $\epsns=10^{-8}$), and running at the average-power optimum costs up to $25\%$, $20\%$ and $12\%$ in peak power for $\epsns=10^{-6}$, $10^{-7}$ and $10^{-8}$. The three optima differ by at most $8\%$ of the $g_\mathrm{fac}$ sweep range, which reflects the flatness of the cost along the iso-fidelity contours (see \Cref{app:further}). For the hardware model considered here, minimizing the average power or the energy is therefore interchangeable, while a peak-power constraint selects its own, slightly different, operating point.

The analysis above provides a basis for selecting the hardware operating point once the binding constraint is known. Which constraint binds, i.e., a fixed circuit fidelity, a peak-power limit set by the wiring and cooling budget, or an average-power or energy budget, depends on the application and on the platform. In each case, the procedure is the same: the iso-fidelity contours of \Cref{fig:MNR_1} identify the control parameters that minimize the resource for a target fidelity, while \Cref{fig:MNR_2} indicates which attenuation configuration achieves that target at the lowest cost. \Cref{tab:BB_iswap_MNR_data} collects the outcome of this procedure for several target fidelities, across the three attenuation configurations and the three values of $\epsns$. For each target, it reports the minimum average power, the minimum energy per query, the corresponding optimal $(g_{\mathrm{fac}}, \tau_{\mathrm{1qb}})$, and the resulting peak power. The same data for a larger BB-QRAM with $n = 20$, $m = 1$ are reported in \Cref{app:large_QRAM}, where several targets become unreachable: none for $\epsns = 10^{-6}$ and $\Fid \geq 0.99$ for $\epsns = 10^{-7}$. The tables can therefore be read as a lookup: given a target fidelity and a power or energy budget, they show whether a configuration meets all the requirements and at which control parameters. Since the optimal control parameters for a given target fidelity differ between configurations, the choice of attenuation profile and of the operating point should be made jointly.

\Cref{tab:BB_iswap_MNR_data,tab:BB_iswap_MNR_data_n20} also show that, at a fixed target fidelity, the resource cost changes by one to two orders of magnitude between consecutive values of $\epsns$. For $\Fid \geq 0.9$ with the \cfgGeo{} configuration in \Cref{tab:BB_iswap_MNR_data}, for instance, the minimum average power drops from $22.1\,$W at $\epsns = 10^{-6}$ to $0.256\,$W at $\epsns = 10^{-7}$. This difference originates from the error budget. Since the idling error is charged to every qubit at every layer, a larger $\epsns$ consumes a larger share of the infidelity allowed by the target, leaving a smaller budget for the gate errors. The same circuit fidelity can then be attained only with higher-fidelity gates, which in our model require a stronger drive and hence a higher power.
\begin{table}
	\centering
	\footnotesize
	\caption{Energetic comparison of the attenuation configurations and
		idling-noise strengths for the BB-QRAM with $n = 12$ and $m = 1$.
		$^{\dagger}$The fidelity constraint is inactive: for
		$\epsns = 10^{-8}$ the circuit fidelity exceeds $0.970$ over
		the entire sweep range, so the reported minimum is attained at the boundary
		of the domain, $g_{\mathrm{fac}} = 1$ and
		$\tau_{\mathrm{1qb}} = 130\,\mathrm{ns}$, and is set by the limits of the landscape rather than by the fidelity target.}
	\setlength{\tabcolsep}{2pt}
	\begin{tabular}{@{}p{0.17\columnwidth}
		p{0.33\columnwidth}
		>{\raggedleft\arraybackslash}p{0.14\columnwidth}
		>{\raggedleft\arraybackslash}p{0.14\columnwidth}
		>{\raggedleft\arraybackslash}p{0.14\columnwidth}@{}}

		\toprule
		                                   &                                    & \textsc{Cold-}\newline\textsc{Heavy} & \textsc{Geo-}\newline\textsc{metric} & \cfgLogT{}       \\
		\midrule
		\multicolumn{5}{@{}l}{\textit{$\epsns = 10^{-6}$}}                                                                                                                       \\
		\cmidrule(lr){1-5}
		\multirow{5}{*}{$\Fid \geq 0.9$}   & Min avg.\ $\Pavg$ (W)              & 79.5                                 & 22.1                                 & 37.1             \\
		                                   & Min energy ($\mu$J)                & 2240                                 & 621                                  & 1040             \\
		                                   & Optimal $g_{\mathrm{fac}}$         & 1.55                                 & 1.55                                 & 1.55             \\
		                                   & Optimal $\tau_{\mathrm{1qb}}$ (ns) & 10.1                                 & 10.1                                 & 10.1             \\
		                                   & Peak $\Ppeak$ (W)                  & 842                                  & 233                                  & 393              \\
		\cmidrule(lr){1-5}
		\multirow{5}{*}{$\Fid \geq 0.95$}  & Min avg.\ $\Pavg$ (W)              & 409                                  & 114                                  & 191              \\
		                                   & Min energy ($\mu$J)                & 5670                                 & 1570                                 & 2650             \\
		                                   & Optimal $g_{\mathrm{fac}}$         & 13                                   & 13                                   & 13               \\
		                                   & Optimal $\tau_{\mathrm{1qb}}$ (ns) & 10                                   & 10                                   & 10               \\
		                                   & Peak $\Ppeak$ (W)                  & 6980                                 & 1940                                 & 3260             \\
		\midrule
		\multicolumn{5}{@{}l}{\textit{$\epsns = 10^{-7}$}}                                                                                                                       \\
		\cmidrule(lr){1-5}
		\multirow{5}{*}{$\Fid \geq 0.9$}   & Min avg.\ $\Pavg$ (W)              & 0.903                                & 0.256                                & 0.425            \\
		                                   & Min energy ($\mu$J)                & 238                                  & 66.8                                 & 112              \\
		                                   & Optimal $g_{\mathrm{fac}}$         & 1.45                                 & 1.41                                 & 1.46             \\
		                                   & Optimal $\tau_{\mathrm{1qb}}$ (ns) & 92.6                                 & 90.6                                 & 92.3             \\
		                                   & Peak $\Ppeak$ (W)                  & 10                                   & 2.91                                 & 4.72             \\
		\cmidrule(lr){1-5}
		\multirow{5}{*}{$\Fid \geq 0.95$}  & Min avg.\ $\Pavg$ (W)              & 3.73                                 & 1.06                                 & 1.76             \\
		                                   & Min energy ($\mu$J)                & 485                                  & 136                                  & 227              \\
		                                   & Optimal $g_{\mathrm{fac}}$         & 1.45                                 & 1.4                                  & 1.38             \\
		                                   & Optimal $\tau_{\mathrm{1qb}}$ (ns) & 45.6                                 & 44.4                                 & 44.4             \\
		                                   & Peak $\Ppeak$ (W)                  & 41.5                                 & 12.1                                 & 20.4             \\
		\cmidrule(lr){1-5}
		\multirow{5}{*}{$\Fid \geq 0.99$}  & Min avg.\ $\Pavg$ (W)              & 96.2                                 & 27.3                                 & 45.4             \\
		                                   & Min energy ($\mu$J)                & 2470                                 & 693                                  & 1160             \\
		                                   & Optimal $g_{\mathrm{fac}}$         & 1.98                                 & 2.04                                 & 2.01             \\
		                                   & Optimal $\tau_{\mathrm{1qb}}$ (ns) & 10.1                                 & 10.1                                 & 10.1             \\
		                                   & Peak $\Ppeak$ (W)                  & 1040                                 & 298                                  & 493              \\
		\midrule
		\multicolumn{5}{@{}l}{\textit{$\epsns = 10^{-8}$}}                                                                                                                       \\
		\cmidrule(lr){1-5}
		\multirow{5}{*}{$\Fid \geq 0.9$}   & Min avg.\ $\Pavg$ (W)              & 0.342$^{\dagger}$                    & 0.0948$^{\dagger}$                   & 0.16$^{\dagger}$ \\
		                                   & Min energy ($\mu$J)                & 147                                  & 40.7                                 & 68.6             \\
		                                   & Optimal $g_{\mathrm{fac}}$         & 1                                    & 1                                    & 1                \\
		                                   & Optimal $\tau_{\mathrm{1qb}}$ (ns) & 130                                  & 130                                  & 130              \\
		                                   & Peak $\Ppeak$ (W)                  & 5.1                                  & 1.41                                 & 2.38             \\
		\cmidrule(lr){1-5}
		\multirow{5}{*}{$\Fid \geq 0.95$}  & Min avg.\ $\Pavg$ (W)              & 0.342$^{\dagger}$                    & 0.0948$^{\dagger}$                   & 0.16$^{\dagger}$ \\
		                                   & Min energy ($\mu$J)                & 147                                  & 40.7                                 & 68.6             \\
		                                   & Optimal $g_{\mathrm{fac}}$         & 1                                    & 1                                    & 1                \\
		                                   & Optimal $\tau_{\mathrm{1qb}}$ (ns) & 130                                  & 130                                  & 130              \\
		                                   & Peak $\Ppeak$ (W)                  & 5.1                                  & 1.41                                 & 2.38             \\
		\cmidrule(lr){1-5}
		\multirow{5}{*}{$\Fid \geq 0.99$}  & Min avg.\ $\Pavg$ (W)              & 2.52                                 & 0.798                                & 1.25             \\
		                                   & Min energy ($\mu$J)                & 399                                  & 118                                  & 192              \\
		                                   & Optimal $g_{\mathrm{fac}}$         & 2.45                                 & 2.51                                 & 2.36             \\
		                                   & Optimal $\tau_{\mathrm{1qb}}$ (ns) & 68                                   & 64.2                                 & 65               \\
		                                   & Peak $\Ppeak$ (W)                  & 28.5                                 & 9.11                                 & 14               \\
		\cmidrule(lr){1-5}
		\multirow{5}{*}{$\Fid \geq 0.999$} & Min avg.\ $\Pavg$ (W)              & 265                                  & 85.3                                 & 132              \\
		                                   & Min energy ($\mu$J)                & 4340                                 & 1320                                 & 2110             \\
		                                   & Optimal $g_{\mathrm{fac}}$         & 7.19                                 & 8.79                                 & 7.86             \\
		                                   & Optimal $\tau_{\mathrm{1qb}}$ (ns) & 10                                   & 10                                   & 10               \\
		                                   & Peak $\Ppeak$ (W)                  & 3880                                 & 1310                                 & 1980             \\
		\bottomrule
	\end{tabular}
	\label{tab:BB_iswap_MNR_data}
\end{table}

\subsection{Applications}
\label{sec:applications}

To connect the BB-QRAM's physical cost to an end-to-end application, we consider
a simple quantum machine-learning workflow based on principal-component
analysis. Our aim is to identify the
resources required to operate this workflow at the data scales for which its
quantum query complexity can outperform that of a classical counterpart,
under favorable simplifying assumptions. We translate a
quantum advantage based on the query complexity into more realistic requirements on QRAM size, fidelity, peak power, and energy per query. %

\begin{table}[tbp]
	\centering
	\caption{Parameters necessary for the quantum steps, their descriptions, and chosen values. The top part of the table specifies the parameter for the quantum binary search, and the second part describes the parameter for the quantum PCA extraction. Values marked by $\dagger$ are estimated from data.}
	\label{tab:pcc_runtime_parametersKDD}
	\begin{adjustbox}{max width=\linewidth}
		\begin{tabular}{clc}
			\toprule
			\textbf{Param.}      & \textbf{Description}                        & \textbf{Value} \\
			\midrule
			$\|X\|_F$            & Frobenius norm of the matrix                & 3611$^\dagger$ \\
			$\varepsilon_\theta$ & Error in approximation of singular values   & 1              \\
			$\eta$               & Error in approximation of retained variance & 0.1            \\
			\hline
			$k$                  & Number of principal components              & 6              \\
			$d$                  & Number of features                          & 50             \\
			$N$                  & Number of training data points              & $4\times10^6$  \\
			$\theta$             & Cut-off threshold for retained variance     & 632$^\dagger$  \\
			$\delta$             & $\ell_2$-error on principal components      & 0.1            \\
			$p$                  & Amount of retained variance                 & 0.7            \\
			$\|X\|$              & Spectral norm of the dataset                & 1780$^\dagger$ \\
			$\xi$                & Failure probability of quantum algorithm    & $1/d=0.02$     \\
			\bottomrule
		\end{tabular}
	\end{adjustbox}
\end{table}

A recent work by Bellante \emph{et al.}~\cite{BELLANTE2025104341} proposes a systematic framework for assessing the potential speedups that fault-tolerant quantum machine learning (QML) algorithms might offer in cybersecurity tasks—particularly those involving network intrusion detection. In short, the methodology has three main steps: (i) selecting a quantum algorithm with formal runtime guarantees; (ii) introducing approximation errors in a classical simulation to reflect expected quantum errors; and (iii) measuring key dataset parameters (e.g., matrix norms and condition numbers) that govern the quantum (and classical) algorithm's complexity.
By comparing the quantum and classical complexities in terms of these parameters, the authors pinpoint data scales at which a ``quantum advantage'' may theoretically emerge. This allows practitioners to decide whether the size and structure of their cybersecurity datasets justify adopting quantum methods.
The paper demonstrates this approach using different anomaly detection models derived from PCA, which is obtained using a simulation of qPCA~\cite{bellante2022quantum}. In their work, they show how to tune the parameters of the quantum algorithm to balance runtime efficiency with performance (in their case, accuracy of the classification on anomaly detection). The runtime of the quantum algorithm is specified in terms of the number of queries to a circuit that is a block-encoding of a dataset matrix $X$. It is possible to create this unitary using $2\log(Nd)$ queries to a QRAM storing the data structure derived from the dataset~\cite{gilyen2019quantum}.

\paragraph{Principal Component Classification (PCC).} For our experiments, we focus on a model called \emph{Principal Component Classification}. The PCC computes
\begin{align*}
	C_1=\sum_{i=1}^k\frac{y_i^2}{\lambda_i} \quad\text{and}\quad C_2=\sum_{i=r-q+1}^r\frac{y_i^2}{\lambda_i}
\end{align*}
where $y_i=\boldsymbol{e}_i^\top\boldsymbol{z}$ and $\boldsymbol{z}$ is a new test point (appropriately standardized and scaled), $\lambda_i$ are the eigenvalues of the covariance matrix (ordered decreasingly), $\boldsymbol{e}_i$ are the eigenvectors of the covariance matrix and $k$ and $r$ are threshold on the largest and smallest components. An observation is classified as anomalous if $C_1>c_1$ or $C_2>c_2$, with thresholds derived from a false-alarm rate $\alpha \in (0,1)$~\cite{PcaMajorMinor}.
Fitting the PCC classifiers reduces to fitting a PCA model, which in the quantum setting is done via QPCA~\cite{bellante2022quantum}. Following~\cite{BELLANTE2025104341}, we focus on a model for the major components that only uses $C_1>c_1$ (i.e., $C_2$ is ignored). The thresholds $c_1$ and $c_2$ are computed on a validation set as follows: for a given false-alarm rate $\alpha$, the quantile $q = \sqrt{1-\alpha}$ is computed and $c_1$ (respectively $c_2$) is set to the $q$-th quantile of the empirical distribution of $C_1$ (respectively $C_2$) values computed on the validation set~\cite{PcaMajorMinor}. As $\alpha$ increases, $q$ decreases, leading to lower thresholds and more observations being classified as attacks.
The value of $k$ was chosen in order to retain 70\% of the variance in the major components. The dataset uses numerical features (normalized and with constant features removed). The first step of the algorithm is the quantum binary search for $\theta$ (the threshold value of the eigenvalues that are retaining 70\% of the variance in the major components), and the second step is the quantum principal component extraction of such components. The two steps have a runtime, respectively, of
\begin{align*}
	                & \widetilde{O}\left(\frac{\|X\|_F\log(\|X\|_F/\varepsilon_\theta)}{\varepsilon_\theta\eta}\right)                  \\
	\text{and}\quad & \widetilde{O}\left(\frac{dk\lVert X\rVert\|X\|_F\log(k)\log(d)}{\theta\sqrt{p}\varepsilon_\theta\delta^2}\right).
\end{align*}
Here $\widetilde{O}(\cdot)$ hides polylogarithmic factors in the number of samples $N$ and number of features $d$.
Moreover, $p$ represents the user-specified retained variance, $\varepsilon_\theta$ is the approximation error in performing singular value estimation (a generalization of phase estimation), and $\delta$ is the $\ell_2$-error which we tolerate in the reconstruction of the singular vectors $\boldsymbol{e}_i$, while $\eta$ is the approximation error in the retained variance during the binary search (see \Cref{tab:pcc_runtime_parametersKDD} for a description of parameters relevant to the problem).
The runtimes are expressed in number of queries to the oracles preparing the block-encodings. The failure probability of the algorithms is chosen to be $1/d$.
The authors compared this to randomized classical PCA with runtime $\widetilde{O}(Nkd\log k)$ and observed a query-complexity advantage only for large datasets (around $N \approx 4\times 10^6$ samples and $d \approx 50$ features).

The above thresholds refer to the query-complexity crossing, i.e., they show where the quantum algorithm's asymptotic gate/query count becomes formally smaller than its classical counterpart. Real-world implementations must also account for quantum error correction overhead and slower quantum clock speeds vs.\ classical, meaning that thresholds can be higher. If these physical-layer costs dominate, a dataset may need to be even larger in order to realize a true ``wall-clock time'' advantage.

To translate this application-scale threshold into physical resources, we
evaluate one BB-QRAM query using the superconducting
full-stack model and the fixed-point assumptions of \Cref{sec:fixed_point}.
Specifically, we use the \cfgLogT{} attenuation allocation,
$g_{\mathrm{fac}}=1$, and $f_{\mathrm{fac}}=4$, while retaining the other
fixed-point parameters. The compiled query uses $3.15\times 10^{6}$ qubits and contains $5.31\times 10^7$ single-qubit gates and $4.4\times 10^7$ two-qubit gates.
Its total depth is $2094$, with at most $1.31\times 10^6$ active qubits in any layer.
The one- and two-qubit gate errors
are $3.09\times10^{-8}$ and $1.24\times10^{-7}$, respectively, yielding a query
infidelity $1-\Fid=0.13$ ($\Fid=0.87$) and propagation correction
$\delta=0.027$. The query reaches a peak dynamic power of
$1.65\times10^{4}\,\mathrm{W}$ and consumes
$9.01\times10^{-2}\,\mathrm{J}$ of dynamic energy. These values characterize
one application-scale QRAM query; the aggregate cost of the complete QML
routine additionally depends on the number of block-encoding oracle calls
specified by the complexities above, the cost of preparing a block-encoding, and the architecture of the QRAM for $m>1$.

\section{Discussion and conclusions}
\label{sec:discussion}

In this work, we explored the energetic cost of querying a QRAM circuit on a superconducting hardware, in particular, a modified version of the bucket-brigade architecture based on (controlled) iSWAP gates. The dynamic energetic cost was studied through the MNR framework~\cite{Fellous-Asiani2023Oct}, a methodology that analyzes the minimum resource cost, like energy or power in our case, required in a quantum computation while ensuring that a performance metric, like fidelity, is maintained above a certain threshold. Within the MNR framework, we adopt (i) an energetic model that relates the device's energetic footprint to the energetics of quantum gates, and (ii) a mixed-unitary error model describing the effect of noise on qubits. The former model, based on~\cite{Fellous-Asiani2023Oct}, describes the trade-off between gate performance (fidelity) and gate cost (energy or power) as a function of several system properties, e.g., temperature stages, attenuation profiles, qubit frequency, spontaneous emission rate, and gate duration, some of which are more controllable than others. It is this energetic model that ultimately sets the single- and two-qubit gate errors. On the other hand, the latter model describes how these gate infidelities actually affect the hardware qubits when running a QRAM circuit. We have further complemented the error model with an environment noise $\epsns$ affecting all qubits, regardless of whether they are idling or not. The final QRAM circuit fidelity depends on its architecture and has been studied in~\cite{hann2021resilience} for the bucket-brigade architecture.

Under the above models, we calculated the energetic footprint for querying once a BB-QRAM with $n=12$ address qubits and $m=1$ target qubit. As an example, we have found that an energetic cost of $621\,\mu\text{J}$, an average power of $22.1\,\text{W}$, and a peak power of $233\,\text{W}$ are required to keep a circuit fidelity of at least $0.9$ under an environment noise $\epsns = 10^{-6}$ with the \cfgGeo{} attenuation allocation. Such figures can be drastically improved to $40.7\,\mu\text{J}$ (energy), $0.0948\,\text{W}$ (average power), and $1.41\,\text{W}$ (peak power) under a lower environment noise of $\epsns = 10^{-8}$.

Although the energy level required for one BB-QRAM query is relatively modest (as a result of the short superconducting-based circuit duration), its peak power can be extremely high due to the exponential circuit width of the bucket-brigade architecture that concentrates most of its gates at end, shortly before reading the classical memory (see \Cref{fig:bucket-brigade-cswap}). This issue can be mitigated by breaking up layers with many gates into several layers with less gates. As long as the resulting circuit has a depth mildly larger than the original circuit, the fidelity stays mostly unchanged. For example, the peak power of $64\,\text{W}$ in~\Cref{tab:fixed_point} can be reduced down to $16\,\text{W}$ by breaking the few wide layers into multiple steps, which brings the depth up to $1{,}424$ from the original circuit depth of $1{,}214$. For $n=19$ we can reduce the peak power from $8{,}238\,\text{W}$ to $200\,\text{W}$ by increasing the depth from $1{,}984$ to $5{,}350$.

As previously mentioned, the choice for environment noise $\epsns\in\{10^{-6},10^{-7},10^{-8}\}$ is quite optimistic given that state-of-the-art physical experiments can only reach error levels of $10^{-5}$. On the other hand, the value of $\epsns$ has a massive impact on the energetic resources, since the MNR model pushes the requirement of meeting the minimum fidelity onto smaller gate errors and, consequently, onto the enabling technologies. This evidences the need for an environment noise as small as possible. Apart from the obvious and non-trivial solution of demanding better superconducting qubits, an easier alternative is to refine the fidelity analysis of the bucket-brigade, either by improving the arguments of Hann \emph{et al.}~\cite{hann2021resilience} or focusing on very specific error channels. The latter is mostly appropriate to better analyze the parameter $\delta$ from \Cref{subsec:bb-cswaps}, which was upper bounded in a very loose manner. Another solution is to enhance the noise resilience of the bucket-brigade by error correcting it. Recent works~\cite{dalzell2025distillation} suggest that light-weight quantum error codes might suffice to properly increase its noise resilience without the need for full quantum error correction.

A better noise resilience of the bucket-brigade architecture would not only improve the energetic resources but also allow for larger QRAM sizes. As evident from the results on a $20$-address-qubit BB-QRAM from \Cref{app:large_QRAM} (see \Cref{tab:BB_iswap_MNR_data_n20}), the energetic resources grow with the memory size $2^n$: at a fixed operating point they scale linearly, by $2^8=256$, while at a fixed target fidelity the growth depends strongly on $\epsns$ (about $170\times$ in average power for $\Fid\geq0.9$ at $\epsns=10^{-8}$, but about $2700\times$ at $10^{-7}$), and some targets become unreachable. This renders a QRAM physically unfeasible for large datasets. This issue stems from the standard bucket-brigade architecture being \emph{active}, in that $O(2^n)$ routers are energetically charged during a query. Recent works~\cite{jain2026scalable} have started to consider feasible \emph{passive} architectures, but their end-to-end resources like proper gate counts and depth and their precise error resilience properties are still poorly understood. The bucket-brigade QRAM is fairly well understood at this point, but no realistic passive version has been proposed yet, which could be the ideal solution for the energetic scalability with $n$.

\section{Future directions}

A first important direction is to generalize the present study beyond the regime $m=1$ and analyze memory words of size $m>1$. In that setting, the width of the target register, the number of simultaneous routing operations, and the decomposition cost of the data-loading subroutines all change non-trivially.

It will also be important to broaden the comparison to other QRAM architectures and layout choices. Besides the multiplexer and bucket-brigade families considered here, one could study fan-out, constant-depth, approximate, or hardware-aware variants under the same native-gate and MNR assumptions. A unified comparison across architectures and platforms would help identify which proposals remain competitive once circuit width, control overhead, and physical energy costs are accounted for on equal footing.

Another natural extension of this work is to move from QRAM queries to the Quantum Random Access Gate (QRAG), and extend the energetic cost to creating block-encodings and amplitude encodings from data stored in QRAM, and ultimately to performing end-to-end applications, using a similar framework as described in \Cref{sec:applications}.

\begin{acknowledgments}
	We thank Yidong Liao and Konstantina Koteva for useful discussions. JFD is supported by the Lendület “Momentum” program of the Hungarian Academy of Sciences under grant agreement no.\ LP2025-8/2025. AL is supported by the National Research Foundation, Singapore through the National Quantum Office, hosted in A*STAR, under its Advanced Quantum Computing Infrastructure Funding Initiative (S25Q8D9001) and its Centre for Quantum Technologies Funding Initiative (S24Q2d0009). HV is supported by France 2030 under the French National Research Agency grant numbers ANR-22-QMET-0002 (BACQ project) as part of the MetriQs-France program and grant number ANR-22-EXES-0013, as well as under the OECQ project financed by BPI France with France 2030 and Next Generation EU via France Relance), and by the National Research Foundation, Singapore through the National Quantum Office, hosted in A*STAR, under its Centre for Quantum Technologies Funding Initiative (S24Q2d0009).
\end{acknowledgments}

\section*{Artificial Intelligence Disclaimer}

AI has been used for coding, checking and improving the resource counts from \Cref{tab:resources_table}, and scanning for typos and mistakes. Ideas, proofs, and writing are the product of the authors alone.

\bibliographystyle{unsrt}

\bibliography{biblioz.bib}

\appendix

\section{MNR for cross-resonance gates}
\label{sec:MNR_CR_gates}

As an example of the abstract relations introduced in \Cref{sec:MNR_superc}, the approximate forms of $g_{\mathrm{fac}}$ and $f_{\mathrm{fac}}$ can be derived for a class of two-qubit gates called cross-resonance (CR) gates. In such gates, the control qubit is driven near the target qubit's frequency, generating an effective two-qubit interaction mediated by the exchange coupling~\cite{Rigetti2010,Chow2011}. To leading order in the drive amplitude, the rate of the desired interaction scales as $\Omega J/\Delta$, where $J$ is the qubit-qubit coupling and $\Delta=\omega_A-\omega_B$ is the detuning~\cite{Sheldon2016,Malekakhlagh2020}. This form applies in the perturbative regime, while higher-order corrections in the drive amplitude become important for stronger drives~\cite{Malekakhlagh2020}. The CR gate therefore provides a microscopic realization of the assumptions underlying \Cref{eq:gfac_duration} in the regime where the effective interaction is approximately linear in the applied control amplitude.

We first consider the decoherence-limited gate. During a two-qubit operation both qubits are exposed to decoherence, so that, under the same Markovian noise assumptions used for the single-qubit gate, the leading contribution to the infidelity is
\begin{equation}
\epstwo^{\mathrm{dec}}
\approx
2\gamma(1+\bar{n})\tau_{2\mathrm{qb}} .
\label{eq:cr_eps_dec}
\end{equation}
The factor of two accounts for the two qubits participating in the operation. This approximation neglects differences between the coherence properties of the control and target qubits and additional drive-induced noise, but retains the duration dependence relevant to the MNR phenomenology.

The power consumption follows from the same logic as for single-qubit gates, with one modification. Achieving a $\pi/2$ rotation of an effective interaction of strength $\Omega J/\Delta$ in time $\tau_{2\mathrm{qb}}$ requires, within the conventions used here,
\begin{equation}
\Omega
=
\frac{\pi\Delta}{2J\tau_{2\mathrm{qb}}}.
\end{equation}
Substituting this amplitude into the MNR power-amplitude relation gives
\begin{equation}
P_{2\mathrm{qb}}
\approx
\frac{\hbar\omega\pi^{2}\Delta^{2}}
{16\gamma J^{2}\tau_{2\mathrm{qb}}^{2}} .
\label{eq:cr_power}
\end{equation}
Compared to single-qubit gates, the factor $\Delta^{2}/J^{2}$ reflects the indirect nature of the driven interaction: weaker exchange coupling or larger detuning requires a stronger drive to achieve the same gate duration, increasing the power requirement relative to a resonantly driven single-qubit operation. The heat evacuated by a cryostat operating at finite efficiency is governed by \Cref{eq:stages_stack}, and therefore the corresponding dissipated power can be obtained by substituting $P_{\mathrm{prim}}$ therein.

We may now compare these expressions with the phenomenology. Dividing \Cref{eq:cr_power} by \Cref{eq:single_qubit_power}, and \Cref{eq:cr_eps_dec} by \Cref{eq:single_qubit_infidelity}, gives
\begin{equation}
g_{\mathrm{fac}}^{\mathrm{CR}}
=
\frac{\Delta^{2}}{4J^{2}}
\left(
\frac{\tau_{1\mathrm{qb}}}{\tau_{2\mathrm{qb}}}
\right)^{2},
\qquad
f_{\mathrm{fac}}^{\mathrm{CR}}
=
2\frac{\tau_{2\mathrm{qb}}}{\tau_{1\mathrm{qb}}}.
\label{eq:cr_factors}
\end{equation}
The first expression reproduces the quadratic duration dependence of \Cref{eq:gfac_duration}, with $\Delta/2J$ setting the effective relative efficiency with which the applied drive generates the two-qubit interaction. The second differs from \Cref{eq:ffac_duration} by a factor of two because both qubits decohere during the gate, whereas the single-qubit reference contains the contribution from only one. This factor changes the value of the effective hardware parameter but not the structure of the phenomenology. In particular, the duration ratio cancels in the product,
\begin{equation}
g_{\mathrm{fac}}^{\mathrm{CR}}
\left(f_{\mathrm{fac}}^{\mathrm{CR}}\right)^2
=
\frac{\Delta^{2}}{J^{2}}
\equiv
\alpha_{\mathrm{CR}}^{2}.
\label{eq:cr_alpha}
\end{equation}
Thus, in the perturbative and decoherence-limited regime,
\begin{equation}
\alpha_{\mathrm{CR}}
\approx
\left|\frac{\Delta}{J}\right|,
\end{equation}
and the CR gate inherits the duration-independent energy-infidelity product assumed in \Cref{eq:gfac_ffac_link}. The microscopic model therefore modifies the value assigned to $\alpha$ while preserving the power-infidelity scaling on which the phenomenology is based.

Real CR gates contain additional error mechanisms that delimit this correspondence. Of particular relevance is residual static coupling between the qubits. A residual $\mathrm{Z}\mathrm{Z}$ interaction may be written schematically as
\begin{equation}
H_{\mathrm{res}}
=
\frac{\zeta}{2}\mathrm{Z}\mathrm{Z} ,
\label{eq:cr_residual_hamiltonian}
\end{equation}
and generates an unwanted coherent phase $\phi_{\mathrm{res}}=\zeta\tau_{2\mathrm{qb}}$. Residual dispersive interactions of this type are an important source of coherent error and crosstalk in superconducting processors and can modify the effective CR rotation~\cite{Ku2020,Krinner2020}. Unlike Markovian decoherence, however, a static residual Hamiltonian does not by itself define an incoherent error rate. For a small coherent rotation, the accumulated phase is linear in time while its contribution to average gate infidelity is quadratic in that phase~\cite{Sheldon2016}. We may therefore write, to leading order,
\begin{equation}
\varepsilon_{\mathrm{static}}
\approx
C_{\zeta}
\left(\zeta\tau_{2\mathrm{qb}}\right)^2 ,
\label{eq:cr_static_eps}
\end{equation}
where $C_{\zeta}$ is an order-unity coefficient determined by the precise residual Hamiltonian, target operation, and fidelity convention. This quadratic accumulation is the characteristic behavior of a systematic coherent error, in contrast to the linear accumulation of non-unitary error under the Markovian approximation~\cite{Sheldon2016}.

Including this leading coherent correction gives
\begin{equation}
\epstwo
\approx
2\gamma(1+\bar{n})\tau_{2\mathrm{qb}}
+
C_{\zeta}
\left(\zeta\tau_{2\mathrm{qb}}\right)^2
+\cdots ,
\label{eq:cr_eps}
\end{equation}
where the $\dots$ includes other coherent terms in the effective CR Hamiltonian, leakage, spectator interactions, drive-induced noise, and higher-order corrections in the drive amplitude~\cite{Malekakhlagh2020,Sundaresan2020}. Echoed CR sequences and additional rotary drives are commonly used to suppress such unwanted Hamiltonian and spectator terms~\cite{Sundaresan2020}.

\Cref{eq:cr_eps} also makes explicit the regime in which the phenomenology is faithful to the CR hardware. Defining $\Gamma=\gamma(1+\bar{n})$, the ratio between the residual coherent contribution and the decoherence contribution is
\begin{equation}
\frac{\varepsilon_{\mathrm{static}}}
{\varepsilon_{\mathrm{dec}}}
\approx
\frac{C_{\zeta}\zeta^{2}\tau_{2\mathrm{qb}}}
{2\Gamma}.
\label{eq:cr_error_ratio}
\end{equation}
The phenomenological relation \Cref{eq:gfac_ffac_link} is recovered when
\begin{equation}
C_{\zeta}\zeta^{2}\tau_{2\mathrm{qb}}
\ll
2\gamma(1+\bar{n}),
\label{eq:cr_decoherence_limited}
\end{equation}
so that the gate is decoherence limited and $\epstwo\propto\tau_{2\mathrm{qb}}$. Outside this regime the coherent term introduces an additional power of the gate duration into $f_{\mathrm{fac}}$, and the cancellation in \Cref{eq:cr_alpha} no longer holds exactly. The resulting departure from a constant $\alpha$ is therefore not a failure of the microscopic CR description, but marks the boundary of the decoherence-limited approximation assumed in constructing the phenomenology.

Fluctuations of the residual coupling require a separate distinction. If $\zeta$ varies in time or between experimental realizations, $\zeta(t)=\bar{\zeta}+\delta\zeta(t)$, its fluctuating component can generate dephasing rather than a deterministic coherent rotation. The corresponding duration dependence is then determined by the noise spectrum and need not follow the quadratic law of \Cref{eq:cr_static_eps}. Similarly, coupling to uncontrolled spectator degrees of freedom can produce an effectively non-unitary channel after those degrees of freedom are traced out. Such effects may be incorporated into an effective decoherence rate when their dynamics are approximately Markovian, but are not represented by the static parameter $\zeta$ in \Cref{eq:cr_residual_hamiltonian}. This distinction is important because coherent and non-unitary errors exhibit different accumulation with repeated operations and can be separated experimentally~\cite{Sheldon2016}; spectator and residual-interaction errors in CR gates likewise require explicit characterization and suppression~\cite{Sundaresan2020}.

It is instructive to gauge the magnitude of the effective parameter in the decoherence-limited regime. From \Cref{eq:cr_alpha}, $\alpha_{\mathrm{CR}}\approx|\Delta/J|$, which for $\Delta/2\pi\in[80,300]\,\mathrm{MHz}$ and $J/2\pi\in[1.5,5]\,\mathrm{MHz}$~\cite{Sheldon2016,kjaergaard2020superconducting} spans approximately $16$--$200$. This range should not be interpreted as a universal value characteristic of CR gates: the effective interaction depends on detuning, anharmonicity, drive amplitude, pulse construction, and higher-order corrections neglected in the leading-order expression~\cite{Malekakhlagh2020}. Nor do we import these values directly into the main text, where $\alpha$ is fixed from the reference operating point of \Cref{sec:results}. The purpose of this exercise is instead to establish which features of the phenomenology emerge from a microscopic gate model. In the perturbative and decoherence-limited regime, CR gates reproduce the quadratic duration dependence of $g_{\mathrm{fac}}$, the linear duration dependence of $f_{\mathrm{fac}}$, and consequently the duration-independence of their constrained product. Residual coherent interactions and other non-decoherence-limited errors provide corrections to this relation and explicitly delimit the regime in which the single-parameter phenomenology applies.

\section{Further experiments}
\label{app:further}

The analysis in \Cref{sec:results} fixes all hardware parameters at the values of \Cref{tab:fixed_values} and optimizes over the two runtime control parameters $(g_{\mathrm{fac}},\tau_{1\mathrm{qb}})$ alone. Some questions remain in the analysis though. The first is which part of the error model sets the attainable fidelity, since the iso-fidelity contours of \Cref{fig:MNR_1}  span a narrow range and do not reveal whether the variation stems from the gate errors or from idling. The second is whether the choice of resource cost---average power, peak power, or total energy---changes the operating point. The third is how the conclusions change when parameters held fixed by the cryostat and the fabrication, namely the qubit temperature and the qubit frequency, are allowed to move. We address these questions with three sets of experiments, one per pair of parameters, each shown for the \cfgLogT{} allocation, and %
for $\epsns = 10^{-7}$ and $10^{-8}$.

\subsection{Runtime controls: sources of error and choice of resource}
\label{app:runtime}

The fidelity bound of \Cref{sec:qrams} combines four contributions: the idling error $\epsns$ charged to every router and incident mode at every layer and the one- and two-qubit gate errors $\epsone$ and $\epstwo$ charged only at the layers in which a gate acts. %
Among these, only the gate errors depend on the control parameters through \Cref{eq:single_qubit_infidelity}, while the idling contribution is fixed by $\epsns$ and the layer durations. At the benchmark point of \Cref{tab:fixed_point}, the gate errors $\epsone = 3.09 \times 10^{-8}$ and $\epstwo = 1.24 \times 10^{-7}$ are comparable to the idling error $\epsns(l) = 10^{-7}$ of a two-qubit layer. The gate errors, however, are charged only at the layers in which a router or its incident mode is acted upon, whereas the idling error is charged at every layer, and it enters the propagation correction $\delta$ with the weight $d-1$ of \Cref{eq:delta_bound}. The idling contribution therefore accumulates over far more router-layer pairs than the gate contributions do. This is the origin of the observation in \Cref{sec:results} that the idling noise, rather than the attenuation allocation or the drive control parameters, sets the band of accessible fidelities. On the other hand, shortening the gates reduces the duration of every layer, and hence the idling error, but the power required to do so grows as $\tau_{\mathrm{1qb}}^{-2}$. The trade-off navigated by the MNR optimization is then effectively between the drive power and the idling error.

\begin{figure*}[tbp]
	\centering
	\includegraphics[width=0.49\textwidth]{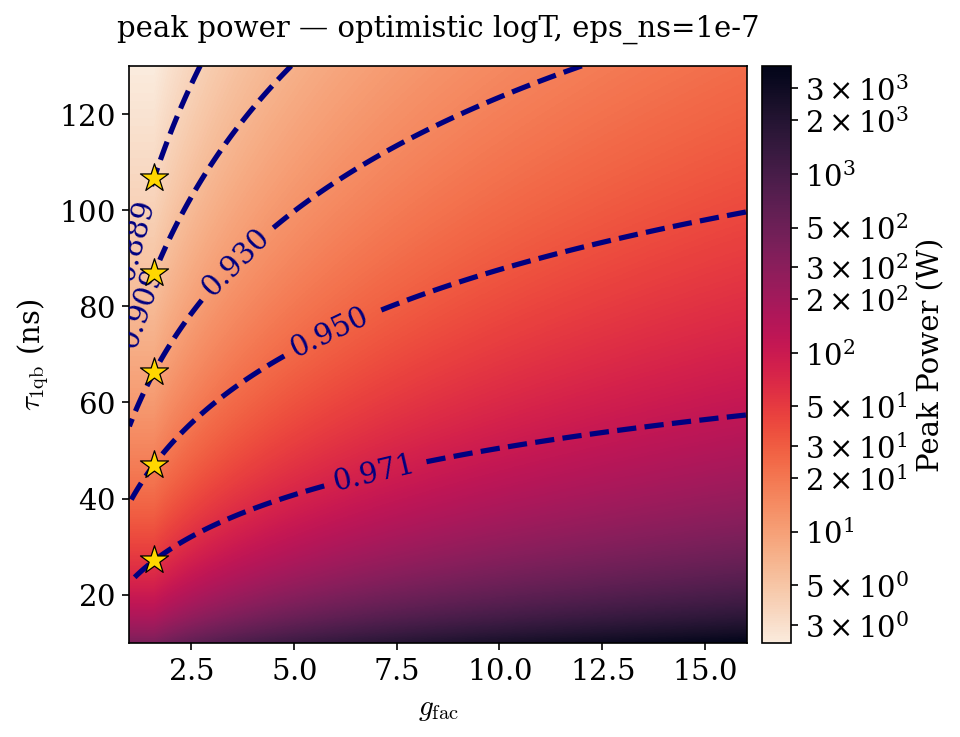}\hfill
	\includegraphics[width=0.49\textwidth]{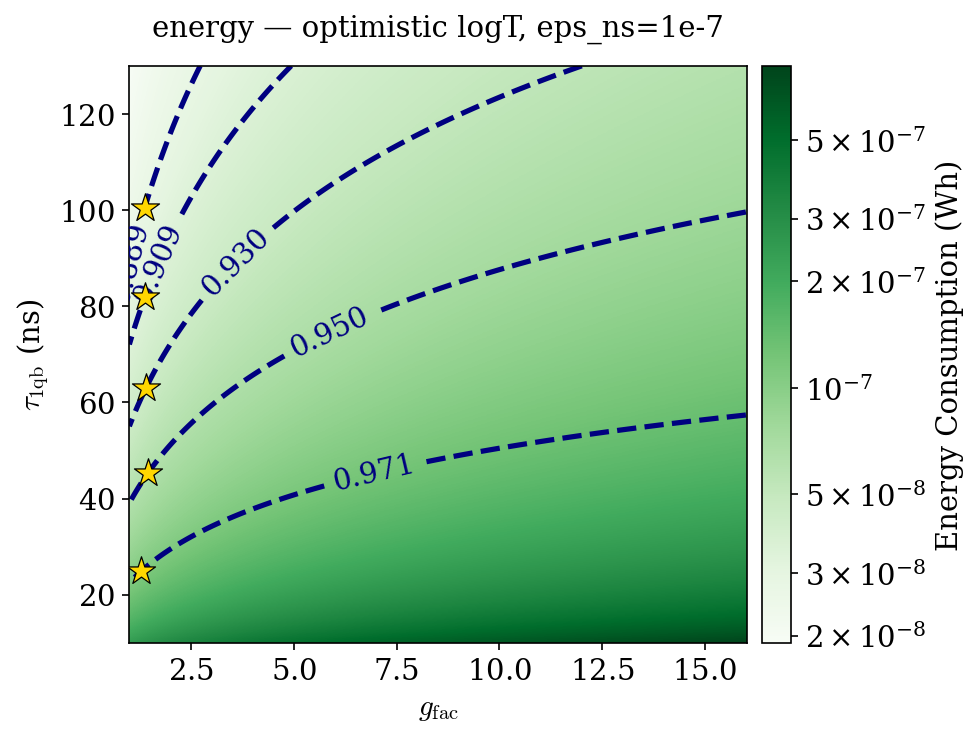}\\[1ex]
	\includegraphics[width=0.98\textwidth]{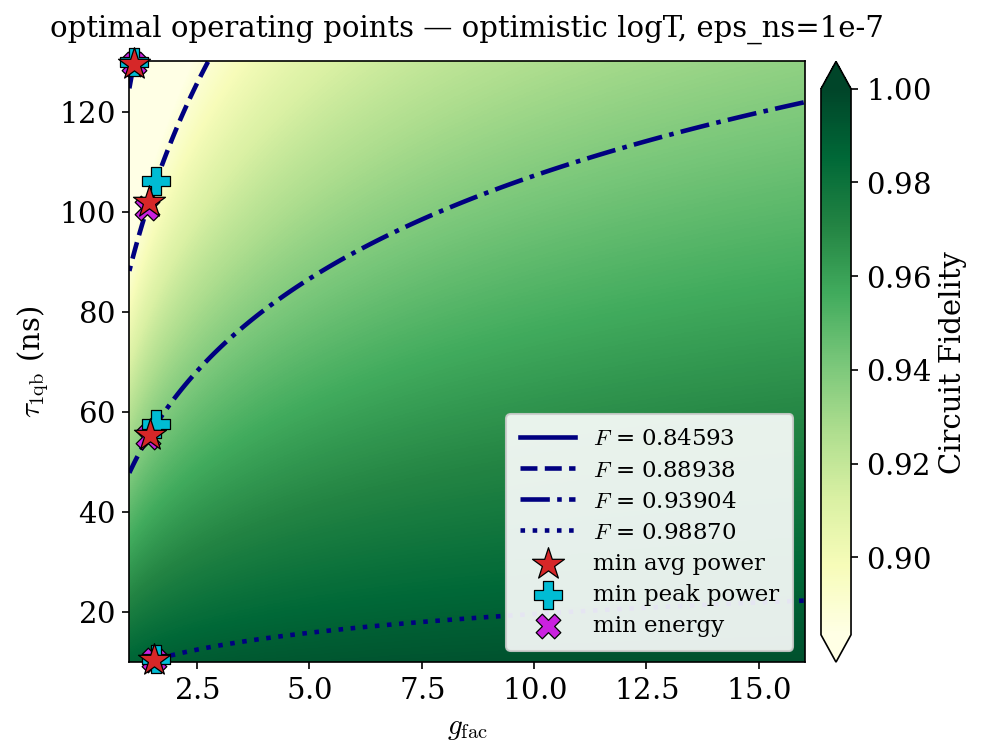}
	\caption{Run-time controls $(g_{\mathrm{fac}}, \tau_{1\mathrm{qb}})$ for the BB-QRAM query ($n = 12$, $m = 1$), \cfgLogT{} allocation, and $\epsns = 10^{-7}$. Top panel, left (a) peak power $\Ppeak$ and right (b) energy per query $\Edyn$ over the control plane, overlaid with the iso-fidelity contours; stars mark the minimum of each resource along each contour. Bottom panel: Optimal $g_{\mathrm{fac}}$ (left) and $\tau_{1\mathrm{qb}}$ (right) as a function of the target fidelity, for the minimum average power (star), minimum peak power ($+$) and minimum energy ($\times$). The average power and the energy select nearly the same operating point at every target; the peak power selects $g_{\mathrm{fac}}\approx 1.6$, where the widest two-qubit layer starts to set the peak. Below $\Fid \approx 0.84$ the optima lie on the boundary of the range of $g_{\mathrm{fac}} \in [1, 16]$ and $\tau_{1\mathrm{qb}} \in [10, 130]\,\mathrm{ns}$.}
	\label{fig:app_gfac_tau}
\end{figure*}

The three resources weigh the gate count and the schedule differently. The average power $\Pavg$ rewards a weak drive, the energy $\Edyn$ additionally rewards a short wall-clock time, and the peak power $\Ppeak$ is fixed by the widest layer of the schedule rather than by any average over it. It is therefore not evident a priori that the three are minimized at the same operating point. In the $(g_{\mathrm{fac}}, \tau_{1\mathrm{qb}})$-plane the average power and the energy are, and the peak power nearly is. The peak-power and energy landscapes of \Cref{fig:app_gfac_tau}(a,b) carry the same iso-fidelity contours as the average-power landscape of \Cref{fig:MNR_1}(a); the energy optima fall on the same points, while the peak-power optima sit slightly to the right of them, at $g_{\mathrm{fac}}\approx 1.6$: below this value the peak is set by the $2^n$ single-qubit gates of the leaf level and does not depend on $g_{\mathrm{fac}}$, so the peak-power optimum trades a larger $g_{\mathrm{fac}}$ for a longer gate. \Cref{fig:app_gfac_tau}(c) makes this explicit by tracing the operating point selected by each resource as a function of the target fidelity. The optimal gate duration decreases almost linearly with the target, while the optimal $g_{\mathrm{fac}}$ for the average power remains within $1.3$--$1.6$ for $\epsns = 10^{-7}$. At low targets, $\Fid \lesssim 0.84$, the fidelity constraint is inactive and all three optima sit at the corner $g_{\mathrm{fac}} = 1$, $\tau_{1\mathrm{qb}} = 130\,\mathrm{ns}$ of the optimization landscape. For this configuration, optimizing the average power while the hardware is constrained by the energy, or vice versa, costs less than $0.1\%$ of the constrained resource, and optimizing the peak power costs less than $0.3\%$ in either; optimizing the average power under a peak-power constraint costs up to $16\%$ ($25\%$ across all configurations and idling-noise values; see \Cref{sec:results}). This is consistent with the separation argument of \Cref{sec:results}: at fixed $g_{\mathrm{fac}}$, $\Pavg$ and $\Ppeak$ share the same $\tau_{1\mathrm{qb}}^{-2}$ dependence, and along an iso-fidelity contour the gate duration is pinned closely enough that the $\tau_{1\mathrm{qb}}^{-1}$ dependence of $\Edyn$ does not move the optimum.

\subsection{Qubit temperature}
\label{app:temperature}

The main text fixes the qubit temperature and frequency, which are set by the cryostat and the fabrication. Both, however, pull the fidelity and the resource cost in opposite directions. Although they are design-time choices, they indicate which platform parameters are worth improving.

\begin{figure*}[tbp]
	\centering
	\includegraphics[width=0.49\textwidth]{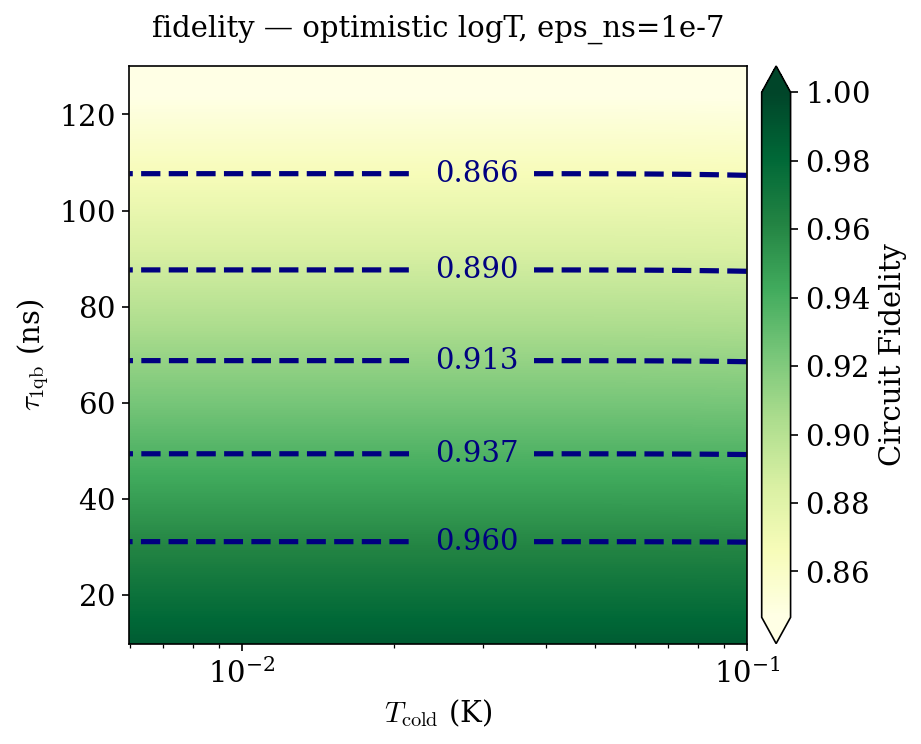}\hfill
	\includegraphics[width=0.49\textwidth]{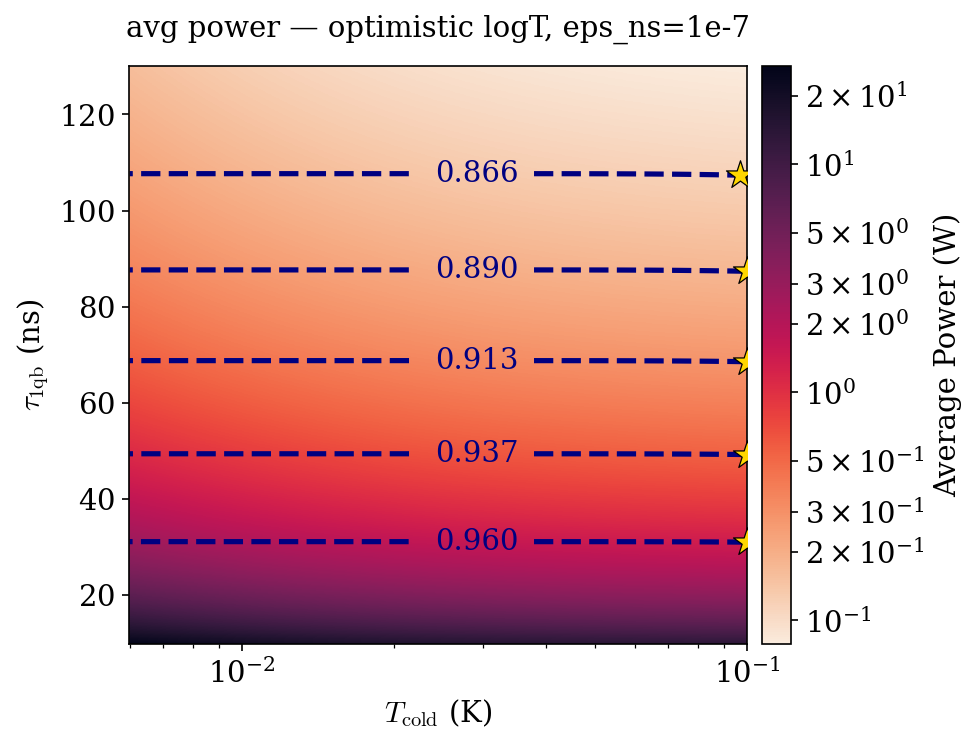}\\[1ex]
	\includegraphics[width=0.49\textwidth]{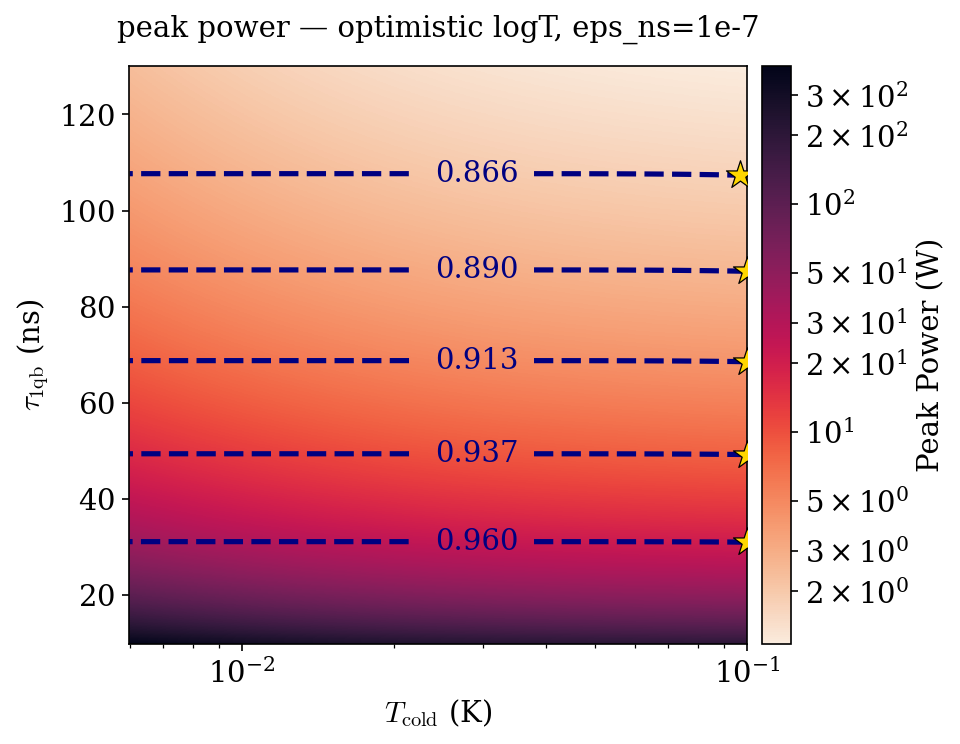}\hfill
	\includegraphics[width=0.49\textwidth]{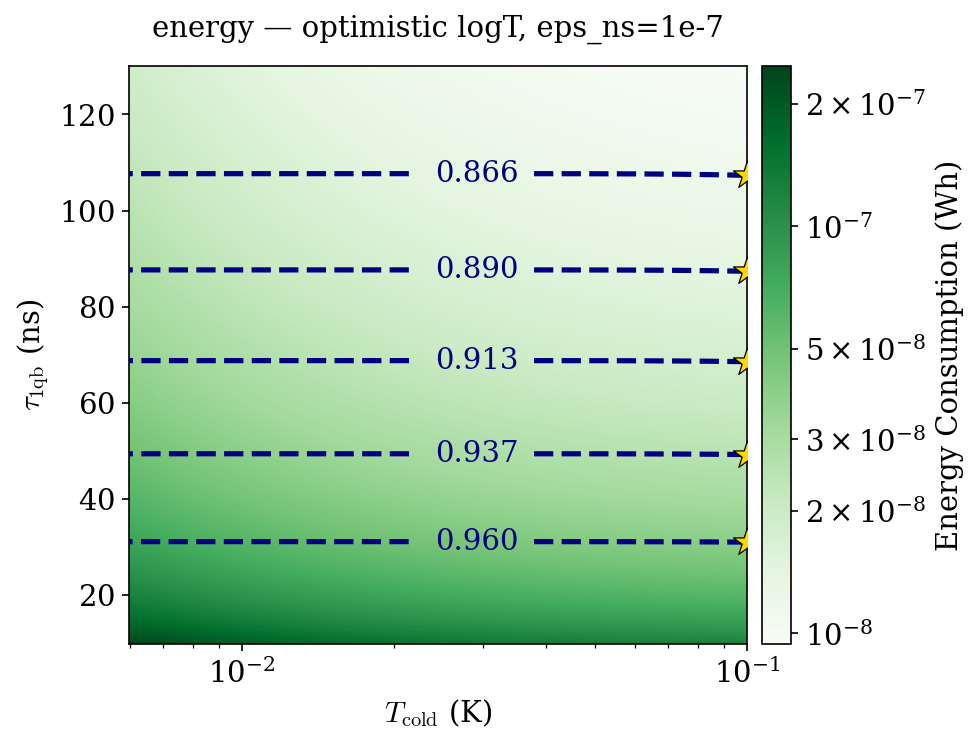}
	\caption{Temperature $T_1$ of the coldest stage against the single-qubit gate duration $\tau_{1\mathrm{qb}}$, for the \cfgLogT{} allocation, $g_{\mathrm{fac}} = 1$, and $\epsns = 10^{-7}$. (a) Circuit fidelity with iso-fidelity contours and the operating points of minimum average power (star), minimum peak power ($+$) and minimum energy ($\times$) on each contour. (b) Average power, (c) peak power, and (d) energy per query, overlaid with the same contours; stars mark the minimum of each resource along each contour. The limits of the optimization parameters are $6\,\mathrm{mK} \le T_1 \le T_2 = 100\,\mathrm{mK}$ with all other stage temperatures fixed. The contours are flat and the resources decrease monotonically with $T_1$, so that the optima lie at the upper limit. This reflects the temperature independence of $\gamma$ and $\epsns$ in the model and should not be read as an optimal operating temperature.}
	\label{fig:app_temp_tau_eps7}
\end{figure*}

\begin{figure*}[tbp]
	\centering
	\includegraphics[width=0.49\textwidth]{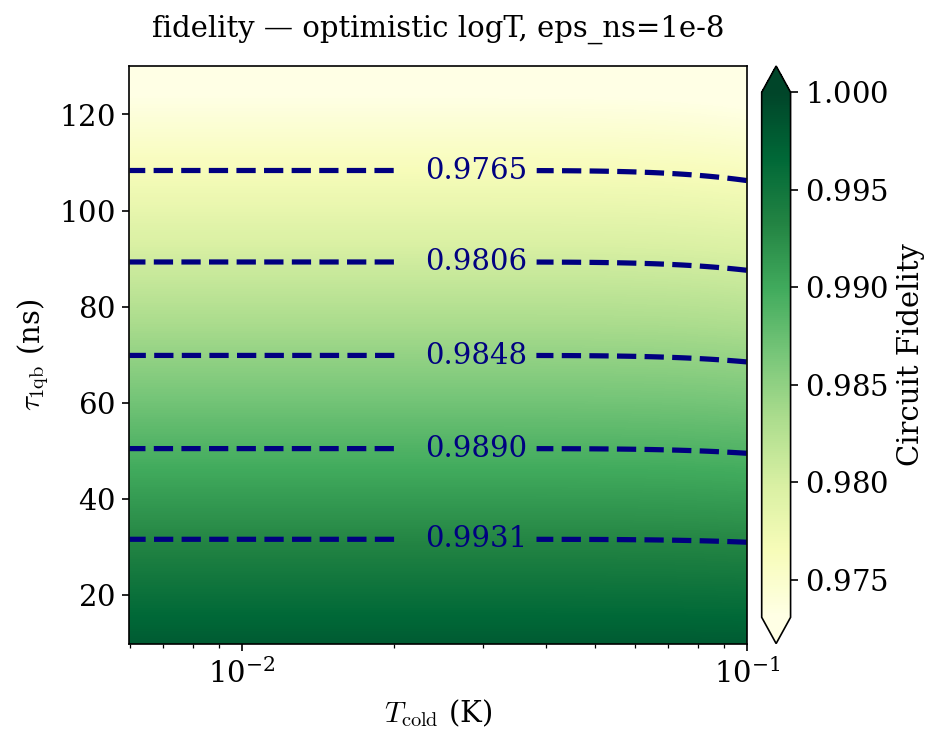}\hfill
	\includegraphics[width=0.49\textwidth]{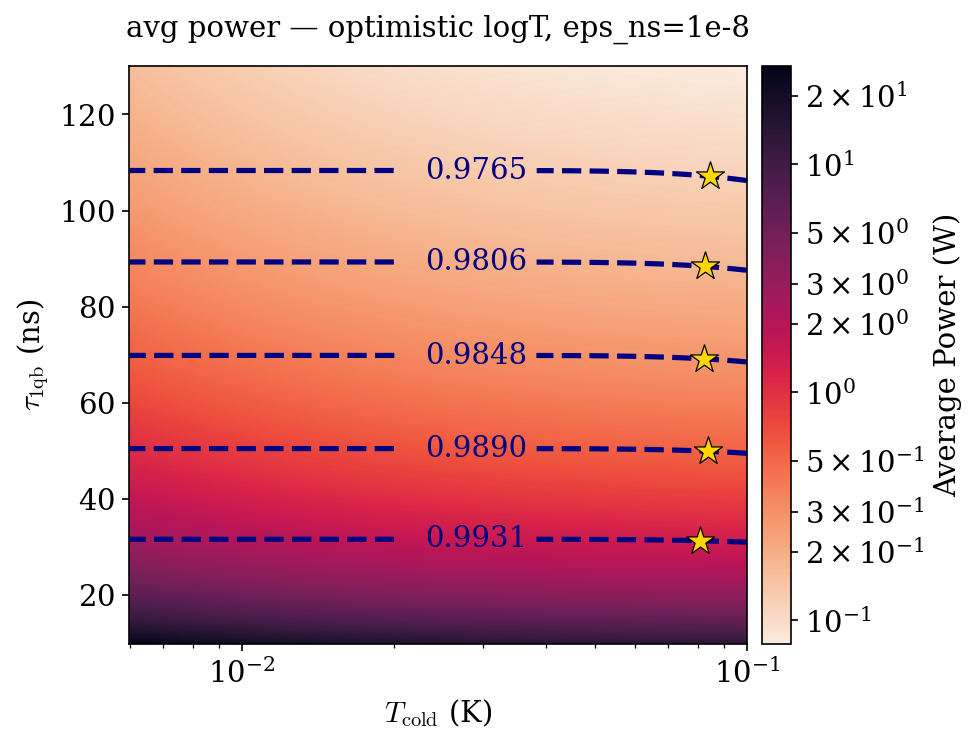}\\[1ex]
	\includegraphics[width=0.49\textwidth]{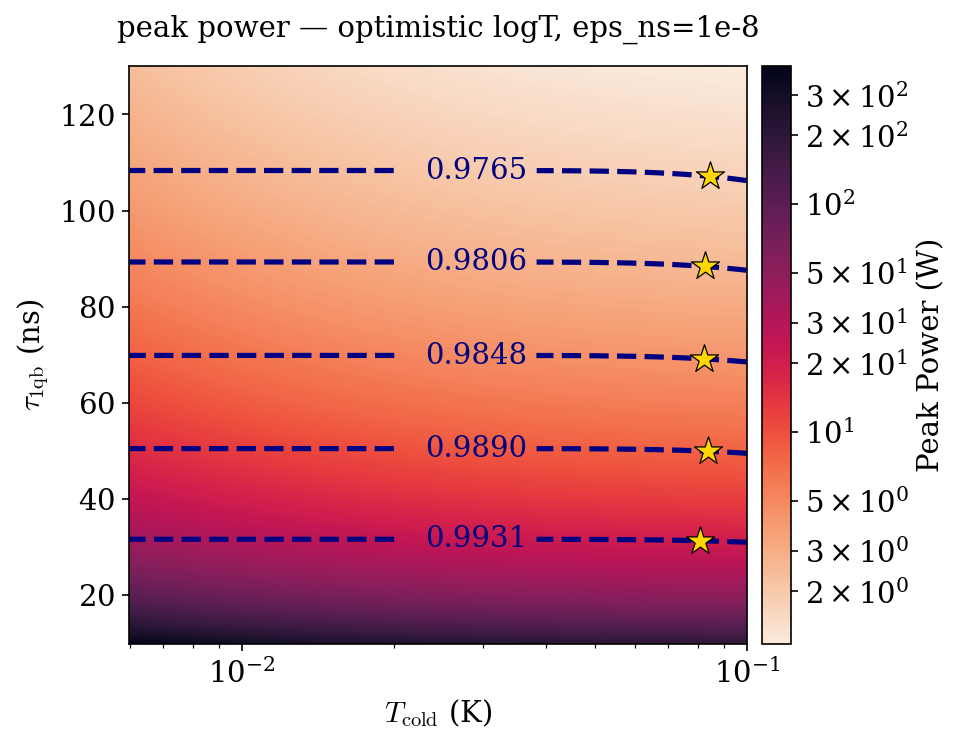}\hfill
	\includegraphics[width=0.49\textwidth]{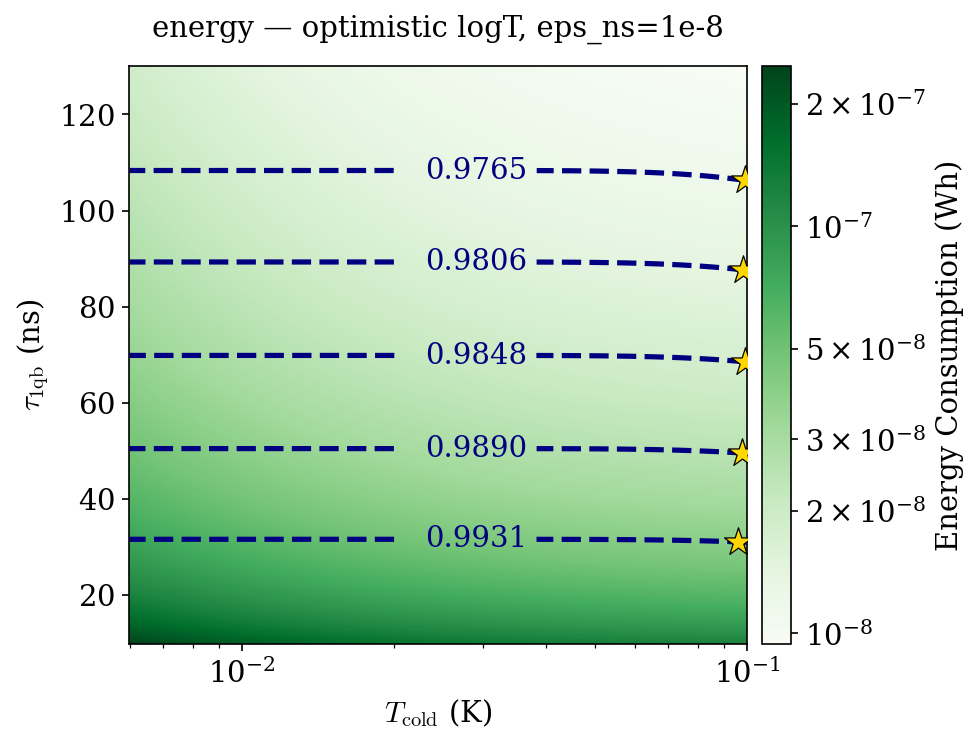}
	\caption{Same as \Cref{fig:app_temp_tau_eps7} for $\epsns = 10^{-8}$. Top left: Circuit fidelity with iso-fidelity contours and optima. Top right: average power. Bottom left: peak power. Bottom right: energy per query. With the idling error reduced, the gate errors weigh more in the error budget and the contours bend slightly above approximately $60\,\mathrm{mK}$. The power optima move marginally inside the landscape, to approximately $85\,\mathrm{mK}$, while the energy optimum remains close to the upper limit ($94$--$99\,\mathrm{mK}$).}
	\label{fig:app_temp_tau_eps8}
\end{figure*}

Lowering the temperature of the coldest stage suppresses the thermal occupation $\bar{n}$ of \Cref{eq:effective_noise} and hence the gate infidelity, while inflating the Carnot factors of \Cref{eq:stages_stack}. We sweep the temperature $T_1 \equiv T_{\mathrm{qb}}$ of the coldest stage between $6\,\mathrm{mK}$ and the temperature of the next stage, $T_2 = 100\,\mathrm{mK}$, keeping the \cfgLogT{} allocation and all other stage temperatures fixed. \Cref{fig:app_temp_tau_eps7,fig:app_temp_tau_eps8} show that within this range the iso-fidelity contours are flat for both idling-noise values: at $\omega/2\pi = 6\,\mathrm{GHz}$ the thermal occupation of the coldest stage remains small compared with the photons leaking from the warmer stages, and the idling error, which carries no temperature dependence, dominates the error budget. The dressed power decreases monotonically with $T_1$, since the Carnot factor of the coldest attenuator falls from $4.9 \times 10^4$ at $6\,\mathrm{mK}$ to $2.9 \times 10^3$ at $100\,\mathrm{mK}$. The average power at fixed control parameters is thereby reduced to $0.53$ times its benchmark value at $50\,\mathrm{mK}$ and to $0.50$ times at $100\,\mathrm{mK}$, with no loss of circuit fidelity. For $\epsns = 10^{-7}$ all optima therefore sit at the upper limit in the landscape (\Cref{fig:app_temp_tau_eps7}). For $\epsns = 10^{-8}$, shown in \Cref{fig:app_temp_tau_eps8}, the bending of the contours near the higher temperature limit $\sim 100\,\mathrm{mK}$ requires a slightly shorter gate to hold the fidelity, which moves the power optima marginally inside the landscape, to approximately $85\,\mathrm{mK}$, while the energy optimum, less sensitive to the gate duration, remains at $100\,\mathrm{mK}$.

\subsection{Qubit frequency}
\label{app:frequency}

\begin{figure*}[tbp]
	\centering
	\includegraphics[width=0.49\textwidth]{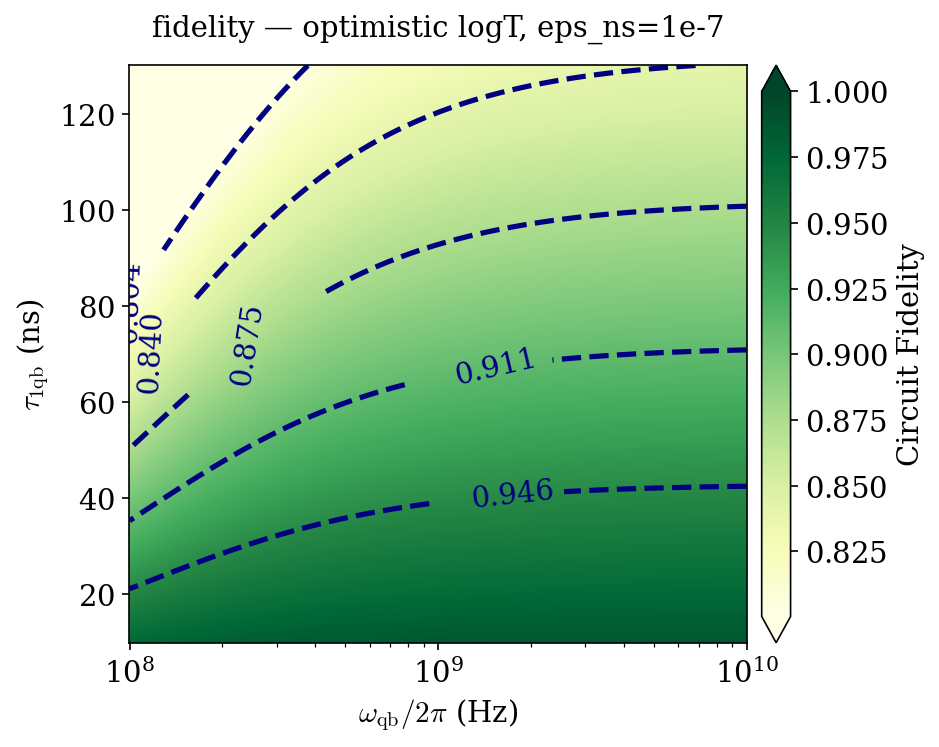}\hfill
	\includegraphics[width=0.49\textwidth]{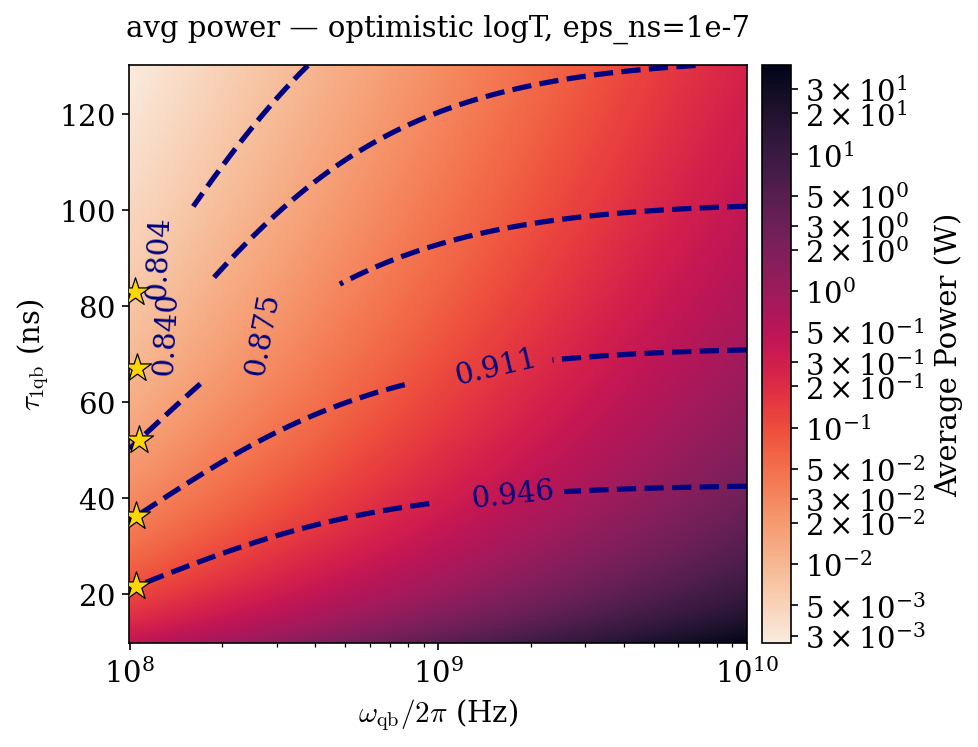}\\[1ex]
	\includegraphics[width=0.49\textwidth]{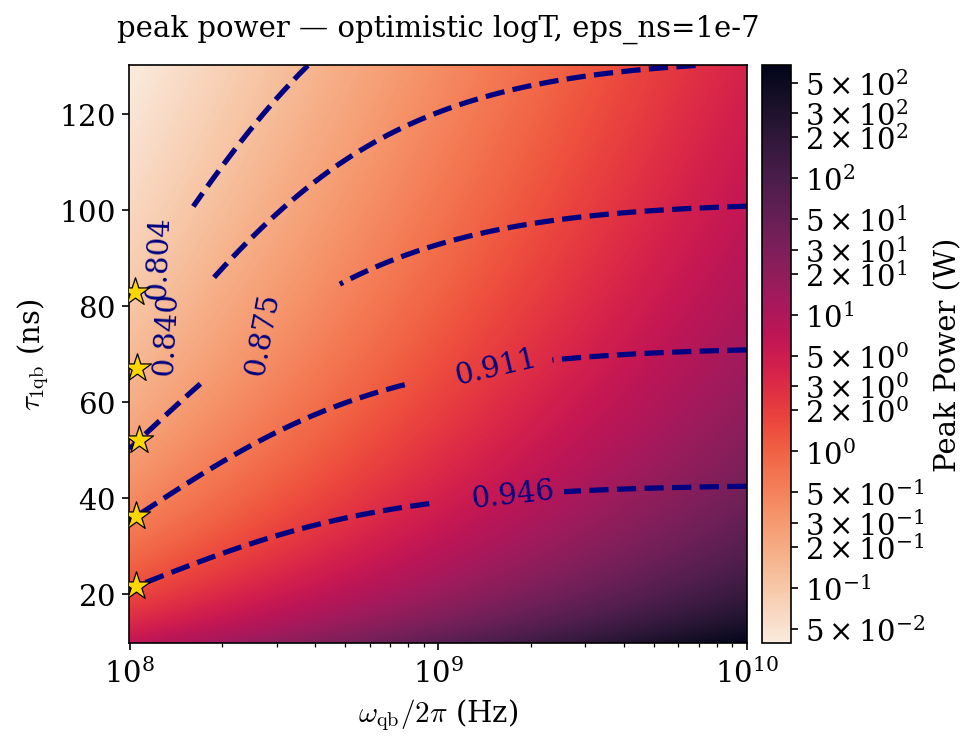}\hfill
	\includegraphics[width=0.49\textwidth]{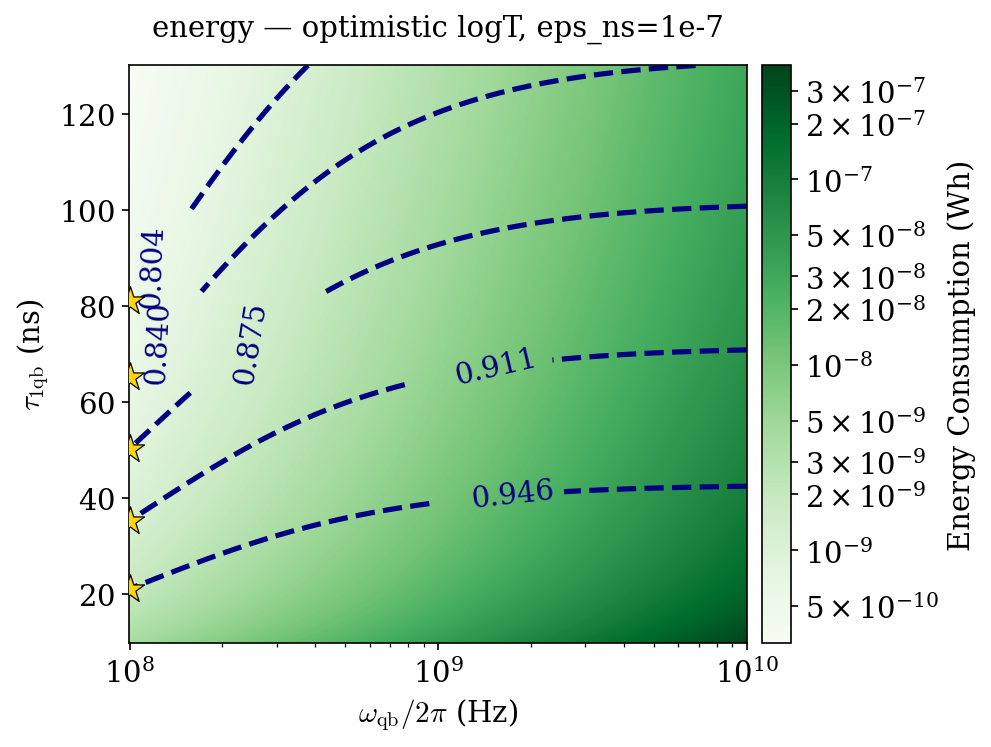}
	\caption{Qubit frequency $\omega/2\pi$ against the single-qubit gate duration $\tau_{1\mathrm{qb}}$, for the \cfgLogT{} allocation, $g_{\mathrm{fac}} = 1$, and $\epsns = 10^{-7}$. Top left: Circuit fidelity with iso-fidelity contours and the operating points of minimum average power (star), minimum peak power ($+$) and minimum energy ($\times$) on each contour. Top right: Average power. Bottom left: peak power. Bottom right: energy per query, overlaid with the same contours; stars mark the minimum of each resource along each contour. The fidelity is insensitive to $\omega$ above $\approx 1\,\mathrm{GHz}$, where the idling error dominates, and all optima lie at the lower limit in the landscape. The spontaneous emission rate $\gamma$ is held independent of $\omega$.}
	\label{fig:app_omega_tau_eps7}
\end{figure*}

Raising the qubit frequency suppresses $\bar{n}$ through the Bose-Einstein distribution while increasing the energy $\hbar\omega$ of every drive photon. We sweep $\omega/2\pi$ between $0.1$ and $10\,\mathrm{GHz}$, keeping the \cfgLogT{} allocation and $g_{\mathrm{fac}} = 1$. For $\epsns = 10^{-7}$ the average power grows linearly with $\omega$, following \Cref{eq:single_qubit_power}, while the circuit fidelity is insensitive to $\omega$ above approximately $1\,\mathrm{GHz}$, where the idling error dominates (see \Cref{fig:app_omega_tau_eps7}). The fidelity degrades only below this frequency, where $\hbar\omega$ approaches $k_B T$ at the colder stages and $\bar{n}$ grows, so that the optimal operating points sit at the lowest frequencies in the optimization landscape. Within the transmon range of $4$--$8\,\mathrm{GHz}$, the cost therefore simply scales with the qubit frequency and a lower frequency is preferred for as long as the thermal occupation remains negligible.

\begin{figure*}[tbp]
	\centering
	\includegraphics[width=0.49\textwidth]{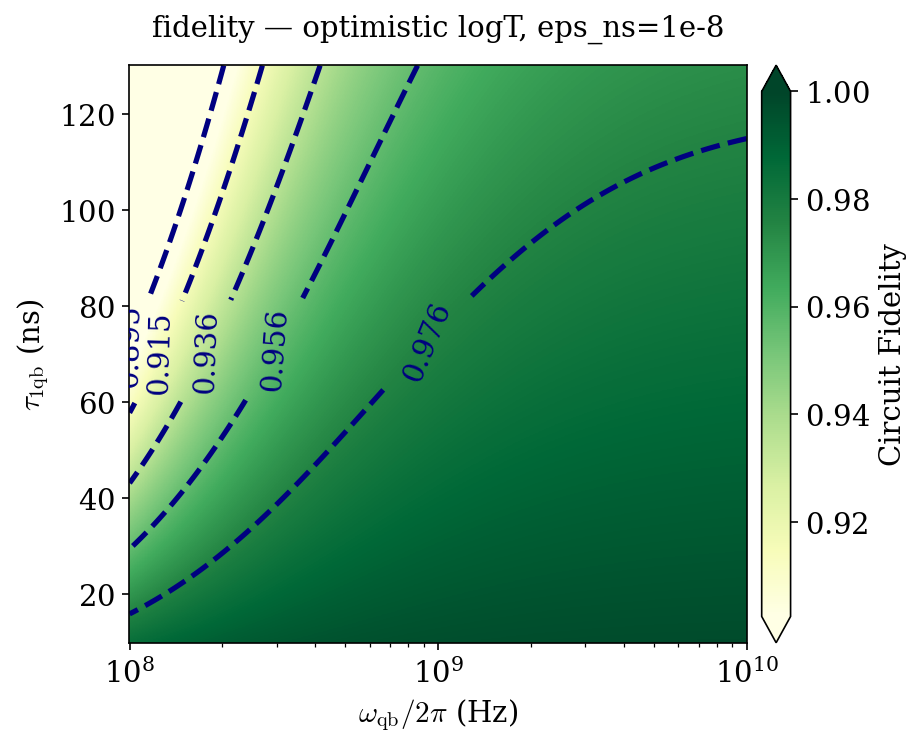}\hfill
	\includegraphics[width=0.49\textwidth]{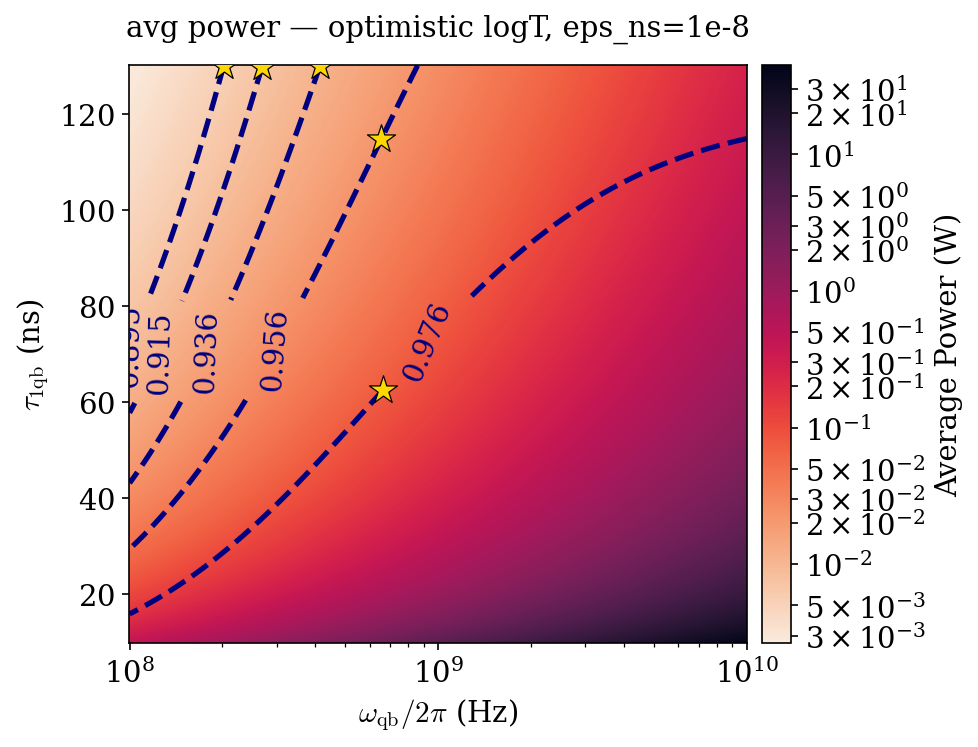}\\[1ex]
	\includegraphics[width=0.49\textwidth]{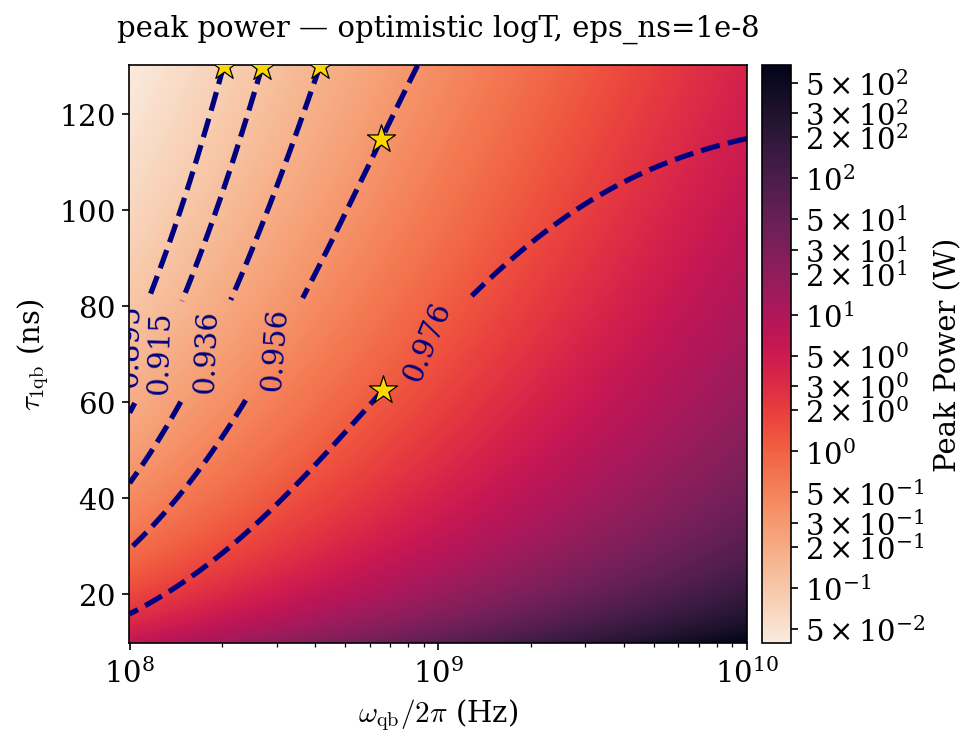}\hfill
	\includegraphics[width=0.49\textwidth]{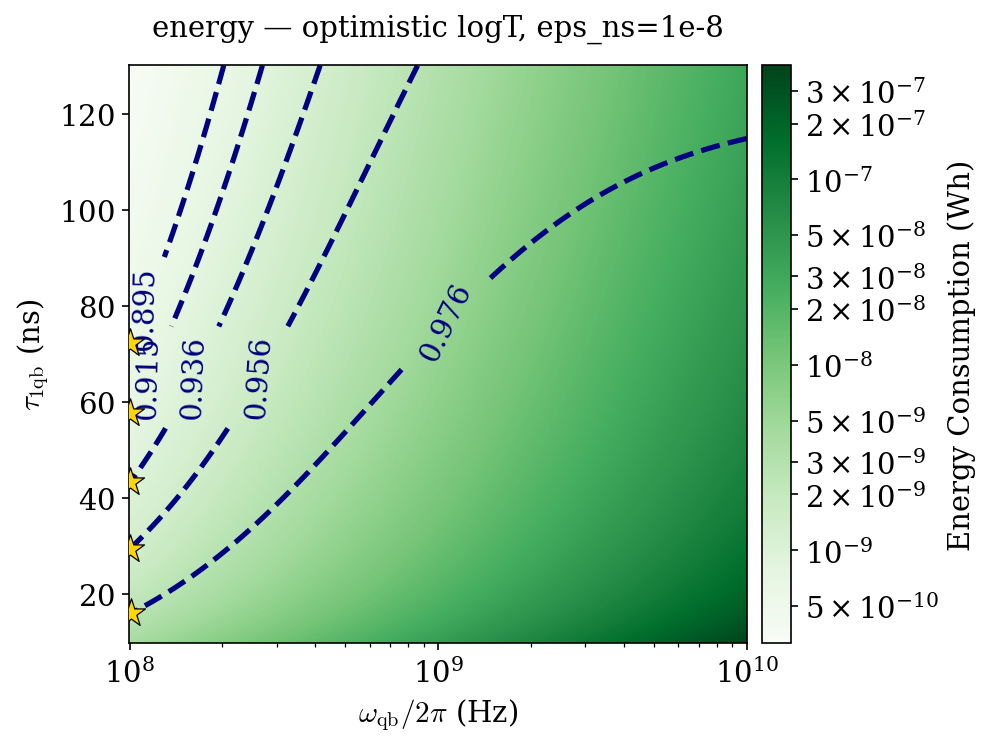}
	\caption{Same as \Cref{fig:app_omega_tau_eps7} for $\epsns = 10^{-8}$. Top left: Circuit fidelity with iso-fidelity contours and optima. Top right: average power. Bottom left: peak power. Bottom right: energy per query. With the idling error reduced, the energy optimum separates from the power optima: since $\Pavg \propto \omega/\tau_{1\mathrm{qb}}^{2}$ while $\Edyn \propto \omega/\tau_{1\mathrm{qb}}$, the energy favors lower frequencies and shorter gates. On the $\Fid = 0.976$ contour the minimum average power lies at $\omega/2\pi \approx 0.67\,\mathrm{GHz}$ and $\tau_{1\mathrm{qb}} \approx 63\,\mathrm{ns}$, and the minimum energy at $\omega/2\pi = 0.1\,\mathrm{GHz}$ and $\tau_{1\mathrm{qb}} \approx 16\,\mathrm{ns}$.}
	\label{fig:app_omega_tau_eps8}
\end{figure*}

The frequency as a control parameter also separates the energy optimum from the power optima, which in the $(g_{\mathrm{fac}}, \tau_{1\mathrm{qb}})$-plane coincide (see \Cref{sec:results}); the average and peak power, which share the same $\tau_{1\mathrm{qb}}^{-2}$ dependence, continue to select the same point. The dressed single-qubit power of \Cref{eq:single_qubit_power} scales as $\omega/\tau_{1\mathrm{qb}}^2$, so that at fixed $g_{\mathrm{fac}}$ the average power scales as $\omega/\tau_{1\mathrm{qb}}^{2}$ while the energy scales as $\omega/\tau_{1\mathrm{qb}}$. Lowering the frequency reduces both, but also raises $\bar{n}$, which must be compensated by a shorter gate to remain on the same iso-fidelity contour. Since the energy penalizes the shorter gate less than the average power does, the energy optimum moves to lower frequencies and shorter gates. The effect is clearly visible for $\epsns = 10^{-8}$ in \Cref{fig:app_omega_tau_eps8}, where the gate errors weigh more in the error budget. From \Cref{fig:app_omega_tau_eps8}, it can be seen that on the $\Fid = 0.976$ contour, the minimum average power lies at $\omega/2\pi \approx 0.67\,\mathrm{GHz}$ and $\tau_{1\mathrm{qb}} \approx 63\,\mathrm{ns}$, while the minimum energy lies at the lower edge of the limit, $\omega/2\pi = 0.1\,\mathrm{GHz}$ and $\tau_{1\mathrm{qb}} \approx 16\,\mathrm{ns}$. The choice of constraint---an application fidelity target, a wiring and cooling budget, or a total energy budget---then determines the operating point. We note that both optima lie below the frequency range of transmon qubits, and that the model keeps $\gamma$ independent of $\omega$; the optimization landscape is thus an illustration of how the resources separate once a design-time parameter enters, rather than a recommendation for the qubit frequency, which would require a more detailed model.

\section{Larger QRAM sizes}
\label{app:large_QRAM}

To assess how the conclusions of \Cref{sec:results} extend to larger memories, we repeat the MNR optimization for a BB-QRAM with $n = 20$ address qubits and $m = 1$, keeping all hardware parameters of \Cref{tab:fixed_values} and the control parameters' range unchanged. The results are collected in \Cref{tab:BB_iswap_MNR_data_n20}. The qualitative picture of \Cref{tab:BB_iswap_MNR_data} persists: the \cfgGeo{} configuration is the least expensive at every reachable target, followed by \cfgLogT{} and \cfgCold{}, and the optimal $g_{\mathrm{fac}}$ remains within $1$--$2.5$ across all cases.

The larger memory, however, shrinks the set of attainable fidelities and raises the resource cost substantially. The circuit contains roughly $2^{8} = 256$ times more gates and qubits than for $n = 12$, and a deeper schedule, so that both the idling and the gate errors accumulate over many more router-layer pairs. At $\epsns = 10^{-6}$, this accumulation exceeds the error budget of every target considered, and at $\epsns = 10^{-7}$ the target $\Fid \geq 0.99$ can no longer be reached. Where a target remains reachable, the reduced budget for gate errors requires shorter, higher-power gates, while the larger number of gates executed per layer adds to the power drawn. For $\Fid \geq 0.9$ with the \cfgGeo{} configuration and $\epsns = 10^{-7}$, for instance, the optimal gate duration decreases from $90.6\,$ns to $21.3\,$ns, and the minimum average power increases from $0.256\,$W to $700\,$W, i.e., by a factor of about $2700$. Reaching the fidelities required by applications at this scale therefore depends even more strongly on reducing the idling noise than on the choice of attenuation profile or operating point.

\begin{table*}[ht]
  \centering
\caption{Energetic comparison of the attenuation configurations and idling-noise strengths for the BB-QRAM with $n = 20$, $m = 1$, obtained with the same procedure as \Cref{tab:BB_iswap_MNR_data}. For each target fidelity, the table reports the minimum average power and the minimum energy per query over the sweep control parameters $g_{\mathrm{fac}} \in [1, 16]$, $\tau_{\mathrm{1qb}} \in [10, 130]\,\text{ns}$, together with the corresponding optimal $(g_{\mathrm{fac}}, \tau_{\mathrm{1qb}})$ and the resulting peak power. Entries marked ``not reachable'' indicate that no point in the sweep domain attains the target fidelity: this is the case for every target at $\epsns = 10^{-6}$, and for $\Fid \geq 0.99$ at $\epsns = 10^{-7}$.}

  \begin{tabular}{llrrr}
    \toprule
       &  & \cfgCold{} & \cfgGeo{} & \cfgLogT{} \\
    \midrule
    \multicolumn{5}{l}{\textit{$\epsns = 10^{-6}$}} \\
    \cmidrule(lr){1-5}
    $\Fid \geq 0.9$ & \multicolumn{4}{l}{not reachable} \\
    \cmidrule(lr){1-5}
    $\Fid \geq 0.95$ & \multicolumn{4}{l}{not reachable} \\
    \cmidrule(lr){1-5}
    $\Fid \geq 0.99$ & \multicolumn{4}{l}{not reachable} \\
    \midrule
    \multicolumn{5}{l}{\textit{$\epsns = 10^{-7}$}} \\
    \cmidrule(lr){1-5}
    \multirow{5}{*}{$\Fid \geq 0.9$} & Min avg.\ power (W) & 2480 & 700 & 1170 \\
     & Min energy ($\mu$J) & $2.63\times10^{5}$ & $7.36\times10^{4}$ & $1.23\times10^{5}$ \\
     & Optimal $g_{\mathrm{fac}}$ & 1.4  & 1.45 & 1.47 \\
     & Optimal $\tau_{\mathrm{1qb}}$ (ns) & 21.2 & 21.3 & 21.5 \\
     & Peak power (W) & $4.92\times10^{4}$ & $1.35\times10^{4}$ & $2.22\times10^{4}$ \\
    \cmidrule(lr){1-5}
    \multirow{5}{*}{$\Fid \geq 0.95$} & Min avg.\ power (W) & $1.03\times10^{4}$ & 2890 & 4820 \\
     & Min energy ($\mu$J) & $5.35\times10^{5}$ & $1.49\times10^{5}$ & $2.50\times10^{5}$ \\
     & Optimal $g_{\mathrm{fac}}$ & 1.29 & 1.45 & 1.39 \\
     & Optimal $\tau_{\mathrm{1qb}}$ (ns) & 10.1 & 10.5 & 10.4 \\
     & Peak power (W) & $2.15\times10^{5}$ & $5.57\times10^{4}$ & $9.59\times10^{4}$ \\
    \cmidrule(lr){1-5}
    $\Fid \geq 0.99$ & \multicolumn{4}{l}{not reachable} \\
    \midrule
    \multicolumn{5}{l}{\textit{$\epsns = 10^{-8}$}} \\
    \cmidrule(lr){1-5}
    \multirow{5}{*}{$\Fid \geq 0.9$} & Min avg.\ power (W) & 53.3 & 16.4 & 26 \\
     & Min energy ($\mu$J) & $3.86\times10^{4}$ & $1.13\times10^{4}$ & $1.84\times10^{4}$ \\
     & Optimal $g_{\mathrm{fac}}$ & 1.05 & 1.21 & 1.13 \\
     & Optimal $\tau_{\mathrm{1qb}}$ (ns) & 130 & 130 & 130 \\
     & Peak power (W) & 1310 & 364 & 609 \\
    \cmidrule(lr){1-5}
    \multirow{5}{*}{$\Fid \geq 0.95$} & Min avg.\ power (W) & 216 & 66.8 & 106 \\
     & Min energy ($\mu$J) & $7.77\times10^{4}$ & $2.28\times10^{4}$ & $3.72\times10^{4}$ \\
     & Optimal $g_{\mathrm{fac}}$ & 2.18 & 2.53 & 2.32 \\
     & Optimal $\tau_{\mathrm{1qb}}$ (ns) & 85.4 & 85.7 & 85.6 \\
     & Peak power (W) & 4110 & 1320 & 2040 \\
    \cmidrule(lr){1-5}
    \multirow{5}{*}{$\Fid \geq 0.99$} & Min avg.\ power (W) & 5540 & 1710 & 2710 \\
     & Min energy ($\mu$J) & $3.94\times10^{5}$ & $1.15\times10^{5}$ & $1.88\times10^{5}$ \\
     & Optimal $g_{\mathrm{fac}}$ & 2.42 & 2.24 & 2.15 \\
     & Optimal $\tau_{\mathrm{1qb}}$ (ns) & 17.6 & 16.1 & 16.4 \\
     & Peak power (W) & $1.08\times10^{5}$ & $3.29\times10^{4}$ & $5.15\times10^{4}$ \\
    \bottomrule
  \end{tabular}
  \label{tab:BB_iswap_MNR_data_n20}
\end{table*}

\section{Why the query leaves no trace in the routing tree}
\label{app:clean_query}

The bucket-brigade query of \Cref{fig:bucket-brigade-cswap} reads the memory in the phase basis: a Hadamard on the bus before the routing and after the unrouting, and a virtual $\mathrm{Z}^{m_i}$ on every leaf mode in place of the bit flips $\mathrm{X}^{m_i}$ of the original construction~\cite{giovannetti2008quantum,hann2021resilience}. This appendix explains, in steps, why the bit-flip read is not a valid query once the modes of the tree are qubits, and why the phase read is.

\paragraph{Step 1: the state after routing.}
After the $n$ address qubits have been routed, the router at level $d$ on the path holds $a_d$, every other router holds $|0\rangle$, and the bus has been carried along the path to the leaf mode of address $a$. Every other mode of the tree, in particular the $2^n-1$ other leaf modes, is in $|0\rangle$: the forward routing $V$ maps $|a\rangle|b\rangle|0\cdots0\rangle$ to $|a\rangle\otimes|b\rangle_{\text{leaf }a}\otimes|0\rangle_{\text{rest}}$, up to the content of the routers and the phase $i$ collected by a $|1\rangle$ at each iSWAP hop.

\paragraph{Step 2: what a bit-flip read does to the empty modes.}
The read must change the bus and nothing else. A gate $\mathrm{X}^{m_i}$ acting on leaf mode $i$ cannot know whether the bus is there: it also flips the empty modes with $m_i=1$. Unrouting with $V^\dagger$ cannot undo this, since $V^\dagger$ only reverses what $V$ produced; worse, a two-level router in $|0\rangle$ routes its left mode, so during the unrouting these excitations are carried upwards through the idle part of the tree. The pattern left behind depends on $a$. For $n=1$ and memory $(m_1,m_2)=(1,1)$, address $0$ leaves the right mode in $|1\rangle$ and address $1$ the left one in $|1\rangle$, so a query on $(|0\rangle+|1\rangle)/\sqrt2$ ends in
\begin{equation*}
    \tfrac{1}{\sqrt2}\big(|0\rangle|1\rangle_{\mathtt{T}}|0\,1\rangle_{\text{modes}} + |1\rangle|1\rangle_{\mathtt{T}}|1\,0\rangle_{\text{modes}}\big),
\end{equation*}
whose address register is a mixture of purity $1/2$ (purity $2^{-n}$ for an all-ones memory): the tree has recorded the address.

\paragraph{Step 3: read in the phase basis.}
Prepare the bus in the $\mathrm{X}$ basis, $\mathrm{H}|b\rangle=(|0\rangle+(-1)^b|1\rangle)/\sqrt2$, and apply $\mathrm{Z}^{m_i}$ to every leaf mode. Since $\mathrm{Z}|0\rangle=|0\rangle$, the empty modes are untouched whatever the memory contains, while on the leaf holding the bus $\mathrm{Z}^{m_a}$ multiplies the $|1\rangle$ component by $(-1)^{m_a}$, i.e., $\mathrm{H}|b\rangle\mapsto\mathrm{H}|b\oplus m_a\rangle$.

\paragraph{Step 4: the unrouting is exact.}
After the read, the state differs from the routed state of Step~1 only by a phase on the $|1\rangle$ component of the leaf holding the bus; it is therefore still in the image of $V$. Hence $V^\dagger$ returns every router and every mode to $|0\rangle$ and brings $\mathrm{H}|b\oplus m_a\rangle$ back to the bus. The hop phases cancel because the same component travels the same path down with iSWAP and up with iSWAP$^\dagger$ (with iSWAP in both halves they combine to a fixed power of $\mathrm{S}$ on the bus, a virtual correction). A final Hadamard yields $|a\rangle|b\oplus m_a\rangle$, the map of \Cref{eq:qram}, with the tree in the same product state for every $a$.

\paragraph{Step 5: cost.}
The $\mathrm{Z}^{m_i}$ are virtual gates (\Cref{ass:native_gates}): the $2^{n-1}$ NOT gates of the read and their layer disappear from the energetic cost and from the error budget, since the level-$(n+1)$ modes no longer receive a gate error. The price is one physical $\mathrm{R}_x(\tfrac{\pi}{2})$ per Hadamard ($\mathrm{H}=\mathrm{S}\,\mathrm{R}_x(\tfrac{\pi}{2})\,\mathrm{S}$) on the bus, in the first and the last layer of the circuit. For $m>1$ the same argument holds bit by bit, with $\mathrm{Z}^{x_{i,j}}$ on the $j$-th data mode of leaf $i$. The failure in Step~2 is not specific to two-level routers: any read that acts on a mode as if the bus were present creates an excitation in an empty mode; two-level routers merely spread it.

\section{Resource counts and schedule of the BB-QRAM}
\label{app:bb_counts}

In this appendix we describe how the resource counts from \Cref{tab:resources_table} were obtained and how the circuit is scheduled. A controlled-iSWAP is always the gadget of \Cref{fig:controlled_iswap}, and the root of the routing tree is level $0$. The closed forms were checked against the schedule compiled for the numerical results of this paper for $m=1$ and every $n\in\{2,\dots,14\}$ and $n=20$. That compiler routes a single bus qubit, so the dependence on $m$ is derived by hand, treating every additional bus qubit like an additional address qubit.

\paragraph*{Two-qubit gates.} A simple counting argument based on \Cref{fig:bucket-brigade-cswap} tells that the number of SWAP gates is
\begin{align*}
    \underbrace{2\sum_{i=0}^{n-1} 2^i}_{\rm route\,address} = 2^{n+1} - 2,
\end{align*}
while the number of CZ gates is
\begin{align*}
    \underbrace{2\cdot 5\sum_{i=2}^{n} (2^i-2)}_{\rm route\,address} + \underbrace{2\cdot 5\cdot m(2^{n+1}-2)}_{\rm route\,bus} \\
    = 20(m+1)2^n - 20(n+m) - 20,
\end{align*}
to a total of $(20m+22)2^n - 20(n+m) - 22$ two-qubit gates.

\paragraph*{$\mathrm{R}_x$-gates.} A controlled-iSWAP has $6$ ``internal'' $\mathrm{R}_x$ gates, in between the first and the last $\mathrm{CZ}$ of the gadget, and $2$ ``outer'' ones: $\mathrm{S}\,\mathrm{R}_x(\frac{\pi}{2})$ before and $\mathrm{R}_x(-\frac{\pi}{2})\,\mathrm{S}^\dagger$ after, which act on the same target $t_1$. The gadget is symmetric under the exchange of its two targets, so $t_1$ may be the \emph{incident} mode of the router (``in'', as drawn in \Cref{fig:controlled_iswap}) or the \emph{output} mode the routed qubit moves to (``out''); the other target carries the four $\mathrm{R}_x(\pm\frac{\pi}{4})$ and the control $\mathrm{CZ}$. Two outer pairs on the same mode cancel, $\mathrm{R}_x(-\frac{\pi}{2})\,\mathrm{S}^\dagger\,\mathrm{S}\,\mathrm{R}_x(\frac{\pi}{2})=\mathrm{I}$, whenever no other gate acts on that mode in between; idle time is irrelevant. A routing step at level $\ell$ consists of an open-controlled and a plain controlled-iSWAP on the same incident mode, towards the left and the right output mode of the router, respectively; the unrouting is the mirror image of the routing, and the bus, the last qubit routed, is the first one unrouted.

We place the outer pair on the output mode for the gadgets of even levels and on the incident mode for those of odd levels, with one exception: the last routing step of every qubit (the step right before an address qubit is loaded into the routers, and, for the bus, the step at the leaves) always uses the incident mode. The surviving outer rotations are then counted per mode:
\begin{itemize}
    \item A mode of an odd level $j$ receives the pairs of the gadgets that route a qubit into it (level $j-1$, even, pair on the output) and of those that route a qubit out of it (level $j$, odd, pair on the incident mode). These pairs are consecutive on the mode from the first routing step through it to the last unrouting step, so only the first leading and the last trailing rotation survive: $2$ per mode. The same holds for the wire of $a_2$ at level $0$, whose only step is its last one, and for the modes of level $n-1$ whatever the parity of $n$, since only the bus passes them.
    \item A mode of an even level $2\leq j\leq n-2$ is touched by no outer pair, except by the last routing step of the qubit $a_{j+2}$, which is parked at level $j+1$ and whose pair therefore sits on the incident mode. This pair is isolated, so its leading and trailing rotation survive in both halves of the circuit: $4$ per mode.
\end{itemize}
Adding the $2$ Hadamard rotations per bus qubit, the $\mathrm{R}_x$-count is
\begin{align*}
     &\underbrace{2\cdot 6 \sum_{i=2}^{n}(2^i-2)}_{\text{route\,address,\,internal}\,\mathrm{R}_x} + \underbrace{2\cdot 6\cdot m (2^{n+1}-2)}_{\text{route\,bus,\,internal}\,\mathrm{R}_x} \\
     &+ \underbrace{2\sum_{j=0}^{n-1} 2^j}_{\text{outer}\,\mathrm{R}_x} + \underbrace{2\sum_{j=2 ~\text{even}}^{n-2} 2^j}_{\text{outer}~\mathrm{R}_x,~\text{isolated steps}} + \underbrace{2m}_{\rm Hadamards}\\
     &= \left(24m + 26 + \frac{2^{\mathbf{1}[n\,\text{even}]}}{3}\right)2^n - 24n - 22m - \frac{86}{3},
\end{align*}
for instance $196{,}274$ internal gates and Hadamard rotations plus $10{,}918$ surviving outer rotations, $207{,}192$ in total, for $n=12$ and $m=1$. A placement shifted by one level (pair on the output for the odd levels) would remove $(2^n+2)/3-2n$ outer rotations for $m=1$ and even $n\geq 6$ ($1{,}342$ for $n=12$) at the same depth; an exhaustive search over all per-level placements for $n\leq 8$ finds no placement with fewer $\mathrm{R}_x$ gates, or with a smaller depth, than that one. We keep the placement above, which is the one underlying \Cref{tab:resources_table}.

\paragraph*{$\mathrm{R}_z$-gates.} The virtual $\mathrm{R}_z$ gates have four sources. Each open-controlled iSWAP has $1$ internal $\mathrm{Z}$ gate, according to the identity $(\mathrm{X}\otimes\mathrm{I})\mathrm{CZ}(\mathrm{X}\otimes\mathrm{I}) = \mathrm{CZ}(\mathrm{I}\otimes\mathrm{Z})$. Every surviving outer $\mathrm{R}_x$ comes with one $\mathrm{S}$ or $\mathrm{S}^\dagger$. The memory read adds one $\mathrm{Z}$ per $1$-entry; here we assume that half of the memory content is $1$-entries. Finally, each Hadamard $\mathrm{H}=\mathrm{S}\,\mathrm{R}_x(\frac{\pi}{2})\,\mathrm{S}$ contributes one $\mathrm{S}$ gates, $2$ per bus qubit for the two Hadamards. The $\mathrm{R}_z$-count is then
\begin{align*}
     &\underbrace{\sum_{i=2}^{n}(2^i-2)}_{\text{route~address,~open~controls}} + \underbrace{m (2^{n+1}-2)}_{\text{route\,bus,\,open\,controls}} + \underbrace{4m}_{\rm Hadamards} \\
     &+ \underbrace{2\sum_{j=0}^{n-1} 2^j}_{\text{outer}\,\mathrm{S},\mathrm{S}^\dagger} + \underbrace{2\sum_{j=2 \,\text{even}}^{n-2} 2^j}_{\text{outer}\,\mathrm{S},\mathrm{S}^\dagger,~\text{isolated steps}} 
    + \underbrace{m2^{n-1}}_{\rm memory\,read}\\
     &= \left(2.5m + 4 + \frac{2^{\mathbf{1}[n\,\text{even}]}}{3}\right)2^n - 2n + 2m - \frac{20}{3}.
\end{align*}

\paragraph*{Depth and schedule.} The circuit is compiled into layers of fixed content. A controlled-iSWAP takes $9$ layers (CZ; $\mathrm{R}_x$; CZ; $\mathrm{R}_x$; control CZ; $\mathrm{R}_x$; CZ; $\mathrm{R}_x$; CZ, not counting the outer rotations and the virtual gates), a routing step is an open-controlled and a plain controlled-iSWAP in a row ($18$ layers), and an iSWAP takes one layer. The schedule is level-synchronous: all $2^\ell$ routers of level $\ell$ act in the same layers. Consequently a layer holds at most $2^n$ single-qubit gates (the deepest routing step), and at most $\frac{5}{4}2^n$ qubits are active in a layer for $m=1$ and $n\geq 3$: the $2^{n-1}$ iSWAPs that load $a_n$ and, in the middle layer of the same logical step, the $2^{n-3}$ control CZs of the bus at level $n-3$.

As discussed in the main text, as soon as an address qubit has been routed two levels down, the next qubit can enter the root, since routing operations at levels of the same parity act on disjoint qubits. This takes $4$ sequential controlled-iSWAPs, i.e., $36$ layers, and the iSWAP that loads the previous qubit takes place in the middle layer of the first gadget of its successor, at no extra cost. The first half of the circuit, the routing, is therefore made of $20$ layers for $a_1$ and $a_2$ (one layer for the iSWAP that loads $a_1$, during which the Hadamard acts on the bus, $18$ layers for the single routing step of $a_2$, and one layer for its load), $36$ layers for each of $a_3,\dots,a_n$ and for every additional bus qubit, and $18n$ layers for the last bus qubit, which is routed through all $n$ levels. The memory read takes no layer, and the unrouting is the exact mirror image of the routing, which doubles the count.

Each surviving outer rotation is executed in an idle layer of its mode, between the previous gate acting on that mode and the gadget it belongs to (or between the gadget and the next gate), whenever there is one; the compiler prefers a layer without two-qubit gates, so that the widest layers do not grow. If there is none, one layer is inserted. This happens $n-1$ times in each half of the circuit:
\begin{itemize}
    \item once for the leading rotation on the right output mode of the root, which has to follow the load of $a_2$ immediately before $a_3$ enters;
    \item twice for every even level $2\leq j\leq n-2$: for the leading rotation of the step of $a_{j+2}$ on the left modes of level $j$, which the open-controlled iSWAP of the level above has just acted on, and for its trailing rotation on the right modes, which the plain iSWAP of the next qubit reaches immediately;
    \item for odd $n$, once more for the leading rotation of the bus on the left modes of level $n-1$.
\end{itemize}
Altogether,
\begin{align*}
    \mathrm{depth} &= 2\Big(\underbrace{20}_{a_1,\,a_2} + \underbrace{36\,(n+m-3)}_{a_3,\dots,a_n,\,\text{other bus qubits}} \\
    &\qquad\quad + \underbrace{18\,n}_{\text{last bus qubit}} + \underbrace{(n-1)}_{\text{outer}\,\mathrm{R}_x}\Big) \\
    &= 110n + 72m - 178,
\end{align*}
which is $1{,}214$ for $n=12$ and $m=1$.

\end{document}